%% file: ANA-JETM-2022-01-PAPER.tex
\documentclass[PAPER, atlasdraft=false, texlive=2023, UKenglish, cernpreprint, texmf, orcidlogo]{atlasdoc}
 
\usepackage{atlaspackage}
\usepackage{multirow}
\usepackage{atlasbiblatex}
\usepackage{listings}

\usepackage{atlasphysics}

\graphicspath{{logos/}{figures/}}
 
\usepackage{ANA-JETM-2022-01-PAPER-defs}

\AtlasTitle{New techniques for jet calibration with the ATLAS detector}

\AtlasVersion{1.10}
 
\AtlasAbstract{
A determination of the jet energy scale is presented using proton--proton collision data with a centre-of-mass
energy of $\sqrt{s}=13$ TeV, corresponding to an integrated luminosity of 140 \mbox{fb\(^{-1}\)} collected using the ATLAS detector
at the LHC\@. Jets are reconstructed using the ATLAS particle-flow method that combines charged-particle tracks and
topo-clusters formed from energy deposits in the calorimeter cells. The anti-$k_\mathrm{t}$ jet algorithm with radius parameter
$R=0.4$ is used to define the jet. Novel jet energy scale calibration strategies developed for the LHC Run 2 are reported
that lay the foundation for the jet calibration in Run 3. Jets are calibrated with a series of simulation-based corrections,
including state-of-the-art techniques in jet calibration such as machine learning methods and novel \textit{in situ} calibrations to achieve better performance than the baseline calibration
derived using up to 81 \mbox{fb\(^{-1}\)} of Run 2 data. The performance of these new techniques is then examined in the \textit{in situ} measurements by
exploiting the transverse momentum balance between a jet and a reference object. The $b$-quark jet energy scale using particle flow jets is
measured for the first time with around 1\% precision using $\gamma$+jet events.
}

\AtlasRefCode{JETM-2022-01}
 
\PreprintIdNumber{CERN-EP-2023-028}

\AtlasJournal{EPJC}
\AtlasJournalRef{Eur. Phys. J. C (2023) 83:761}
\AtlasDOI{10.1140/epjc/s10052-023-11837-9}

\hypersetup{pdftitle={ATLAS document},pdfauthor={The ATLAS Collaboration}}
 
\begin{document}
 
\maketitle
 
\tableofcontents
 
\section{Introduction}

The energetic proton--proton ($pp$) collisions produced by the Large Hadron Collider (LHC) yield final
states that are predominantly characterised by jets, which are collimated sprays of charged and neutral hadrons and their decay products.
Jets constitute an essential piece of the physics programme carried out using the ATLAS detector,
and a precise understanding of jet reconstruction is critical for a wide variety of processes.
Measurements of both the jet energy scale (JES) and resolution (JER) of these complex objects
are therefore essential for precision measurements of the Standard Model and for searches for new phenomena beyond it.
Several new methods are presented for improving the jet energy scale calibration and evaluating their performance in simulation and data,
paving the way to achieving a better precision on the JES for Run 3 and beyond.
These techniques were developed for jets reconstructed
with the \antikt algorithm~\cite{Cacciari:2008gp,Cacciari:2005hq} with radius parameter $R = 0.4$ using particle flow inputs~\cite{PERF-2015-09,JETM-2018-05}.
Previous  calibration strategies by the ATLAS Collaboration, that used up to 81~\ifb\
data~\cite{PERF-2011-03,PERF-2012-01,PERF-2014-02,PERF-2016-04,JETM-2018-05,JETM-2018-02}, are extended and improved  by
by taking advantage of the full Run 2 data sample of 140~\ifb.
 
The jet energy scale calibration consists of a series of calibration steps.
The first stage of the calibration uses simulation to derive corrections to the jet energy scale to reduce the impact of pile-up, detector effects, and other parameters.
The second stage of the calibration is a residual \insitu calibration, correcting for remaining differences between data and Monte Carlo (MC) simulation,
derived using well-measured reference objects, including photons and $Z$ bosons.
 
The structure of the paper is as follows.
Section~\ref{sec:atlas} describes the ATLAS detector, and Section~\ref{sec:evsel} describes the recorded data and the MC simulation samples,
and the inputs and algorithms used to reconstruct the jets.
Section~\ref{sec:mccalib} describes the methods used and the result of the simulation-based calibration,
Section~\ref{sec:insitu} describes the \textit{in situ} calibration, and conclusions are given in Section~\ref{sec:conclusion}.


\section{The ATLAS detector}
\label{sec:atlas}

\newcommand{\AtlasCoordFootnote}{
ATLAS uses a right-handed coordinate system with its origin at the nominal interaction point (IP)
in the centre of the detector and the \(z\)-axis along the beam pipe.
The \(x\)-axis points from the IP to the centre of the LHC ring,
and the \(y\)-axis points upwards.
Cylindrical coordinates \((r,\phi)\) are used in the transverse plane,
\(\phi\) being the azimuthal angle around the \(z\)-axis.
The pseudorapidity is defined in terms of the polar angle \(\theta\) as \(\eta = -\ln \tan(\theta/2)\).
Angular distance is measured in units of \(\Delta R \equiv \sqrt{(\Delta\eta)^{2} + (\Delta\phi)^{2}}\).}

The ATLAS detector~\cite{PERF-2007-01} at the LHC covers nearly the entire solid angle around the collision point.~\footnote{\AtlasCoordFootnote}
It consists of an inner tracking detector surrounded by a thin superconducting solenoid, electromagnetic and hadron calorimeters,
and a muon spectrometer incorporating three large superconducting air-core toroidal magnets.
 
The inner-detector system (ID) is immersed in a \SI{2}{\tesla} axial magnetic field
and provides charged-particle tracking in the range of  \(|\eta| < 2.5\).
The high-granularity silicon pixel detector covers the vertex region and typically provides four measurements per track,
the first hit normally being in the insertable B-layer (IBL) installed before Run~2~\cite{ATLAS-TDR-19,PIX-2018-001}.
It is followed by the silicon microstrip tracker (SCT), which usually provides eight measurements per track.
These silicon detectors are complemented by the transition radiation tracker (TRT),
which enables radially extended track reconstruction up to \(|\eta| = 2.0\).
The TRT also provides electron identification information
based on the fraction of hits above a higher energy-deposit threshold corresponding to transition radiation.
 
The calorimeter system covers the pseudorapidity range of  \(|\eta| < 4.9\).
In the region \(|\eta|< 3.2\), electromagnetic calorimetry is provided by barrel and
endcap high-granularity lead/liquid-argon (LAr) calorimeters,
with an additional thin LAr presampler covering \(|\eta| < 1.8\)
to correct for energy loss in material upstream of the calorimeters.
Hadron calorimetry is provided by the steel/scintillator-tile calorimeter,
segmented into three barrel structures within \(|\eta| = 1.7\), and two copper/LAr hadron endcap calorimeters.
The solid angle coverage is completed with forward copper/LAr and tungsten/LAr calorimeter modules
optimised for electromagnetic and hadronic energy measurements respectively.
 
The muon spectrometer (MS) comprises separate trigger and
high-precision tracking chambers measuring the deflection of muons in a magnetic field generated by the superconducting air-core toroidal magnets.
The field integral of the toroids ranges between \num{2.0} and \SI{6.0}{\tesla\metre}
across most of the detector.
Three layers of precision chambers, each consisting of layers of monitored drift tubes, cover the region \(|\eta| < 2.7\),
complemented by cathode-strip chambers in the forward region, where the background is highest.
The muon trigger system covers the range of  \(|\eta| < 2.4\) with resistive-plate chambers in the barrel, and thin-gap chambers in the endcap regions.
 
Interesting events are selected by the first-level trigger system implemented in custom hardware,
followed by selections made by algorithms implemented in software in the high-level trigger~\cite{TRIG-2016-01}.
The first-level trigger accepts events from the up-to \SI{40}{\MHz} bunch crossings at a rate below \SI{100}{\kHz},
which is further reduced by the high-level trigger  to record events to disk at about \SI{1}{\kHz}.
 
An extensive software suite~\cite{ATL-SOFT-PUB-2021-001} is used in the reconstruction and analysis of real
and simulated data, in detector operations, and in the trigger and data acquisition systems of the experiment.


\FloatBarrier
 
\section{Data samples and simulated events}
\label{sec:evsel}

\subsection{Data samples}
 
The data used for the calibration study were collected by ATLAS in the $pp$ collisions at $\sqrt{s}=13$~\TeV~ from 2015 to 2018 with all subdetectors operational,
corresponding to an integrated luminosity of \SI{140}{\per\femto\barn}. The proton bunch crossing interval was \SI{25}{\nano\second} during the data taking. The average number of interactions
per bunch crossing (\avint) for the Run 2 data is 34.
These conditions, and those of the detector configuration and reconstruction thresholds, are taken into account in producing and reconstructing simulated data~\cite{DAPR-2018-01}.

\subsection{Monte Carlo simulation}
 
The Monte-Carlo (MC) based calibration uses MC simulated dijet and multijet events.
~\Pythia{ v8.230} ~\cite{Sjostrand:2014zea} is used as the nominal MC generator for simulating dijets.
Samples of $2 \rightarrow 2$ dijet events are simulated using the A14 tune ~\cite{ATL-PHYS-PUB-2014-021}
and the  \NNPDF[2.3] ~\cite{Ball:2012cx} parton distribution function (PDF) set.
Decays of heavy-flavour hadrons are modeled using EvtGen~\cite{Lange:2001uf}.
 
Several alternative samples are used to study the impact of the MC simulation on the calibration, and to determine uncertainties based on this.
Two different dijet samples are simulated using the \SHERPA[2.2.5]~\cite{Bothmann:2019yzt} generator.
The matrix element calculation was included for the $2\rightarrow2$ process at leading-order, and the default \SHERPA parton
shower~\cite{Schumann:2007mg} based on Catani--Seymour dipole factorisation was used for the showering with \pT ordering, using the \CT[14nnlo] PDF set~\cite{Dulat:2015mca}.
The first of these samples made use of the dedicated \SHERPA AHADIC model for hadronisation~\cite{Winter:2003tt}, based on cluster fragmentation.
A second sample was generated with the same configuration but using the \SHERPA interface to the Lund string fragmentation model of \PYTHIA[6]~\cite{Sjostrand:2006za} and its decay tables.
These two sets of samples were used to evaluate uncertainties stemming from the hadronisation modelling.
 
Two sets of samples are simulated using \Herwig~7.1.6~\cite{Bahr:2008pv}~\cite{Bellm:2015jjp}~\cite{Bellm:2017bvx} with the \NNPDF[2.3lo] PDF set.
These samples include the $2 \rightarrow 2$ process at the matrix element level,
and either the default angular-ordered parton shower~\cite{Gieseke:2003rz} or a dipole parton shower~\cite{Platzer:2009jq,Platzer:2011bc}
using splitting kernels based on the Catani--Seymour subtraction scheme~\cite{Catani:1996vz,Catani:2002hc},
and in both the cases, using a cluster hadronisation.
 
Alternative samples of multijet production at NLO accuracy were produced with \POWHEGBOX[v2]~\cite{Nason:2004rx, Frixione:2007vw} interfaced to \PYTHIA[8].
These were generated with the dijet process as implemented in \POWHEGBOX[v2]~\cite{Alioli:2010xd}.
The \pT of the underlying Born configuration was taken as the renormalisation and factorisation scales and the \NNPDF[3.0nlo]~\cite{Ball:2014uwa} PDF was used.
\PYTHIA with the A14 tune and the \NNPDF[2.3lo] PDF was used for the shower and multi-parton interactions.
These samples included per-event weight variations for different perturbative scales in the matrix element,
different parton distribution functions and their uncertainties, and the \PYTHIA perturbative shower uncertainties.
This multijet sample is used as the nominal MC sample for \textit{in situ} $\eta$-intercalibration,
while \POWHEGBOX[v2]~\cite{Nason:2004rx, Frixione:2007vw} interfaced to \Herwig~\cite{Bellm:2017bvx} with an angular-ordered parton shower~\cite{Gieseke:2003rz} and the \NNPDF[2.3lo] PDF is used as an alternative sample.

For the \textit{in situ} $Z+$jet analysis, samples of $Z$ boson  with jets ($Z$+jet)
are produced with~\MADGRAPH~+~\Pythia{8}~\cite{Alwall:2014hca} using
the \NNPDF[3.0nnlo] PDF set~\cite{Bothmann:2019yzt} and the AZNLO set of tuned parameters~\cite{ATL-PHYS-PUB-2013-017}. \Sherpa~2.2.11
is used as the alternative MC sample.
The nominal $\gamma+$jets samples are generated with ~\Pythia~v8.230 ~\cite{Sjostrand:2014zea} using the A14 set of tuned parameters ~\cite{ATL-PHYS-PUB-2014-021}
and \NNPDF[2.3] PDF set. The $\gamma$+jet samples are produced for the direct photon component and
the fragmentation photon component separately. The alternative sample used in $\gamma$+jet events is~\Sherpa~2.2.2 with \NNPDF[3.0nnlo] PDF set~\cite{Bothmann:2019yzt}.
 
All samples are reconstructed using a full detector simulation and superimposed minimum-bias interactions simulated using \Pythia8 with the
A3 set of tuned parameters~\cite{ATL-PHYS-PUB-2016-017} and \NNPDF[2.3lo] PDF set to represent multiple $pp$ interactions during the same or nearby bunch crossings (pile-up).
The distribution of the average number of pile-up interactions in simulation is reweighted during data analysis to match that observed in the Run 2 data.

\subsection{Jet reconstruction}

The jets in these studies are reconstructed with the \antikt algorithm with a radius parameter $R = 0.4$
as implemented in the \Fastjet software package~\cite{Fastjet}.
Four-momentum objects are used as inputs to the algorithm, and may be particles at the generator level of the MC , charged-particle tracks, calorimeter energy deposits,
or algorithmic combinations of the latter two, as in the case of the particle-flow (PFlow) reconstruction technique.
Particles at the MC generator level are referred to as truth particles.
Reconstructed jets use PFlow objects (PFOs) as inputs to jet reconstruction,
which combine measurements from the tracker and the calorimeter to form the input signals for jet reconstruction.
Specifically, energy deposited in the calorimeter by charged particles is subtracted from the observed topo-clusters and replaced by the momenta of tracks that are matched to those topo-clusters,
as described in Ref.~\cite{PERF-2015-09}, and with the updates described in Ref.~\cite{JETM-2018-05}.
These resulting PFlow jets show improved energy and angular resolution, reconstruction efficiency, and pile-up stability compared with jets reconstructed using only calorimeter information.

Charged particle tracks are used for both the PFlow reconstruction and for deriving calibrations.
These tracks are reconstructed in the full acceptance of the inner detector $|\eta| < 2.5$,
and are required to have a $\pt > 500$~\MeV\ unless otherwise stated, and must satisfy criteria based on the number of hits in the ID subdetectors.
In addition, tracks must satisfy $|z_{0} \mathrm{sin} \theta| < 2$~mm, where $z_0$ is the distance of closest approach of the track to the hard-scatter primary vertex along the $z$-axis.
Tracks used in the calibration are matched to jets using ghost association, a procedure that treats them as four-vectors of infinitesimal magnitude during the jet reconstruction
and assigns them to the jet with which they are clustered~\cite{Cacciari:2007fd}.
 
Generator-level jets, referred to as truth jets, are reconstructed using stable final-state particles, defined as those with $c\tau > 10$ ~mm, excluding muons, neutrinos, and particles from pile-up interactions.
Generator-level jets are selected with $\pt > 7$~\GeV\  and $|\eta| < 4.5$, while reconstructed jets used for the MC calibration are selected with $|\eta| < 4.5$.


\FloatBarrier
 
\section{The simulation-based calibration}
\label{sec:mccalib}

 
This section details the simulation-based jet energy scale calibration, which restores the average jet energy to that of truth jets.
The event selection for these steps is described in Section~\ref{sec:mc_evSel}, and the calibration is done in four steps.
The first two steps apply pile-up corrections to remove the excess \pt due to additional $pp$ interactions in the same (in-time) or nearby (out-of-time) bunch crossings.
The first pile-up correction, described in Section~\ref{sec:area_pu},
applies a subtraction based on the median \pt density measured in the event and the jet area (the `pile-up density correction')~\cite{Cacciari:2007fd,Cacciari:2008gn},
minimizing the sensitivity to the model of the pile-up used in the simulation.
Next, a correction for residual dependence on the number of reconstructed primary vertices in the event ($\npv$) and \avint
is applied (the `residual pile-up correction'), based on corrections derived using simulated samples, as described in Section~\ref{sec:residual_pu}.
The third step, the absolute JES calibration detailed in Section~\ref{sec:mcjes}, corrects jets so that they agree, on average, in energy and direction with truth jets from dijet MC events.
Finally, the global calibration improves the jet \pt resolution and related uncertainties by reducing the dependence of the reconstructed jet response
on observables constructed using information from the tracking, calorimeter, and muon chamber detector systems, as introduced in Section~\ref{sec:gnnc}.

\subsection{Event selection}
\label{sec:mc_evSel}
 
All stages of the simulation-based jet energy scale calibration use the same event selection.
The MC simulation is used to determine the energy scale and resolution of jets by comparing PFlow jets with truth jets.
Truth and reconstructed jets are required to satisfy $|\eta| < 4.5$ to be fully contained in the detector acceptance,
and truth jets are additionally required to have $\pt > 7$~\GeV.
Uncalibrated jets have a positive energy, but can become negative in energy after applying the corrections described in Sections~\ref{sec:area_pu}--~\ref{sec:residual_pu}.
Biases are reduced in the determination of the average jet energy response ($E_{\mathrm{reco}}/E_{\mathrm{true}}$) at low energies,
by requiring reconstructed jets to have $\pt > 0$ after the pile-up density correction described in Section~\ref{sec:area_pu},
but making no requirement on the \pt after the correction described in Section~\ref{sec:residual_pu}.
 
Events are required to have at least one reconstructed primary vertex with at least two matched tracks with $\pt > 500$~\MeV.
In simulated reconstructed events, the primary vertex is the reconstructed primary vertex with the largest sum of squared track momentum,
while at the truth level, the primary vertex corresponds to that of the simulated hard-scatter process,
and not the collision with the highest momentum transfer.
This results in some events where the pile-up collision has a larger momentum transfer than the hard-scatter collision.
For MC samples, the reconstructed and truth primary vertices are required to have $z$-positions within $0.2$~mm of each other.
Events are required to have at least two reconstructed jets, and at least one truth jet.
Truth jets are geometrically matched to PFlow jets using the angular distance $\Delta R$ with the requirement $\Delta R < 0.3$.
In addition, truth jets are required to be isolated from all other truth jets by $\Delta R > 1.0$,
while reconstructed jets are required to be isolated from all reconstructed jets by $\Delta R > 0.6$.
To reduce the contribution of events where the pile-up collision has a larger momentum transfer than the hard-scatter collision,
the average \pt of the two leading reconstructed jets is required to be no larger than $1.4 \times \ptscale{truth, leading}$.

 
\subsection{Estimating the median \pt density and the pile-up density correction}
\label{sec:area_pu}
 
Pile-up interactions change the jet energy scale, and jet reconstruction is affected by additional $pp$ interactions in the same or nearby bunch crossings.
The first stage of the jet calibration, referred to as the `pile-up density correction',
subtracts the expected contribution from pile-up based on the area of the jet and the median \pt density in the event~\cite{Cacciari:2007fd}.
To compute the jet area $A$, a dense, uniformly distributed in $\eta\times\phi$ population of infinitesimally soft ghost particles is overlaid on top of the event.
Then, $A$ is defined as the transverse momentum of the sum of the four-momenta of all ghost constituents matched with a given jet after clustering,
normalised by the ghost constituent transverse momentum density.

The median pile-up \pt density $\rho$ is estimated for each event
by the median \pt density ($\pt/A$) of all jets clustered with the $k_t$ algorithm~\cite{Ellis:1993tq,Catani:1993hr} with a radius parameter of $0.4$:
\begin{equation*}
\rho = \text{median} \Bigg\lbrace \frac{p_{\mathrm{T},i}^{\mathrm{jet}}}{A_i^{\mathrm{jet}}} \Bigg\rbrace,
\end{equation*}
where the index $i$ enumerates over the jets.
For this calculation, only jets with $|\eta| < 2$ are used, since $\rho$ falls off steeply beyond this region, due to a combination of physics and detector effects.
 
Assuming the pile-up is a uniform, diffuse background,
the pile-up contribution to the energy of the jet can then be approximated by the product of the jet area times the median \pt density.
The pile-up density-corrected jet \pt, \ptscale{area}, is then defined as
\begin{equation*}
\ptscale{area} = \pt - \rho \times A.
\end{equation*}
The ratio of the \ptscale{area} to the uncorrected jet \pt is applied as a scale factor to the jet four-momentum and does hence not affect its direction.
 
Previously, the inputs to the $\rho$~calculation were the same as those used to build jets:
the neutral \pflow{} objects and the charged \pflow{} objects that satisfy $|z_0\sin\theta|<2$~mm~\cite{JETM-2018-05}.
However, this results in a bias from the inclusion of hard-scatter tracks, shifting the median \pt density to higher values, particularly when the hard-scatter process has a large jet multiplicity.
To prevent such biases, the `pile-up sideband' (PUSB) $\rho$ definition is studied,
which uses neutral \pflow{} objects, and  charged \pflow{} objects that satisfy $2 ~\mathrm{mm}~ <|z_0\sin\theta|<4$~mm  as inputs.
The total amount of pile-up using the sideband cuts is expected to be similar to the nominal criteria, since a similar amount of pile-up will meet these criteria.
This ensures a minimal loss of the event-by-event correlation of the charged pile-up component that is not removed by charged hadron subtraction cuts.

The JES is measured in specific event selections as described in Section~\ref{sec:insitu}, but these calibrations are applied to many final states.
An uncertainty is required to cover potential inadequacy of the model used in the simulation of the difference between the bias in $\rho$~in different event topologies,
specifically the difference between the bias in $\rho$ between dijet and $Z(\rightarrow \mu\mu)+$jets events.
The $Z(\rightarrow \mu\mu)+$jets events are distinctive from dijet events in several ways, including the quark-gluon composition, colour flow, and momentum transfer of the process,
making it a good topology to use to estimate the magnitude of a potential bias.
This bias at a given $\mu$ is estimated by comparing the difference between the value of $\langle\rho{}\rangle$ for the data sample
in two different event topologies to measure potential inadequacy of the model used in the MC simulation of the data:
\begin{equation*}
\Delta^{\rho}(\mu) = \left(\rho(\mu)^{\textrm{dijet}}_{\textrm{MC}} - \rho(\mu)^{\textrm{Z+jet}}_{\textrm{MC}} \right)
- \left(\rho(\mu)^{\textrm{dijet}}_{\textrm{data}} - \rho(\mu)^{\textrm{Z+jet}}_{\textrm{data}} \right),
\end{equation*}
and the bias that is propagated to the jet energy scale uncertainties is the bias determined at the average value of $\mu$ for the data sample:
\begin{equation*}
\textrm{bias [\GeV]} = \Delta_\rho(\langle\mu\rangle).
\end{equation*}
 
The $Z(\rightarrow\mu\mu)+$jets~selection uses the lowest unprescaled single muon trigger, and requires two muons
with $\pT^{\mu1}>30~\GeV, \pT^{\mu2}>25$~\GeV, $80<m^{\mu\mu}<100$~\GeV, and $\pT^{\mu\mu}>25$~\GeV,
and the dijet selection uses the lowest unprescaled single jet trigger, and requires a leading jet with $\pT>500$~\GeV, $|\eta|<2.4$, and
greater than 5\% of the momentum carried by charged particle flow objects.
 
Figure~\ref{fig.rho.rho} shows the dependence of $\langle \rho \rangle$~in the two processes as a function of $\mu$~for data and simulation,
comparing both the $\rho$ definitions described above.
The lower panels compare the values of $\langle \rho \rangle$~in the two processes.
Two different \Sherpa dijet samples are shown: a 2.1.1 sample~\cite{Gleisberg:2008ta} that was used in the previous
calibration~\cite{JETM-2018-05}, and the  2.2.5 sample used now,
while for $Z(\rightarrow\mu\mu)+$jets, only \Sherpa 2.2.1 is used.
The \Sherpa~2.2.X samples include an improvement to the multi-parton interaction (MPI) model, which directly affects the bias in $\rho$.
Significantly larger differences are seen between the dijet \Sherpa~2.1.1 sample and the $Z(\rightarrow\mu\mu)+$jets \Sherpa~2.2.1 sample
than between the dijet \Sherpa~2.2.5 sample and the $Z(\rightarrow\mu\mu)+$jets \Sherpa~2.2.1 sample,
Previously, the bias was determined using the dijet \Sherpa~2.1.1 sample and the $Z(\rightarrow\mu\mu)+$jets \Sherpa~2.2.1 sample, which have different MPI models.
Using the updated dijet \Sherpa sample that uses a consistent MPI model with $Z(\rightarrow\mu\mu)+$jets results in a factor of four reduction in the bias,
showing the importance of MPI modelling in MC simulations.
The new $\rho$ definition, $\rho^{\textrm{PUSB}}$, results in significantly smaller differences between the different topologies, and a better description of the data by the simulation.
Similarly, the improvements to the $\rho$ definition result in almost a factor of three improvement to the uncertainty,
as seen by the difference between data and \Sherpa for the two different $\rho$ definitions.
Together, these improvements reduce the JES uncertainty from the $\rho$ modelling by a factor of nearly seven.

\begin{figure}
\centering
\subfloat[]{\includegraphics[width=0.49\textwidth]{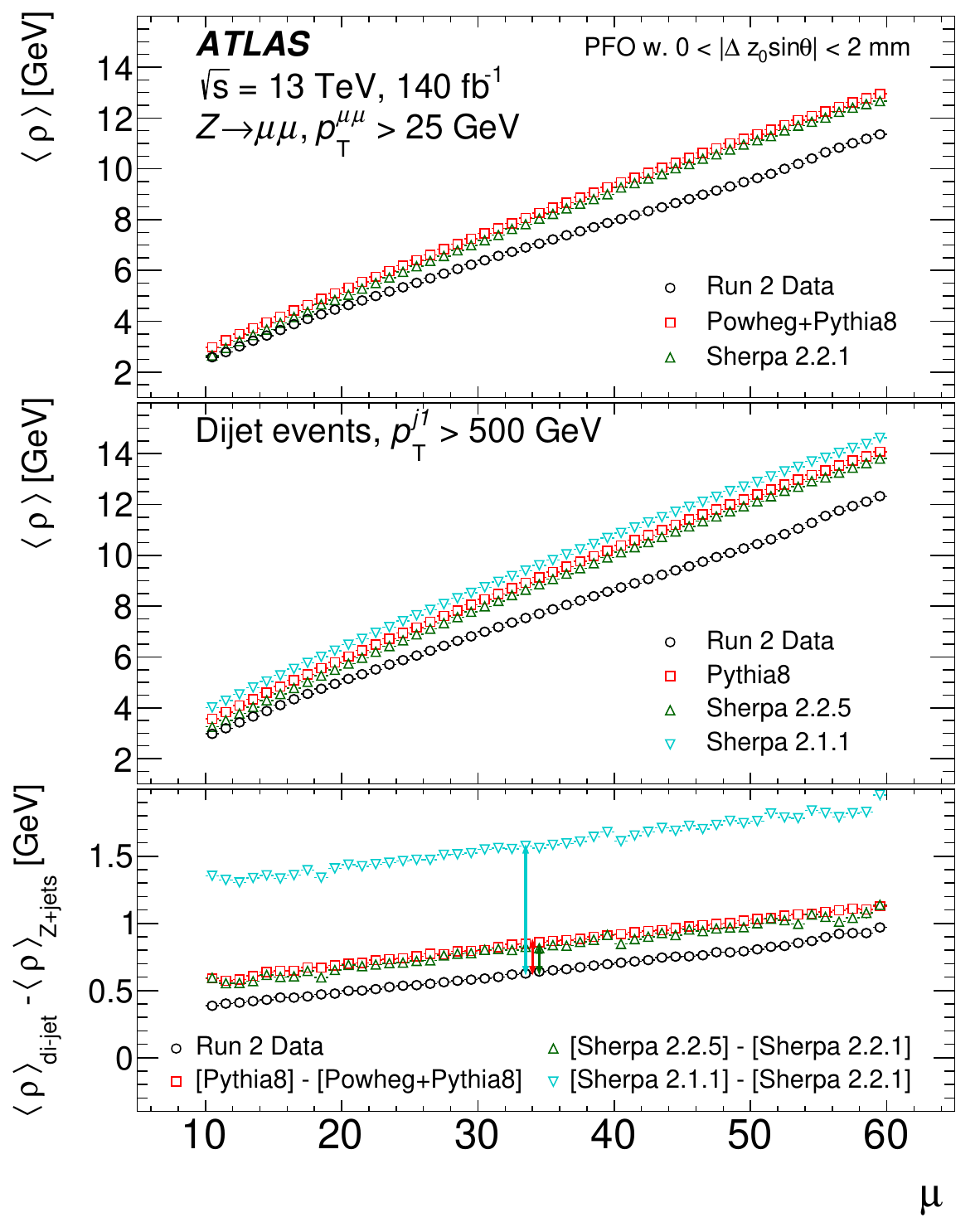}}
\subfloat[]{\includegraphics[width=0.49\textwidth]{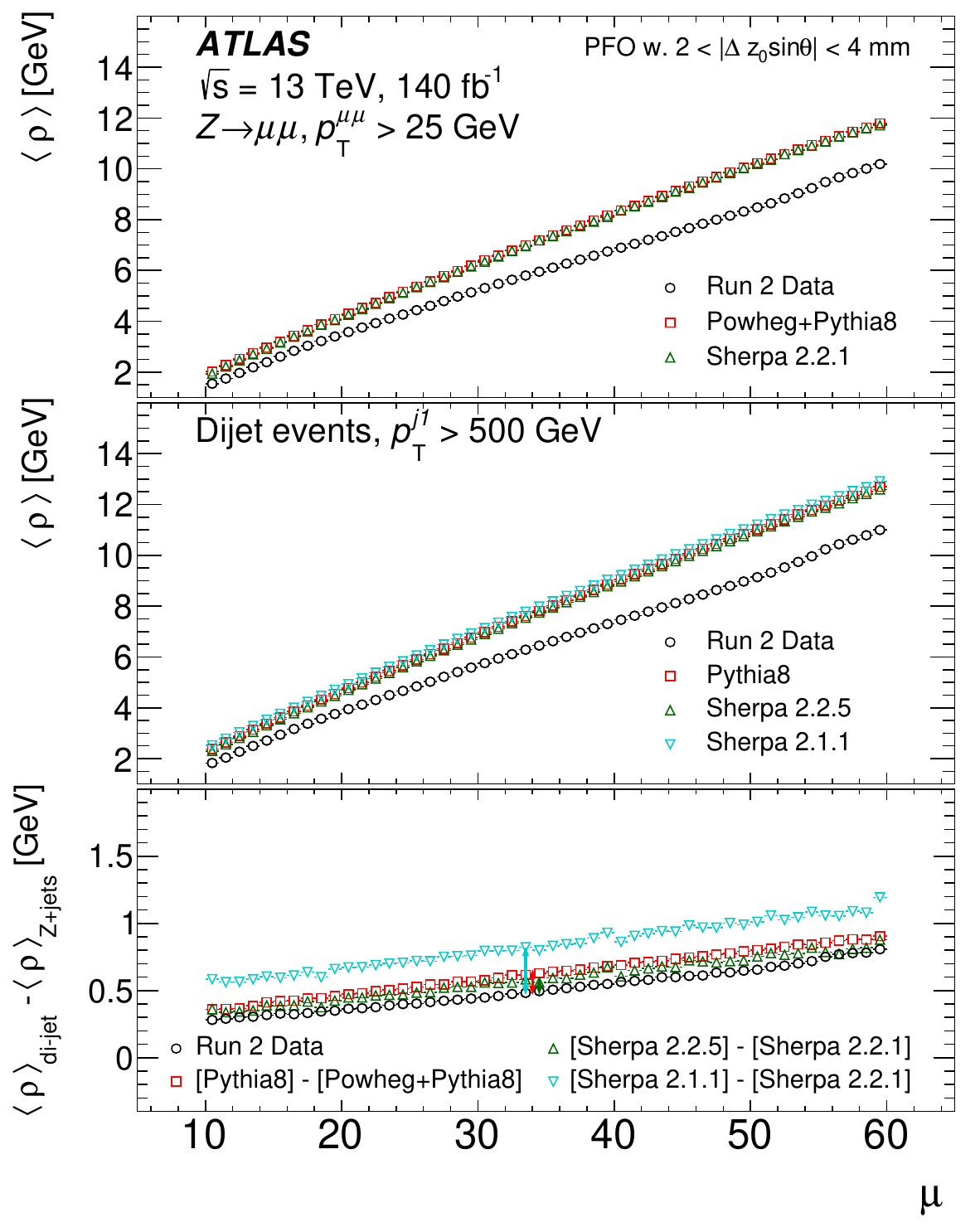}}
\caption{The distribution of $\rho$ as a function of $\avint$ for the (top) $Z(\rightarrow\mu\mu)$+jets and (middle) dijet
selections for data, \Pythia8, \Sherpa~2.2.5, and \Sherpa~2.1.1.
The lower panel shows the difference between the two topologies which is used to determine the uncertainty from
the extrapolation across topologies, indicated by the vertical arrows.
The left plot shows $\rho$ built from the jet constituents: neutral PFOs and charged PFOs with $|z_0\sin\theta|<2$mm,
and the right plot shows $\rho$ built using neutral PFOs and charged PFOs satisfying the new sideband selection.}
\label{fig.rho.rho}
\end{figure}

\subsection{Residual pile-up correction}
\label{sec:residual_pu}
 
To further reduce the impact of pile-up, a residual pile-up correction is applied, based on \npv, $\mu$,  the reconstructed jet $\pt$ (\ptscale{reco}), and
the reconstructed jet $\eta$ (\etascale{reco}).
Due to the fast response of the silicon tracking detectors used to reconstruct the tracks used to find the primary vertices,
\npv is sensitive to the in-time pile-up, while $\mu$ is sensitive to the out-of-time pile-up, since it accounts for the average amount of pile-up around a given bunch crossing.
Typically, in-time pile-up increases the energy of the jet, and out-of-time pile-up decreases it.
The negative dependence of the jet energy scale on $\mu$ for out-of-time pile-up is a result of the liquid-argon calorimeter's pulse shape,
which is negative during the period soon after registering a signal~\cite{Abreu_2010}.
Two options for the residual pile-up correction are compared.
 
\subsubsection{The 1D residual pile-up correction}
 
The first strategy, referred to as the `1D residual pile-up calibration', follows the method outlined in Ref.~\cite{JETM-2018-05},
where additional corrections are applied based on the $\mu$ and \npv of the event, with
\begin{equation*}
\ptscale{1D residual} = \ptscale{area} - (\ptovernpv) \times (\npv-1)- (\ptovermu) \times \mu,
\end{equation*}
where \ptovernpv and \ptovermu are determined as follows.
To determine \ptovernpv, first, the dependence of $\dpt = \ptscale{area}-\ptscale{truth}$ on \npv is fit with a line in bins of $\mu$, \ptscale{truth}, and \pseudorap.
The slope of this function is taken as the dependence \ptovernpv per $\mu$ bin.
The average of these slopes across  $\mu$ is taken to be the \pt dependence on \npv for a given \ptscale{truth} and \pseudorap bin.
For each \pseudorap bin, the average \pt dependence is fit as a function of \ptscale{truth} with a logarithmic function for $20~\GeV~ < \ptscale{truth} < 200$~\GeV.
The value of the logarithmic fit at 25~\GeV~ is taken as the nominal correction, since pile-up effects are most relevant for low-\pt jets.
Finally, a piecewise linear function is fit over the per $|\pseudorap|$ bin values of \ptovernpv,
reducing statistical fluctuations and providing a continuous correction over the full \pseudorap range.
To determine \ptovermu, the same process is repeated with $\npv$ and $\mu$ switched,  to get the dependence of \dpt on $\mu$.

\subsubsection{The 3D residual pile-up correction}
 
The 1D residual pile-up correction does not account for correlations between \avint and \npv, and
does not account for changes in the pile-up contribution as a function of jet \pt.
The 3D residual pile-up correction is designed to include these correlations.
In this calibration, derived in bins of \etascale{reco}, the jet \pt scale is shifted to match the truth jet scale
as a function of (\npv, \avint, \ptscale{area}),
simultaneously correcting for pile-up and detector effects.
The \ptscale{truth} is used as a reference to compute a correction given by \dpt.
For extreme values of $\mu$ and \npv, where there are insufficient events to determine an accurate correction,
the correction is determined using the closest non-empty ($\mu$, \npv) bin (with the same \pt), and the result is smoothed.
This average difference, \dpt, is fit as a function of  \ptscale{area} using a linear plus logarithmic function, in bins of \npv, \avint, and \ptscale{reco},
determined using jets with $10~\GeV~ < \ptscale{truth} < 200$~\GeV.
The corrected value is given by
\begin{equation*}
\ptscale{3D residual} = \ptscale{area} - \dpt(\npv,\mu,\ptscale{area}).
\end{equation*}
 
By construction, this residual pile-up calibration corrects the jet energy scale to the truth jet scale, combining corrections due to pile-up with corrections due to detector effects.
This is contrasted with the 1D residual pile-up correction, which is designed to exclusively remove the impact of pile-up on the jet \pt scale.
Several options to only correct for the pile-up \pt were studied,
but these were found to either increase the pile-up dependence or result in problematic effects such as a large fraction of jets with negative \pt.

\subsubsection{Comparison of the different residual pile-up corrections}
 
A comparison of the different options for the residual pile-up corrections is shown in Figure~\ref{fig:closure}.
As seen in this figure, the residual pile-up calibration is especially useful for improving the pile-up dependence for jets with $|\etascale{reco}| > 2.5$.
Overall, for the 1D residual pile-up correction, the absolute pile-up dependence increases for higher \pt jets, but the relative impact on the \pt response is smaller.
While the 1D residual pile-up correction performs best for the \pt range which it is optimised for (20--30~\GeV), it has a sizeable pile-up dependence at other jet \pt.
In addition, since the 1D residual pile-up correction is optimised for this same bin, its performance appears enhanced by construction,
while a more differential binning would show a worse performance.
The 3D residual pile-up correction significantly reduces the pile-up dependence of the calibration, particularly at high \pt.
Based on these results, the 3D residual pile-up calibration is used for the remainder of the reported studies.

\begin{figure}[ht]
\centering
\subfloat[]{\includegraphics[width=0.49\textwidth]{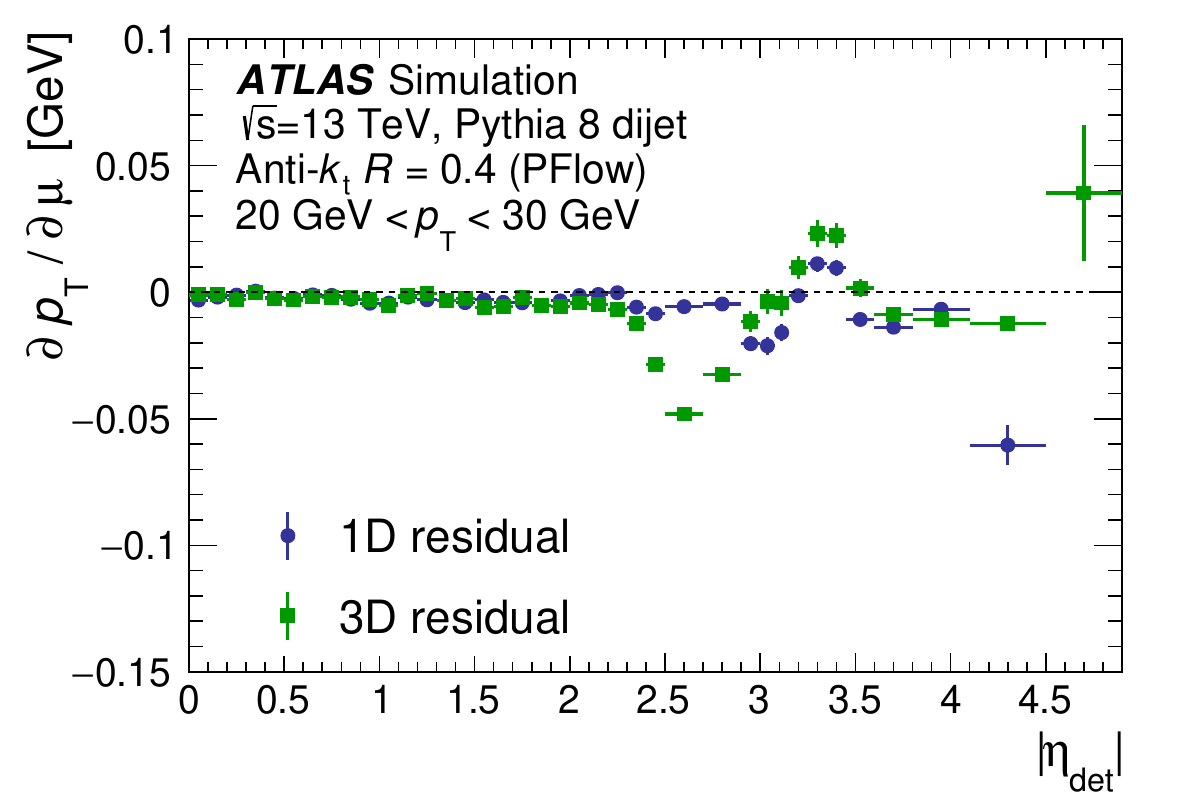}}
\subfloat[]{\includegraphics[width=0.49\textwidth]{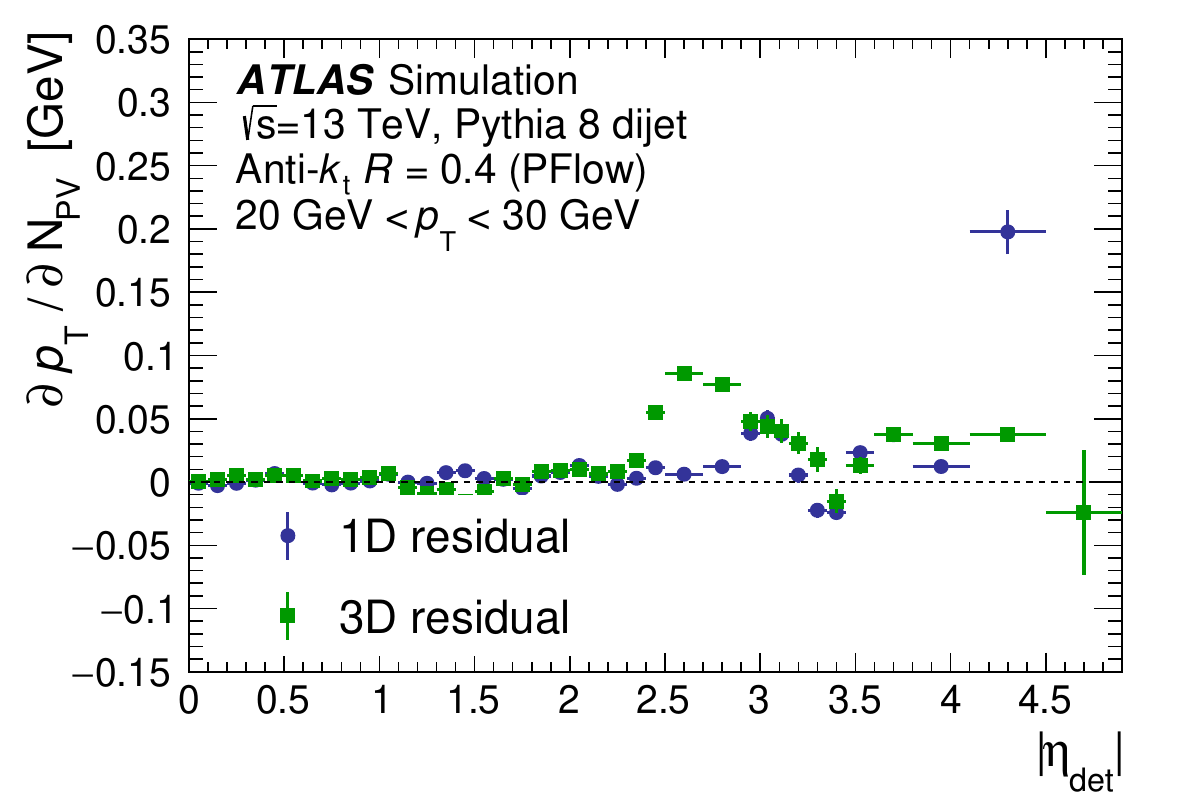}} \\
\subfloat[]{\includegraphics[width=0.49\textwidth]{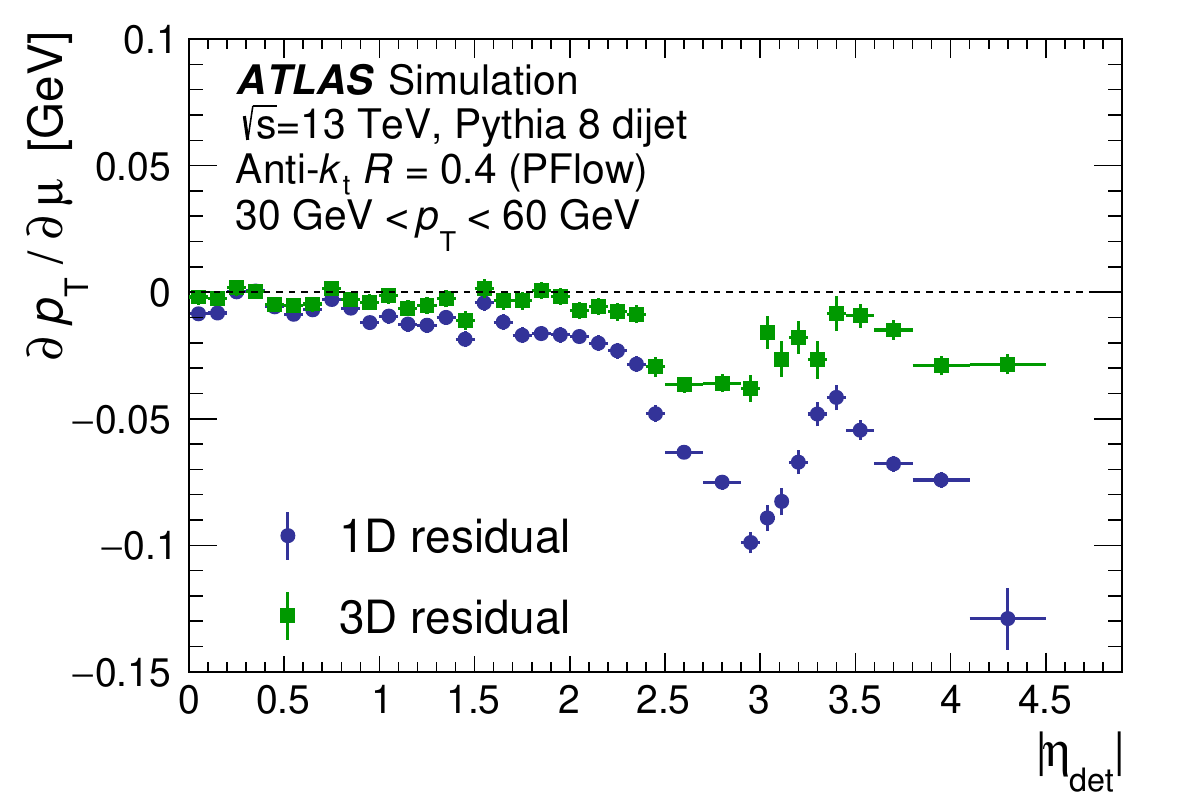}}
\subfloat[]{\includegraphics[width=0.49\textwidth]{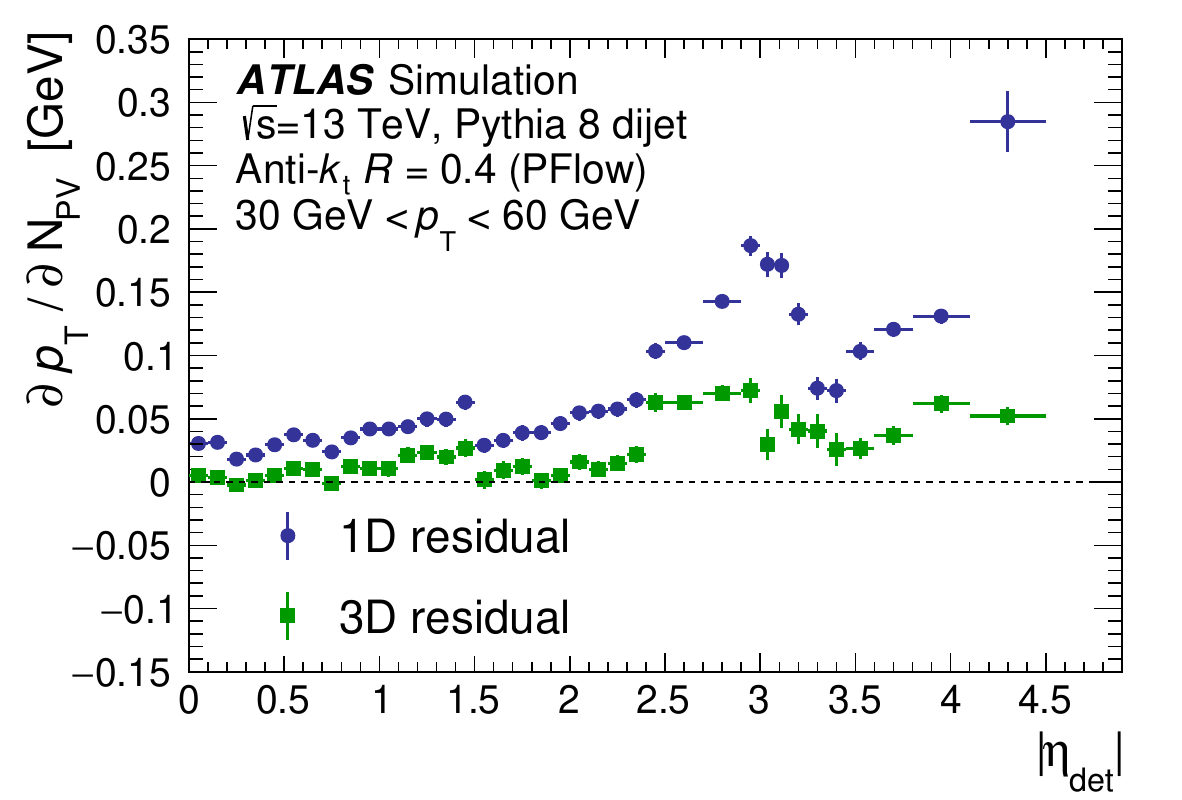}}
\caption{Dependence of \pt on (left) \avint and (right) \npv after the different residual pile-up corrections. The circles indicate the 1D residual pile-up correction, and the squares indicate the 3D residual pile-up correction. This is shown for jets with (top) $20~\GeV < \pt < 30~\GeV$, (bottom) $30~\GeV < \pt < 60~\GeV$.}
\label{fig:closure}
\end{figure}


\FloatBarrier
\subsection{The jet energy scale and $\eta$ calibration}
\label{sec:mcjes}


The absolute jet energy scale (MCJES) and $\eta$ corrections provide calibration functions for the energy and $\eta$
as a function of \etascale{det} and $E^{\mathrm{reco}}$ such that jets agree on average with the truth jet energy and $\eta$.
Since the calorimeters  measure the energy of particles, and not the transverse momenta,
this correction is determined as a function of the jet energy.
The jet energy response $\mathcal{R}$, defined as the mean of a  fit with a Gaussian function
to the core of the $E^{\mathrm{reco}}/E^{\mathrm{true}}$ distribution,
is measured in $E_{\mathrm{true}}$ and $\etascale{det}$ bins,
where $\etascale{det}$ is the jet $\eta$ pointing from the geometric centre of the detector,
which is used to remove any ambiguity about which region of the detector is measuring the jet.
The difference in $\mathcal{R}$ from the expected value of one is referred to as the non-closure,
and regions where the $\mathcal{R}$ has the expected value within the uncertainties are said to demonstrate closure.
 
The jet energy response after the application of the residual calibration is shown as a function
of $E_{\mathrm{true}}$ and $\etascale{det}$ in Figure~\ref{fig:mcjes:uncalib}.
This differs from previous JES calibrations by the ATLAS experiment,
in that the jet energy response is already close to unity, meaning that the correction is relatively small.
This is a feature of the 3D residual pile-up correction, which shifts the energy scale of the jets close to the truth scale,
although there is some significant difference from one at high $\eta$, where the residual calibration
insufficiently captures the behaviour of the energy response.
Since the $\Delta \pt$ term in the 3D residual pile-up correction is determined using jets with $\ptscale{truth} < 200$~\GeV,
the jet energy response shifts away from one at energies corresponding to $\pt > 200$~\GeV.

\begin{figure}[tbh]
\centering
\subfloat[]{\includegraphics[width=0.45\textwidth]{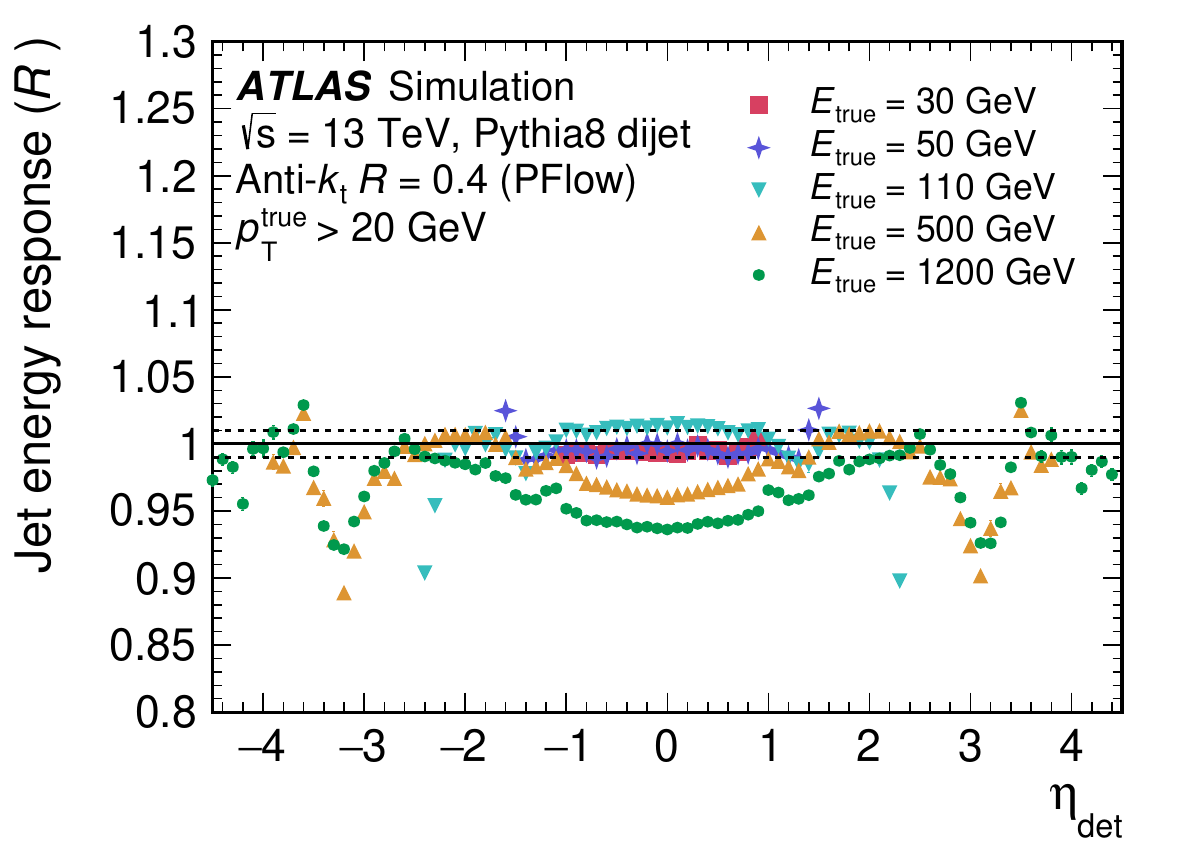}}
\subfloat[]{\includegraphics[width=0.45\textwidth]{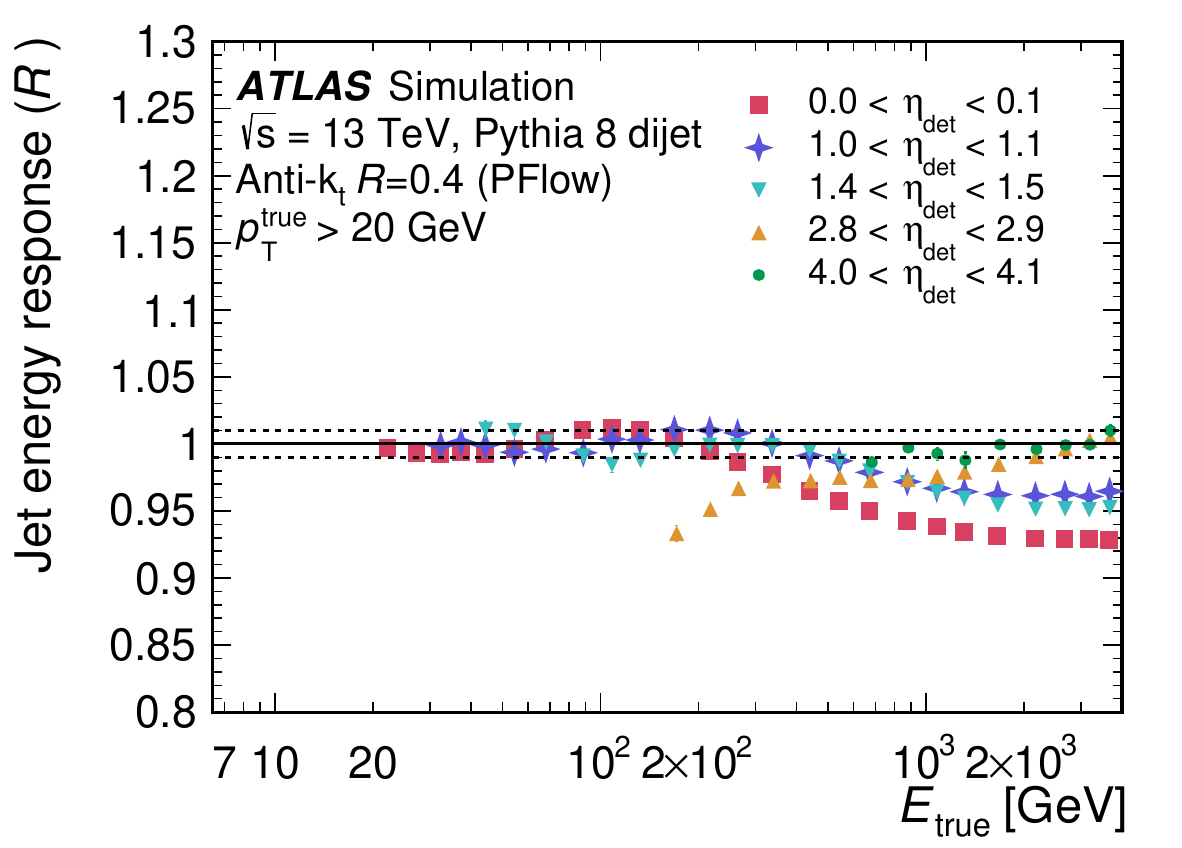}}
\caption{The jet energy response before the MCJES calibration (a) at fixed energies as a function of $\etadet$, and (b) at fixed $\etadet$ as a
function of truth jet energy. (a) The square shows the response for $E_{\text{true}}=30~\GeV$, the plus-sign shows the
response for $E_{\text{true}}=50~\GeV$, the down-triangle shows the response for $E_{\text{true}}=110~\GeV$, the up-triangle
shows the response for $E_{\text{true}}=500~\GeV$, and the circle shows the response for $E_{\text{true}}=1200~\GeV$. (b) The square
shows the response for $0.0 < \etadet < 0.1$, the plus-sign shows the response for $1.0 < \etadet < 1.1$, the down-triangle
shows the response for $1.4 < \etadet < 1.5$, the up-triangle shows the response for $2.8 < \etadet < 2.9$, and the circle shows the response for $4.0 < \etadet < 4.1$.}
\label{fig:mcjes:uncalib}
\end{figure}
 
Directly predicting the jet energy response from $E_{\text{reco}}$ depends on the distribution of $E_{\text{true}}$ used to derive the calibration.
Overall, the distribution of the  response is approximately Gaussian for a given $E_{\text{true}}$, but not for a given $E_{\text{reco}}$~\cite{Cukierman:2016dkb}.
Therefore, the calibration uses a numerical inversion technique~\cite{PERF-2011-03}, where, for each $\eta$ bin, the jet energy response is fit as a function of $E_{\text{true}}$,
and the jet calibration factor as a function of $E_{\textrm{reco}}$ is determined using the inverse of this function.
The two methods of determining the fit function, polynomial fits of order $N$, and penalised splines are compared below.

\subsubsection{Polynomial fits}
 
Following the procedure outlined in Ref.~\cite{JETM-2018-05}, polynomial fits are defined as a function of $\log(E)$, where $N_{\mathrm{max}}=8$ is the maximal order of the fitted polynomials.
Out of the given $N_{\mathrm{max}}$ fit functions, the best fit function is identified using Pearson's $\chi^2$ test~\cite{doi:10.1080/14786440009463897}.
The calibration factors are usually frozen at an $\eta$-dependent energy between $3$ and $4~\TeV$ to reduce statistical fluctuations,
while at for $\pt < 8$~\GeV, a linear extrapolation of the calibration factor is performed.
 
\subsubsection{Penalised splines}
 
In addition to the polynomial fit functions, a new method using penalised splines is studied.
A spline $S(x)$ of degree $N$ is a piecewise polynomial function of degree $N$, where pieces of the spline meet at points called knots,
and the first $N-1$ derivatives are continuous across the knots.
Splines may be defined from b-spline basis functions $B_i(x,t)$~\cite{Curry1966OnPF} via
\begin{equation*}
S(x) = \sum_i^{n-1} a_i B_i(x,t),
\end{equation*}
where $n$ is the number of data fit points, $a_i$ are control points weighting the individual basis functions $B_i$, and $t$ are the knots.
Since $S(x) = 0$ for $x$ outside of the range defined by the knots, an extrapolation to lower (higher) energy values is added using a linear extrapolation
based on the first (last) five points for the low (high) end of the spline.
 
A spline will overfit the data, since the basis function is required to exactly pass through the knots,
where in this case, the knots correspond to the energies where the response is determined.
For a set of points $x_i$ and their corresponding values $y_i$, this can be mitigated by using penalised b-splines (p-splines), which include an additional smoothness penalisation term $P$, minimizing
\begin{equation*}
L = \chi^2 + \alpha P = \underbrace{\sum_{i=0}^{n} \left(y_i - S(x_i) \right)^2}_{\text{least~squares}} + \underbrace{\alpha \int_{a}^{b} \left(S^{\prime \prime}(x)\right)^2 dx}_{\text{penalisation}}
\end{equation*}
with $a$ and $b$ corresponding to the range over which the penalisation term is included, with $a<x_i<b$, and the penalisation parameter $\alpha \geq 0$ is chosen and fixed.
For these studies, the $x$ values correspond to $E_{\mathrm{true}}$, and the $y$ values correspond to the jet energy response.
As $\alpha$ increases from zero to $\infty$, the result moves from a spline to a linear regression,
and this parameter enables a compromise between the curvature penalisation and a close fit to the data.
The penalisation parameter $\alpha$ is defined dynamically for each $\eta$ bin as
\begin{equation*}
\alpha = \frac{\lambda}{n} \cdot \sum_{i=0}^{n} w_i,
\end{equation*}
where $i$ runs over the $n$ data fit points, $\lambda$ is a regulative parameter,
and $w_i$ are the point weights defined as $w=\sigma_y^{-1/2}$, where $\sigma_y$ is the response fit uncertainty from the iterative fit to a Gaussian function.
 
For these studies, the splines are implemented using the Splinter framework~\cite{SPLINTER}, and a spline of degree three is used, with $\lambda$ empirically set to $0.1$.
To check for overfitting, the calibration and the closure test are performed on statistically independent events.

\subsubsection{Comparison of calibrations}
 
A comparison of the MCJES closure for the fitting techniques at different energy values is presented in Figure~\ref{fig:mcjes:splineVpoly}.
Both the strategies can provide closure of 1\% for high energies, while at low energies, the p-spline approach provides better closure than the polynomial fit.
Overall, the p-spline correction provides closure within 1\% across the \pt and $\eta$ range considered,
except for a small number of bins where the calibration becomes difficult due to quickly changing response, and non-Gaussian terms in the energy response.
The rest of the studies use the correction determined from the p-spline fit, since it provides the best overall closure.
 
\begin{figure}[tbh]
\centering
\subfloat[]{\includegraphics[width=0.45\textwidth]{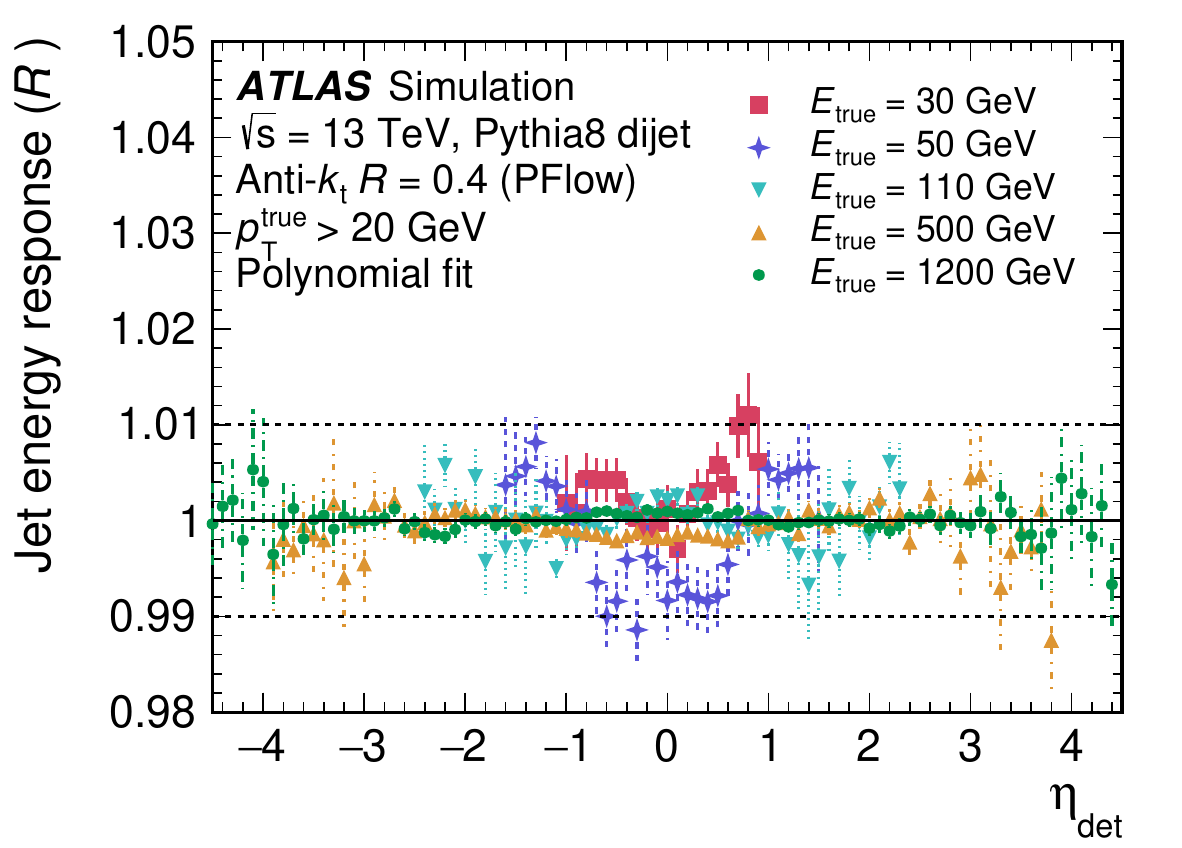}}
\subfloat[]{\includegraphics[width=0.45\textwidth]{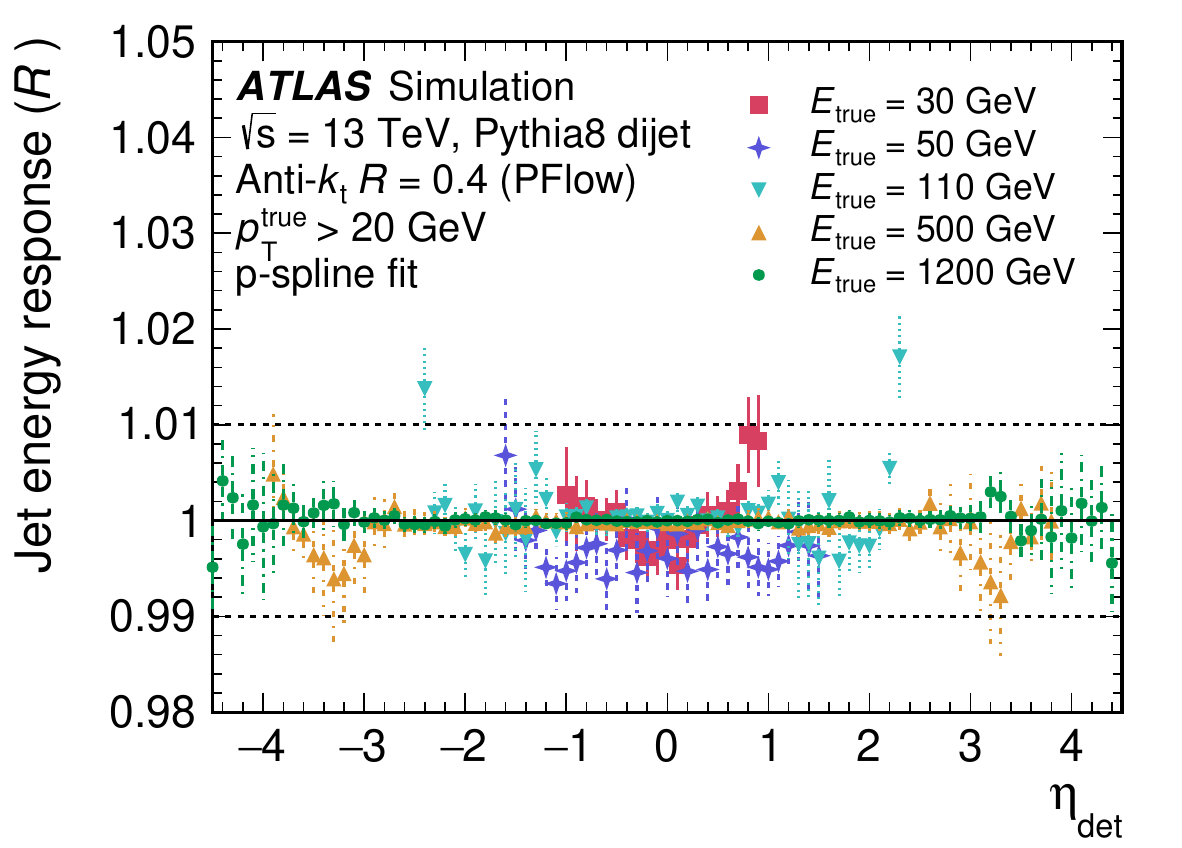}}
\caption{Jet response at fixed energies as a function of $\eta$  (a) after the polynomial MCJES calibration step, and (b) after the p-spline MCJES calibration. The square sign shows the response for $E_{\text{true}}=30~\GeV$, the plus sign shows the response for $E_{\text{true}}=50~\GeV$, the down-pointing triangle shows the response for $E_{\text{true}}=110~\GeV$, the up-pointing triangle shows the response for $E_{\text{true}}=500~\GeV$, and the circle shows the response for $E_{\text{true}}=1200~\GeV$.}
\label{fig:mcjes:splineVpoly}
\end{figure}

\subsubsection{Absolute MC jet $\eta$ calibration}
 
In addition to the jet energy, the jet pseudorapidity $\eta$ is calibrated with a similar approach as the JES calibration to correct for biases in the $\eta$ reconstruction,
following the strategy in Ref.~\cite{JETM-2018-05}.
This bias is most pronounced in the transition region between different parts of the calorimeter, where the discrepant response of the different detectors artificially
shifts the reconstructed energy on one side of the jet, changing the reconstructed $\eta$.
These $\eta$ corrections are particularly needed in the barrel-endcap ($|\eta| \sim 1.4$) and forward-endcap ($|\eta| \sim 3.1$) transitions.
The bias in $\eta$ is defined as $\eta_\mathrm{bias} = \langle \etascale{reco} - \etascale{true} \rangle$, as determined by an iterative
fit to a Gaussian function,
and the correction is performed on a jet-by-jet basis via $\etascale{calib} = \etascale{reco} -  \mathcal{R}_\eta$.
This correction is only applied to the \pt and $\eta$ of the jet, and is parameterised as a function of $E^{\mathrm{true}}$ and $\etascale{det}$.
For this correction, only polynomial fits are studied, using up to an order three polynomial.
There are small correlations between the corrections in $\eta$ and $E$, so this correction is derived simultaneously with the JES.


\FloatBarrier
\subsection{The global property calibration}
\label{sec:gnnc}

The absolute MCJES calibration corrects the jet energy response based on the $E$ and $\etascale{det}$ of the jet.
However, there are many other factors that contribute to the jet response, including the distribution of energy in the jet,
the distribution of energy deposits across different calorimeter layers, and the types of hadrons produced in the jet.
Many of these characteristics depend on whether the jet is quark- or gluon-initiated.
This can be seen in Figure~\ref{fig:gnnc:mcjesflavor},
which shows an example of the jet response distribution for jets with different initiating partons, and the jet \pt response as a function of \ptscale{true},
where the parton label is defined by the highest energy parton ghost-associated with the truth jet.
Not only are there differences between different jet flavours (i.e.\ the flavour of the intiating parton), but the behaviours change with the \pt of the jet.
Quark-initiated jets tend to have fewer hadrons, each with a higher fraction of the jet \pt, which typically results in contributions further into the calorimeter.
In contrast, gluon-initiated jets typically have more, lower-\pt hadrons, leading to a lower calorimeter response and a wider transverse profile.
These behaviours are further complicated by the use of particle flow reconstruction, which adds further dependence based on the charged particles in the jet.
 
The jet \pt response is also impacted by the MC model, as seen by the differences between the jet \pt response shown in Figure~\ref{fig:gnnc:mcjesgenerator}.
Overall, most MC predictions have similar behaviour for quark-initiated jets, while the differences between gluon-initiated jets can be sizeable.
This is due to differences between predictions between MC generators for the amount of soft radiation and its topological distribution in the jet.
There is some separation in the behaviour of models with the Lund string model for hadronisation compared with the other models,
where the Lund string model tends to predict a higher gluon \pt response, with larger differences for jets with $\pt < 100$~\GeV.
This can primarily be attributed to the fraction of jet energy carried by baryons and kaons~\cite{ATL-PHYS-PUB-2022-021}.
 
The global jet property calibration applies further corrections to jets based on their individual characteristics.
While these corrections only have a small effect on the overall closure of the calibration, the closure is significantly improved for different classes of jets, improving the JER\@.
In addition, this calibration reduces differences between MC predictions for the JES, resulting in smaller modelling uncertainties.
Two methods for deriving the global calibration are outlined below: the global sequential calibration (GSC), which was described in previous work~\cite{JETM-2018-05},
and a new method, the global neural network calibration (GNNC)\@.
Both the corrections are derived in $|\etascale{det}|$ bins corresponding to different detector regions, creating a balance
between the statistical uncertainty and the generality of the results.

\begin{figure}[htbp]
\centering
\subfloat[]{\includegraphics[width=0.48\textwidth]{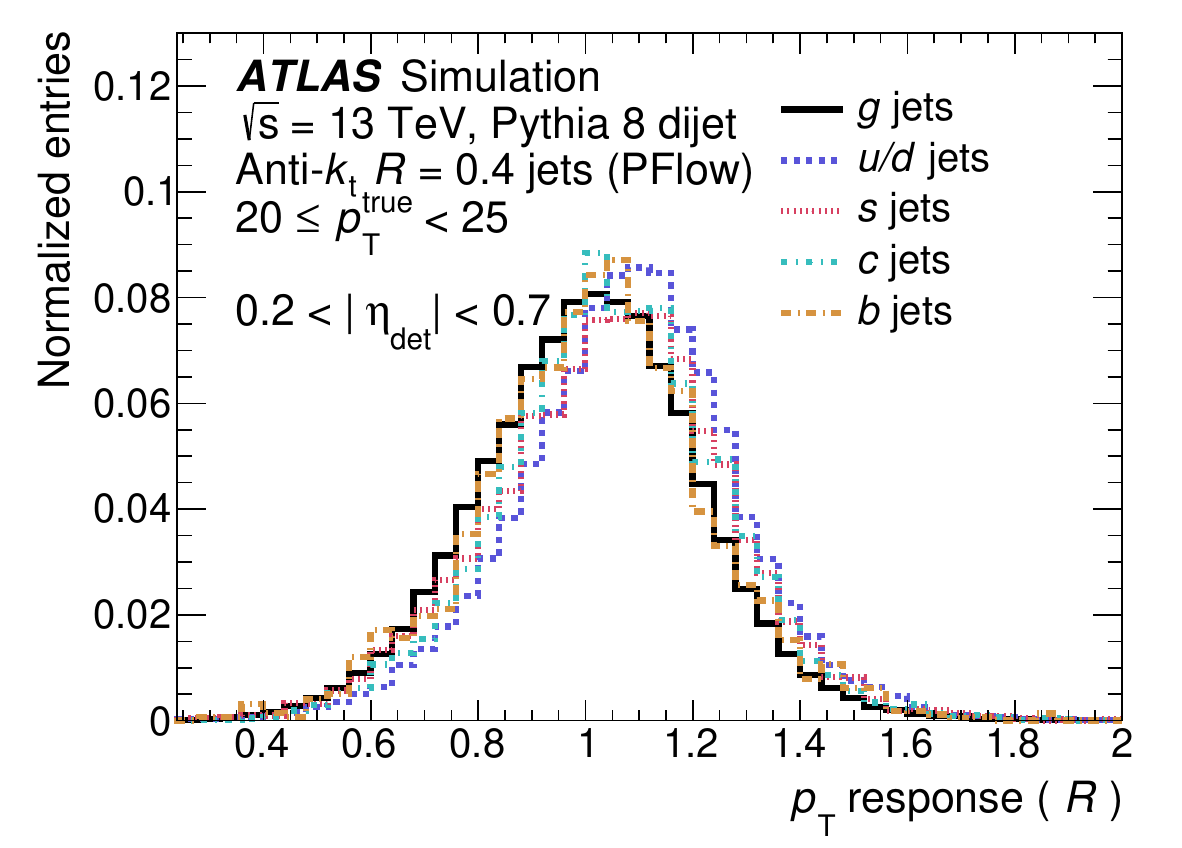}}
\subfloat[]{\includegraphics[width=0.48\textwidth]{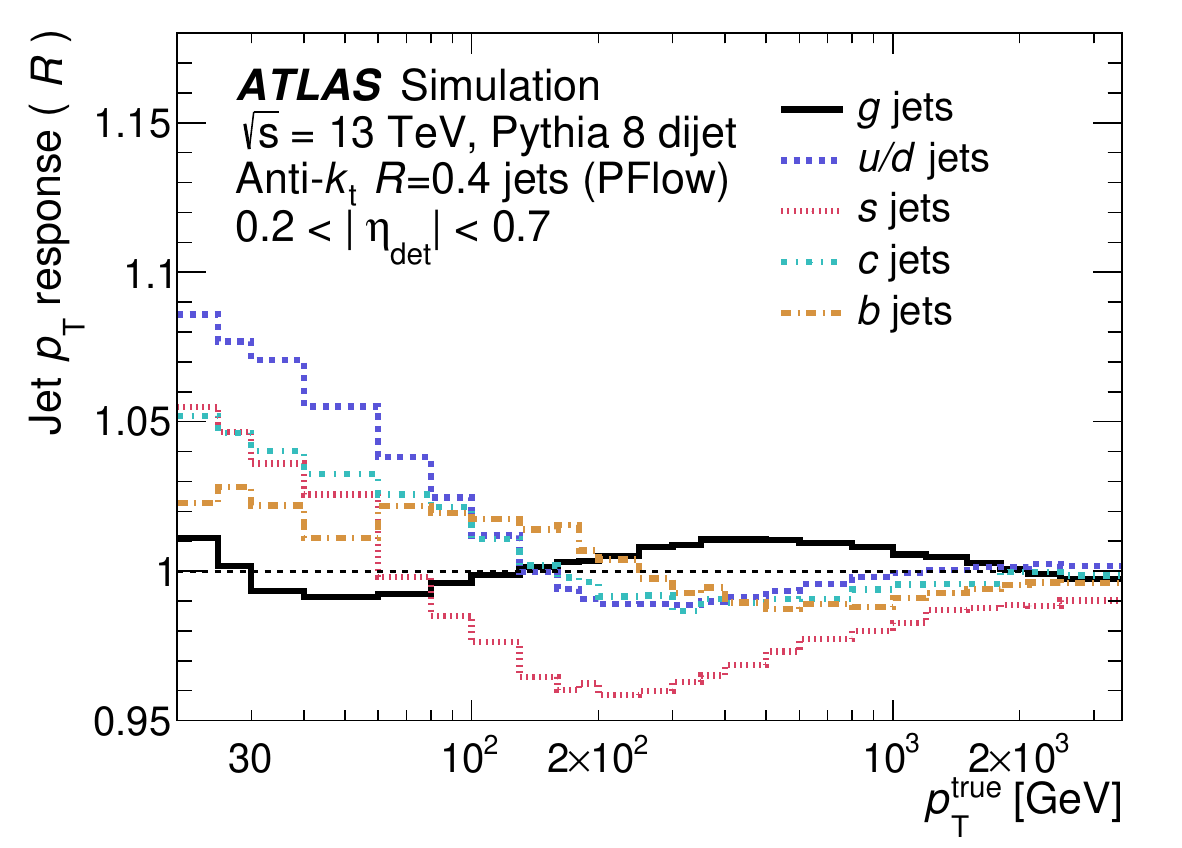}}
\caption{(a) The jet \pt response distribution for different jet flavours for jets with $20~\GeV < \ptscale{true} < 25~\GeV$, and (b) the jet \pt response
for several different flavours of jets as a function of \ptscale{true}. The solid line shows the response for gluon jets, the long dashed line shows the
response for light quark jets, the short dashed line shows strange jets, the alternating medium and short dashed line shows charm jets,
and the alternating long and short dashed line shows bottom jets.}
\label{fig:gnnc:mcjesflavor}
\end{figure}
 
\begin{figure}[htbp]
\centering
\subfloat[]{\includegraphics[width=0.48\textwidth]{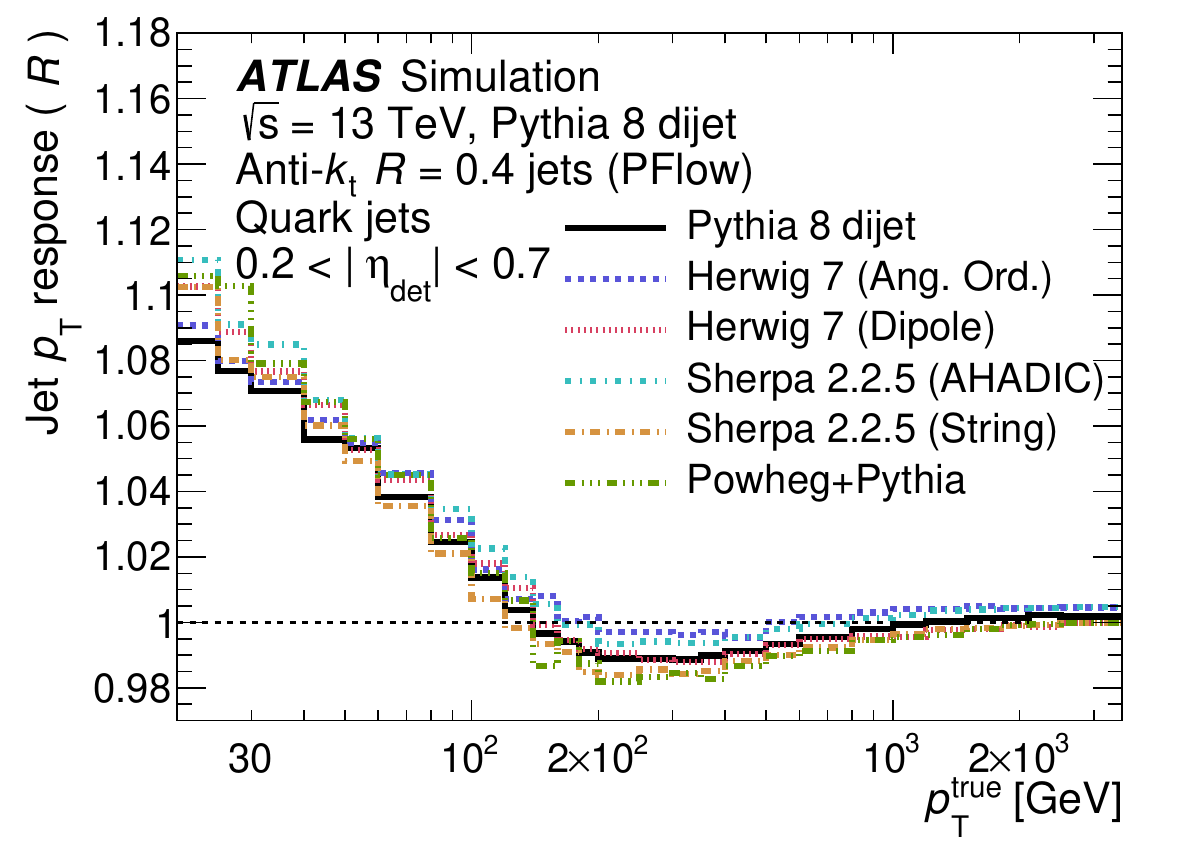}}
\subfloat[]{\includegraphics[width=0.48\textwidth]{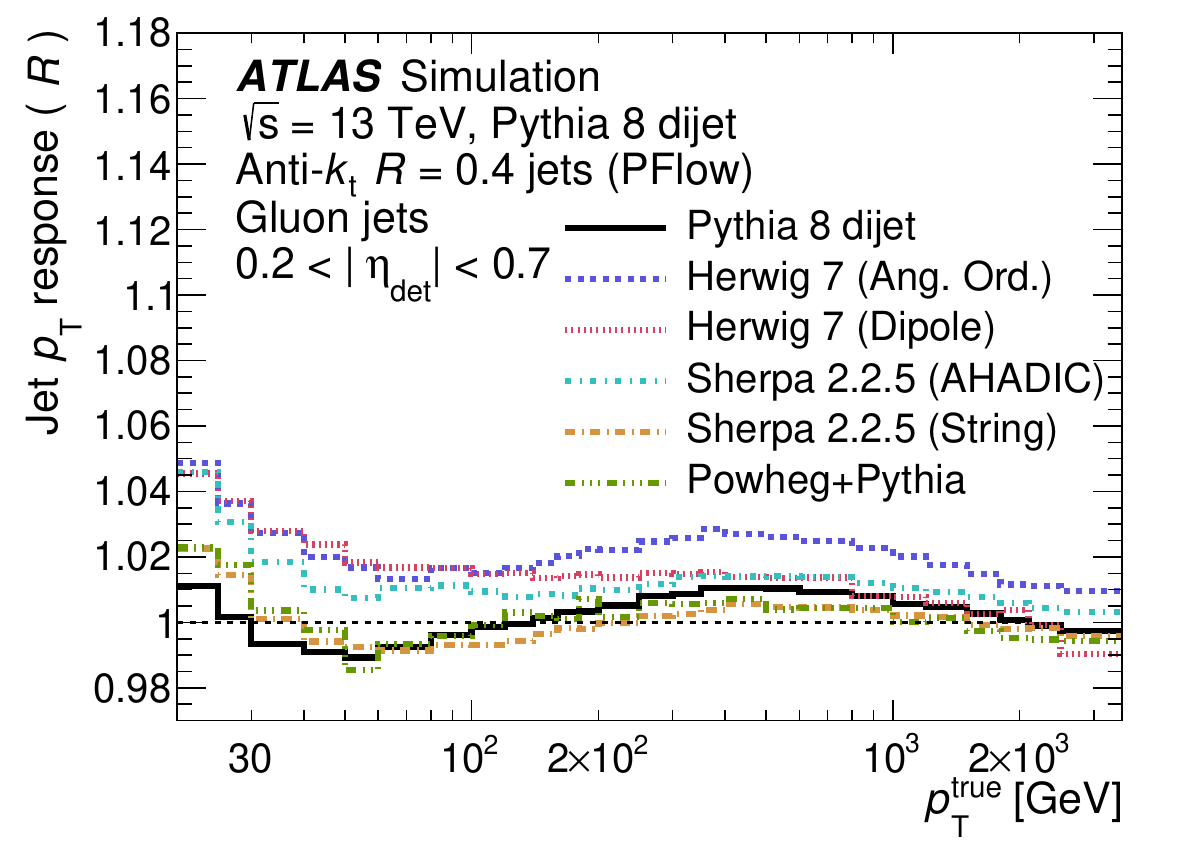}}
\caption{The jet \pt response as a function of \ptscale{true} for several different MC predictions for (a) quark jets, and (b) gluon jets.
The solid line shows \Pythia8, the long dashed line shows \Herwig with an angular ordered parton shower, the short dashed line
shows \Herwig with a dipole shower, the alternating medium and short dashed line shows
Sherpa with the AHADIC hadronisation model, the alternating long and short dashed line shows Sherpa with the string hadronisation model,
and the long dash with three short dashes shows Powheg+\Pythia. }
\label{fig:gnnc:mcjesgenerator}
\end{figure}

\subsubsection{The global sequential calibration}
 
The GSC is a series of multiplicative corrections to account for the differences between the calorimeter response to different types of jets,
which improves the jet resolution without changing the jet energy response.
The GSC is based on global jet observables such as the longitudinal profile of the energy deposits
in the calorimeters, tracking information matched to the jet, and information related to the activity in the muon chambers behind a jet.
Six observables that improve the JER and reduce modelling uncertainties are used as inputs to the GSC\@.
Each GSC correction to the jet four-momentum is derived and applied independently and sequentially, using the following procedure.
First, for a given GSC observable, the jet \pt response distribution is fit for each bin of  $|\eta|$, \ptscale{true}, and the GSC observable,
using the same procedure as in the MCJES calibration.
Next, these fitted values are divided by the inclusive value of the response in a given $|\eta|$ and \ptscale{true} bin to avoid changing the \pt scale of the jet calibration.
Then, for each bin of the GSC observable, the numerical inversion of the jet \pt response is performed after a linear smoothing in a given bin of (\ptscale{true}, $|\etascale{det}|$).
The resulting responses for a given $|\etascale{det}|$ bin are then smoothed simultaneously in \pt and the GSC observable using a Gaussian kernel.
Because the GSC is applied sequentially, it is possible to validate each GSC correction in a systematic way, testing the impact of any mismodelling of the input variables using data.
Such studies were performed  to validate the sequential correction procedure.
 
The six stages of the GSC, in the order of application, are
\begin{itemize}
\item $f_{\mathrm{charged}}$: the fraction of the jet \pt carried by charged particles, as measured using ghost-associated tracks with $\pt > 500~\MeV$, $|\etascale{det} | < 2.5$,
\item $f_{\mathrm{Tile0}}$: the fraction of jet energy ($E_{\mathrm{frac}}$) measured in the first layer of the hadronic tile calorimeter, $|\etascale{det} | < 1.8$,
\item $f_{\mathrm{LAr3}}$: the $E_{\mathrm{frac}}$ measured in the third layer of the electromagnetic LAr calorimeter, $|\etascale{det} | < 3.5$,
\item $N_{\mathrm{track}}$: the number of tracks with $\pt > 1~\GeV$ ghost-associated with the jet, $|\etascale{det} | < 2.5$,
\item $w_{\mathrm{track}}$: also known as track width, the average \pt-weighted transverse distance in the $\eta$-$\phi$ plane,
between the jet axis and all tracks of $\pt > 1~\GeV$ ghost-associated with the jet, $|\etascale{det} | < 2.5$,
\item $N_{\mathrm{segments}}$: the number of muon track segments ghost-associated with the jet, $|\etascale{det} | < 2.8$.
\end{itemize}
The $N_{\mathrm{segments}}$ correction, also known as the punch-through correction, reduces the tails of the response distribution
caused by high-\pt jets that are not fully contained in the calorimeter.
Unlike the other corrections, the $N_{\mathrm{segments}}$ correction is applied as a function of the jet energy instead of the jet \pt,
since this effect is more strongly correlated with energy escaping the calorimeters.
 
The jet \pt response for PFlow jets in MC simulation after each of the GSC corrections is shown in Figure~\ref{fig:gsc:resolution} for one $|\eta|$ bin.
While the jet energy scale is within 1\% at low energies, a small amount of non-closure is introduced when determining the response using \pt instead of energy.
The fractional jet resolution, denoted by  $\sigma$,  is used to determine the magnitude of the fluctuations in the jet energy reconstruction,
where $\sigma$ is the width of the fit to a Gaussian function for the jet \pt response distribution divided by the mean of the fit.
This is shown for PFlow jets with $0.2 < |\etascale{det} | < 0.7$ in MC simulation in Figure~\ref{fig:gsc:resolution}.
As more corrections are applied, the fractional jet resolution improves and the jet response dependence on the jet flavour is reduced
as the calibration is improved for jets with varying features.
The impact of $f_{\mathrm{charged}}$ and $f_{\mathrm{Tile0}}$ are most apparent in Figure~\ref{fig:gsc:resolution},
but the relative impact of the different corrections varies as a function of $|\etascale{det}|$.
In addition, these corrections reduces effects that are less evident in the inclusive case.
For instance, the punch-through correction scales with energy, and so it primarily impacts analyses that are sensitive to high-energy jets,
but its impact is not obvious in the inclusive distribution.

\begin{figure}[htbp]
\centering
\subfloat[]{\includegraphics[width=0.45\textwidth]{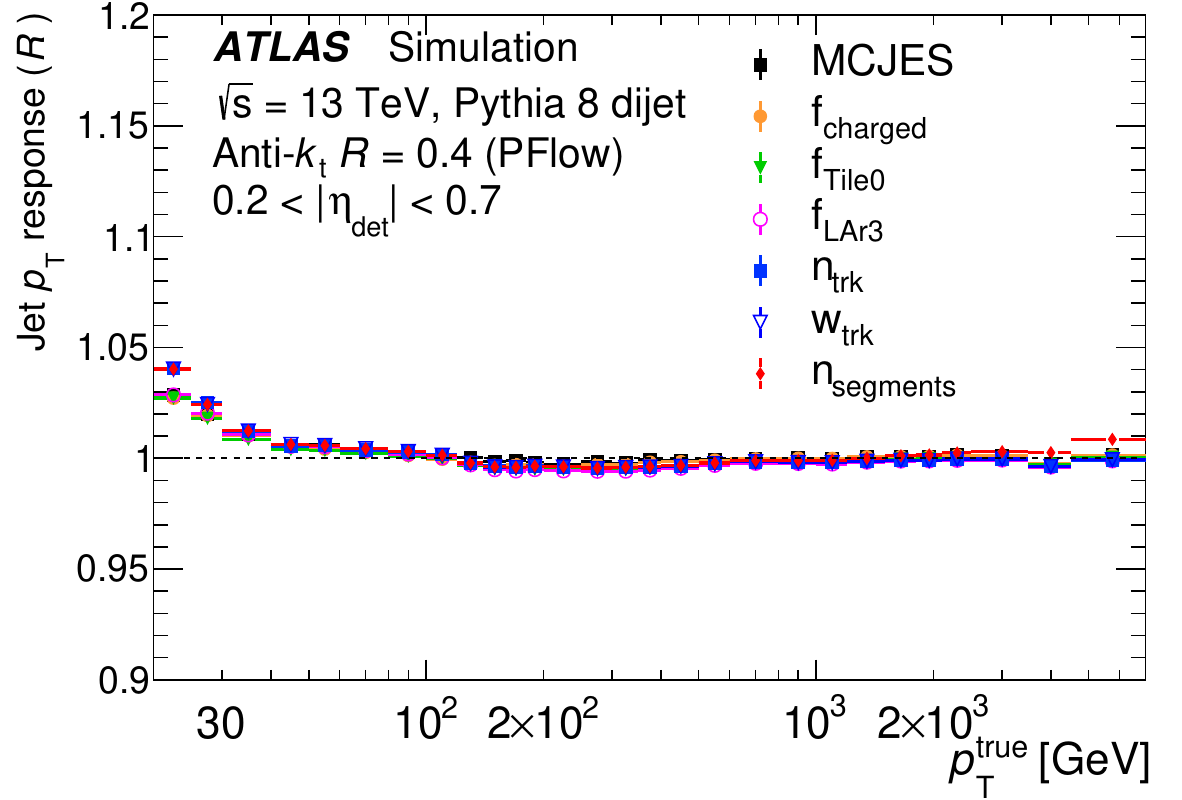}}
\subfloat[]{\includegraphics[width=0.45\textwidth]{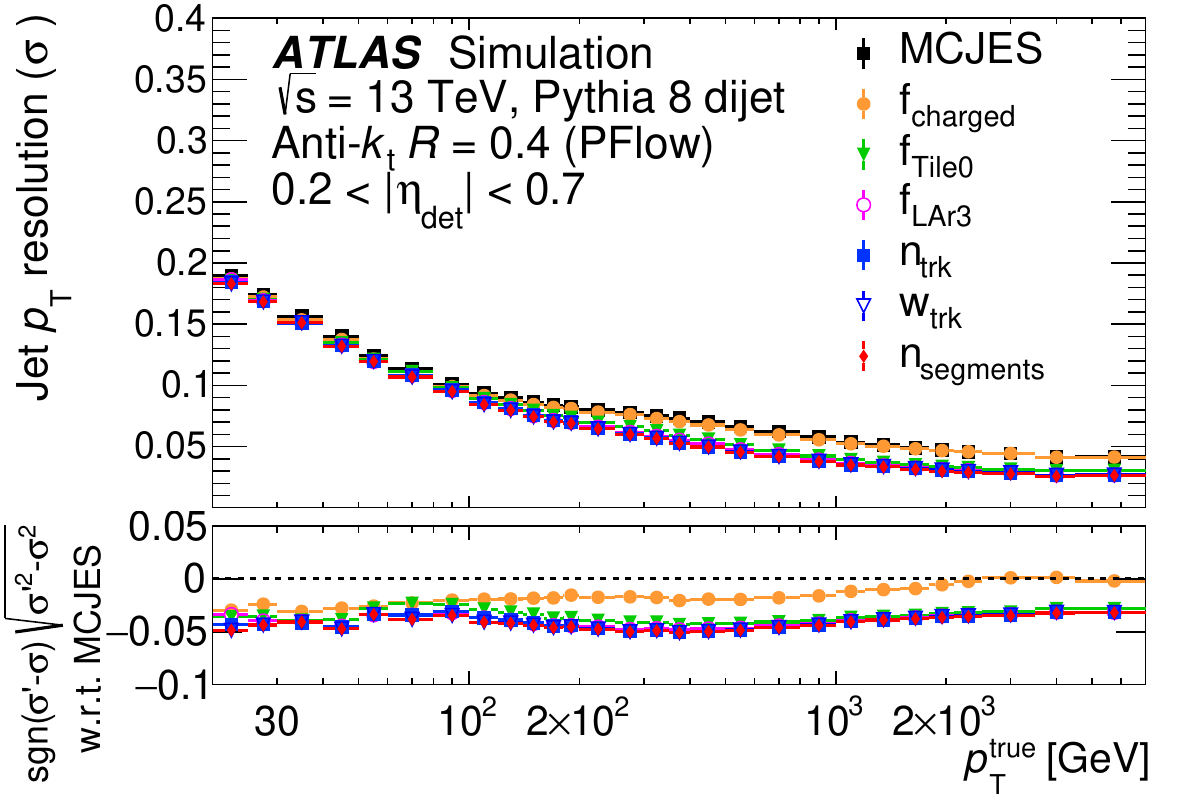}}
\caption{(a) The jet \pt response after each stage of the GSC calibration, and (b) the jet \pt resolution after the MCJES, after the $f_{\mathrm{charged}}$ correction,
after the $f_{\mathrm{Tile0}}$ correction, after the $f_{\mathrm{LAr3}}$ correction, after the $N_{\mathrm{track}}$ correction,
after the  $w_{\mathrm{track}}$ correction, and after the  $N_{\mathrm{segments}}$ correction.}
\label{fig:gsc:resolution}
\end{figure}

\subsubsection{The global neural network calibration}

The GSC is limited to using relatively uncorrelated variables for the correction, since otherwise,
each sequential step would potentially interfere with previous corrections due to correlations between observables.
This constraint is fundamental to the method, limiting the set of corrections that may be applied.
However when adding additional observables, and to account for their correlations, a simultaneous calibration is more appropriate~\cite{ATL-PHYS-PUB-2018-013}.
As an alternative to the sequential calibration, a deep neural network (DNN) is trained to determine a simultaneous correction based on a wide variety of jet properties,
enabling the use of correlated variables for determining the global jet property correction.
Since analyses make selections based on the jet \pt, the DNN is designed to correct the jet \pt response,
in contrast to the GSC, which leaves the energy response unchanged.
 
To improve the performance based on the detector geometry, a DNN is trained for each $|\etascale{det}|$ region used to derive the GSC
to provide a correction to the jet \pt based on various jet- and event-level features.
The DNNs are trained with Keras~\cite{chollet2015keras}, using the Adam~\cite{kingma2017adam} optimisation algorithm.
The network has three hidden layers with swish activation functions~\cite{swish} and a single-node output layer with linear activation.
The number of nodes is optimised for each $|\etascale{det}|$ bin, and ranges between 100 and 300.
The network uses the leaky Gaussian kernel (LGK) loss function~\cite{Cheong:2019upg}
\begin{equation*}
\mathrm{Loss}(x^{\mathrm{target}}, x^{\mathrm{pred}}) =
- \frac{1}{\sqrt{2\pi}} \mathrm{exp}\left( -\frac{(x^{\mathrm{target}} - x^{\mathrm{pred}})^2}{2\alpha^2}\right) + \beta | x^{\mathrm{target}} - x^{\mathrm{pred}} |,
\end{equation*}
where $x^{\mathrm{target}}$ is the jet \pt response, $x^{\mathrm{pred}}$ is the corresponding NN prediction, and $\alpha$ and $\beta$ are tunable parameters.
As $\alpha \rightarrow 0$, the LGK loss learns the mode, and the second term ensures that the gradient of the error function relative to the current weight
does not vanish for large $x^{\mathrm{target}} - x^{\mathrm{pred}}$.
Learning the mode is less biased by cases where the response is not a perfect Gaussian distribution, resulting in
better closure than a loss function that learns the mean of the distribution.
 
The architecture of the network was chosen as the result of a hyperparameter optimisation based on the closure of the result,
where hyperparameters are parameters involving the network structure.
The training is done with a batch size of $10^{4}$ jets, and a learning rate of $10^{-4}$.
For the LGK loss, the parameters are chosen to be $\alpha=10^{-1}$, and $\beta=10^{-6}$ based on the hyperparameter scan.
The training is done to minimise the LGK loss function, and training continues until there are no improvements to the loss for five epochs.
Increasing the patience did not have a noticeable effect on the quality of the results.
Unweighted events are used because this avoids issues in the training due to large differences between the event weights.
Since the target is the \pt response, not the jet \pt itself, the uniform weights do not have a large impact on the final result.
Only the two leading jets in the event are used in the training, since the events were simulated using a dijet process,
and so this avoids potential biases from using jets that originate purely from the parton shower.
For each $|\etascale{det}|$ bin, several networks were trained, and the one with the best closure was chosen for the final result.
 
Several sets of variables were considered as inputs to the NN, and the final list of variables used in the training is given in Table~\ref{table:gnnc:varList}.
This list includes all of the variables used in the GSC calibration, with the addition of more information about the jet kinematics, more granular information about the
energy deposits in different calorimeter layers, and measures of pile-up. While the residual pile-up correction removes most of the pile-up dependence,
some dependence is reintroduced by the absolute MCJES calibration, and so \npv and \avint are included in the training.
Some calorimeter layers are not present for certain $|\etascale{det}|$ regions, in which case their $E_{\mathrm{frac}}$ is set to zero.
Explicitly removing these observables from the list of input variables used in the NN training had a negligible impact on the results,
and so the set of training variables is kept the same for all $|\etascale{det}|$ regions.

\begin{table}[htb]
\begin{center}
\begin{tabular}{| l | l | l |}
\hline
Calorimeter      & $f_{\mathrm{LAr0-3*}}$   & The $E_{\mathrm{frac}}$ measured in the 0th-3rd layer of the EM LAr calorimeter \\
& $f_{\mathrm{Tile0*-2}}$  & The $E_{\mathrm{frac}}$ measured in the 0th-2nd layer of the hadronic tile calorimeter \\
& $f_{\mathrm{HEC,0-3}}$   & The $E_{\mathrm{frac}}$ measured in the 0th-3rd layer of the hadronic end cap\\
&                          &  calorimeter \\
& $f_{\mathrm{FCAL,0-2}}$  &  The $E_{\mathrm{frac}}$ measured in the 0th-2nd layer of the forward calorimeter \\
& $N_{90\%}$   & The minimum number of clusters containing 90\% of the jet energy \\  \hline
Jet kinematics    & \ptscale{JES}*   & The jet \pt after the MCJES calibration \\
& $\etascale{det}$   & The detector $\eta$ \\  \hline
Tracking          & $w_{\mathrm{track}}$*   & The average \pt-weighted transverse distance in the $\eta$-$\phi$ plane \\
&                         & between the jet axis and all tracks of $\pt > 1~\GeV$ ghost-associated \\
&                         & with the jet \\
& $N_{\mathrm{track}}$*   & The number of tracks with $\pt > 1~\GeV$ ghost-associated with the jet \\
& $f_{\mathrm{charged}}$*  &  The fraction of the jet \pt measured from ghost-associated tracks \\  \hline
Muon segments     & $N_{\mathrm{segments}}$*   &  The number of muon track segments ghost-associated with the jet \\ \hline
Pile-up            & $\mu$   & The average number of interactions per bunch crossing \\
& \npv   & The number of reconstructed primary vertices \\
\hline
\end{tabular}
\caption{List of variables used as input to the GNNC\@. Variables with a * correspond to those that are also used by the GSC\@.}
\label{table:gnnc:varList}
\end{center}
\end{table}
 
The jet \pt closure from this calibration is typically better than 1\%,
but it also has some fluctuations, which can sometimes slightly exceed this.
The magnitude of these fluctuations varies with each DNN training but were persistent across different DNN hyperparameters, loss functions, and training targets.
To mitigate this, an additional \pt calibration is derived after the GNNC, using the p-spline method outlined in Section~\ref{sec:mcjes},
but using the truth jet \pt as the target instead of the energy.
This is derived in $|\etascale{det}|$ bins with width of 0.1, which provides better performance than using the same $|\etascale{det}|$ bins as the GNNC correction.
This has a neglible effect on the jet \pt resolution, and only serves to improve the closure and smoothness of the calibration.

\subsubsection{Comparison of the methods}

Figure~\ref{fig:gnnc:comparisonResponse} shows a comparison of the jet \pt response after the MCJES, GSC and GNNC for one representative $|\etascale{det}|$ bins.
As designed, the GSC does not change the energy response of the jets.
Since the JES calibration moves the reconstructed energy scale to match that of the truth scale, this can result in some nonclosure in the jet \pt, which is particularly evident at low \pt.
The GNNC is designed to change the \pt scale of the jets to match the truth jets, and so the closure in \pt is better than that of the GSC closure.
It is worth noting that while the GSC can instead be applied in a way that corrects the jet \pt scale, this does not impact the resolution.
Other $|\etascale{det}|$ bins show similar qualitative features, though the exact nonclosure seen in the \pt response after the MCJES and GSC varies slightly.

Figure~\ref{fig:gnnc:comparisonResolution} show a comparison of the jet \pt resolution
after the MCJES, GSC and GNNC for several representative $|\etascale{det}|$ bins.
In a few cases, the jet \pt resolution becomes worse in the lowest \pt bins, but this is also where the \pt nonclosure is most significant,
making it difficult to have an accurate estimate of the resolution, particularly since the \pt scale of the GNNC is different than that of the MCJES and GSC.
Since the \pt scale of the MCJES and GSC is above one and has a negative slope, the measured resolution is slightly underestimated~\cite{Cukierman:2016dkb} in these bins,
while the GNNC resolution is correctly estimated, since the response closes.
In the $0.2 < \abseta < 0.7$ bin, the GNNC has an average improvement in the jet \pt resolution of over 15\%, and maximum improvements of over 25\%, when compared with the GSC\@.
Other $|\etascale{det}|$ bins show similar average improvements of around 15--25\%, with maximum improvements often over 30\%,
and the improvement generally becomes more pronounced at higher $|\etascale{det}|$,
where the resolution improvements are significant, mostly due to the improvements from the additional detector information.
Studies comparing the GNNC performance with only the GSC observables as inputs find a similar performance to the GSC,
indicating that the improvement in the resolution of GNNC compared with GSC is due to the inclusion of additional observables.
This is made possible by a simultaneous correction that accounts for correlations between observables.
The GNNC provides a larger improvement to the jet energy resolution than the GSC, and so it is used for the remainder of the paper.

\begin{figure}[htbp]
\centering
\includegraphics[width=0.4\textwidth]{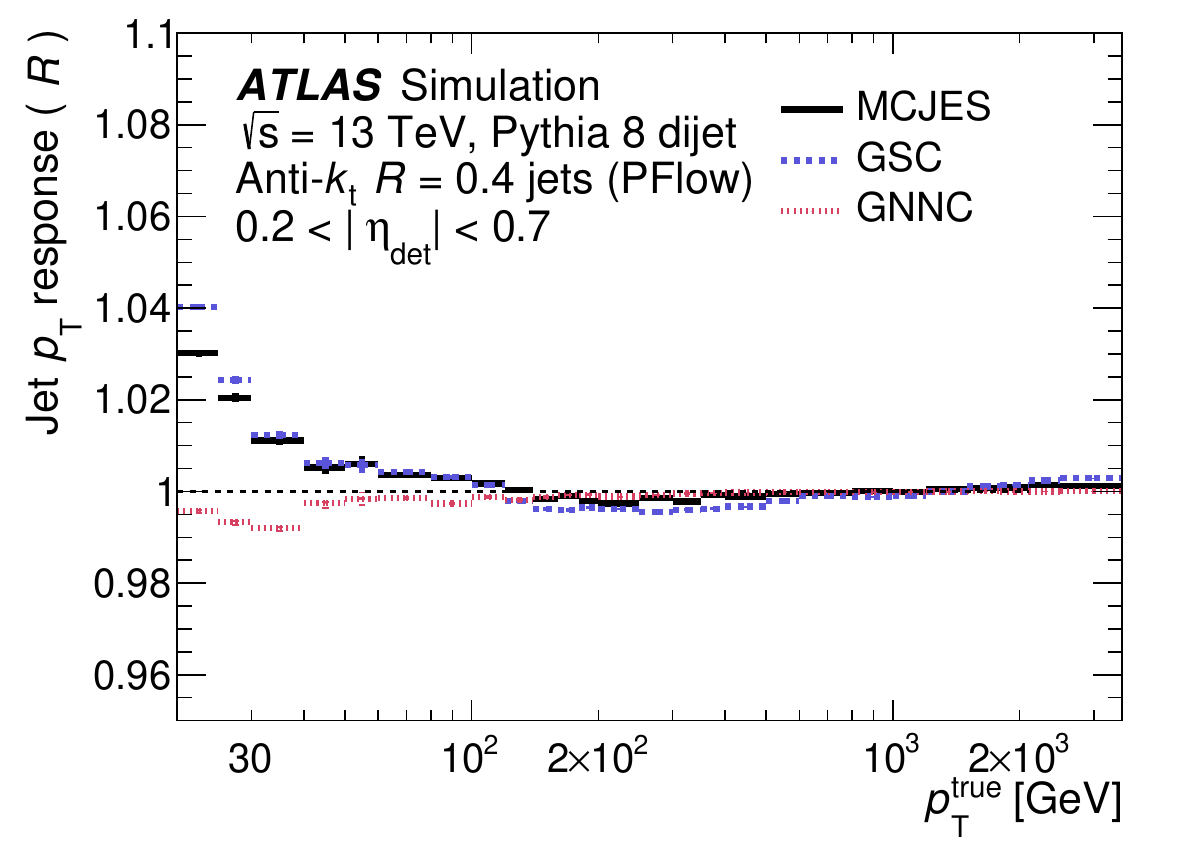}
\caption{The jet \pt closure for $0.2 < |\etascale{det}| < 0.7$. The solid line shows the MCJES, the long dashed line shows the GSC, and the short dashed line shows the GNNC\@.}
\label{fig:gnnc:comparisonResponse}
\end{figure}

\begin{figure}[htbp]
\centering
\subfloat[]{\includegraphics[width=0.4\textwidth]{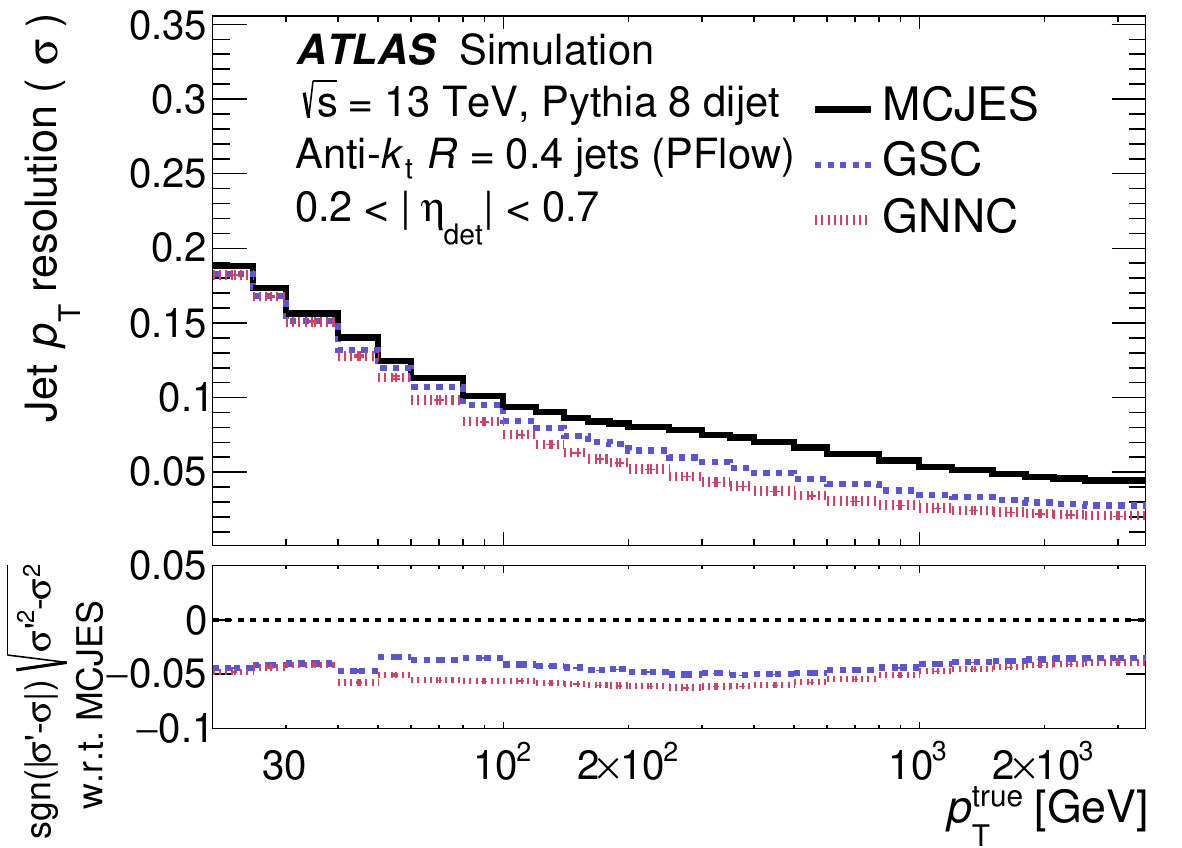}}
\subfloat[]{\includegraphics[width=0.4\textwidth]{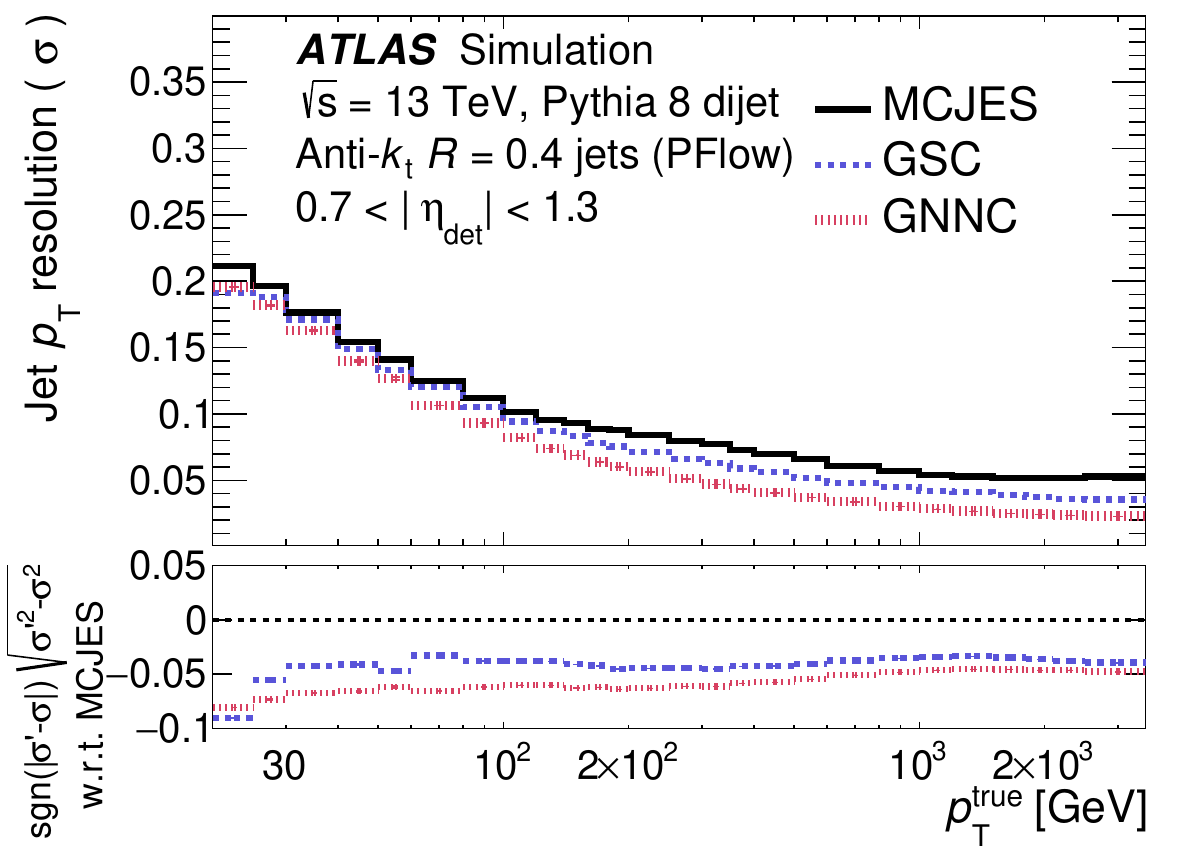}} \\
\subfloat[]{\includegraphics[width=0.4\textwidth]{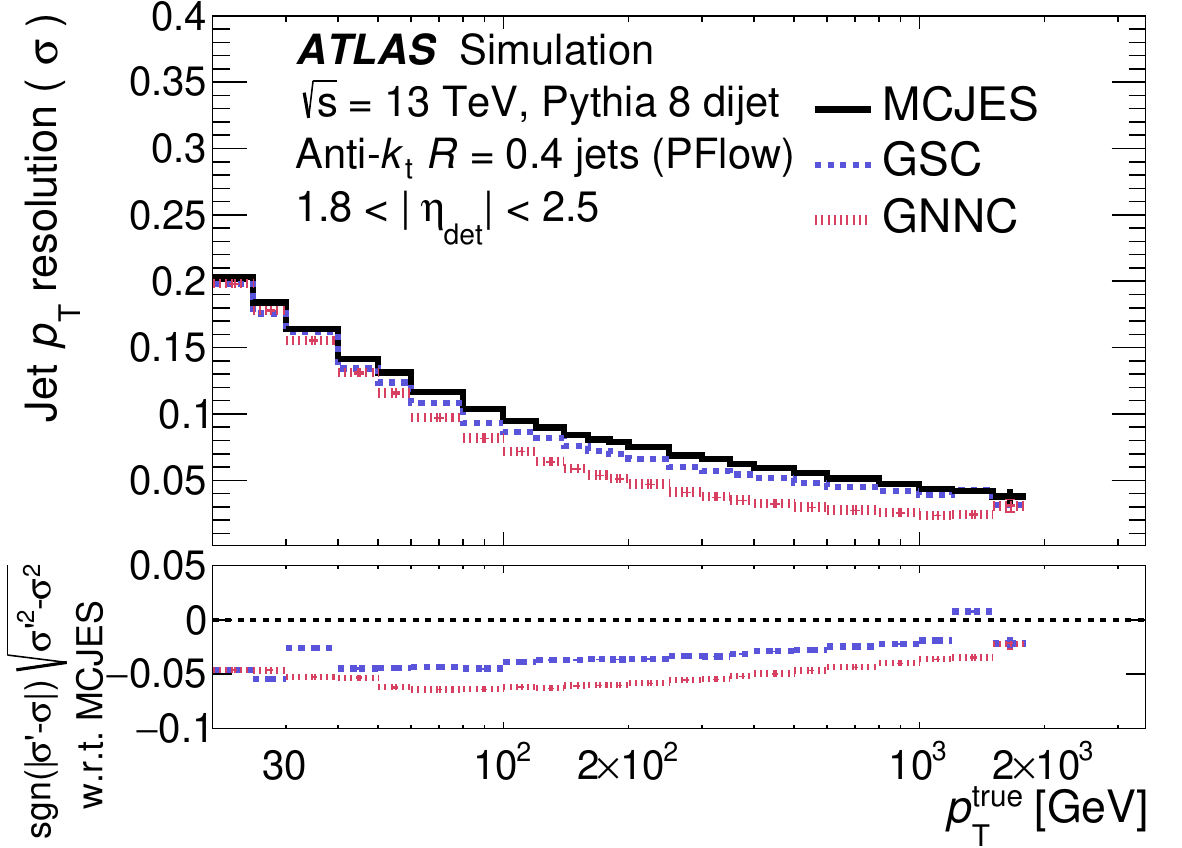}}
\subfloat[]{\includegraphics[width=0.4\textwidth]{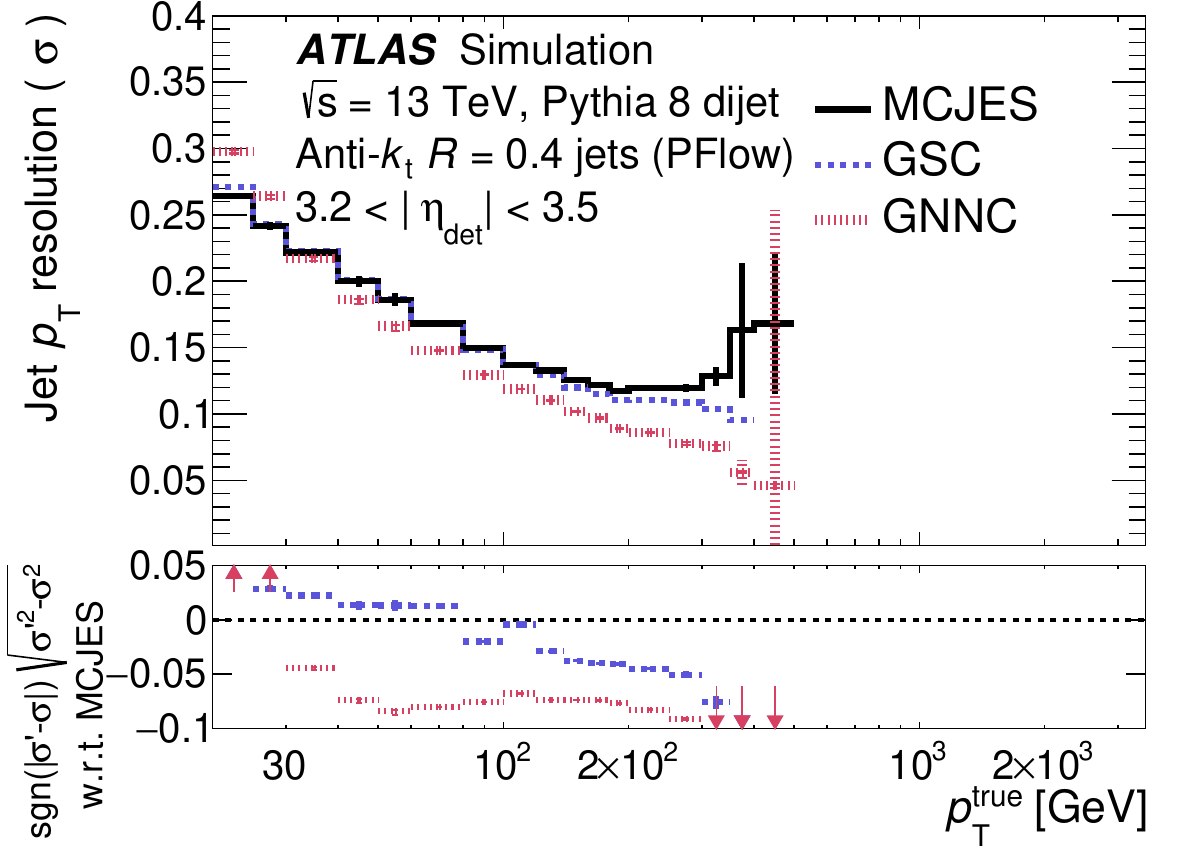}}
\caption{The jet \pt resolution for (a) $0.2 < |\etascale{det}| < 0.7$, (b) $0.7 < |\etascale{det}| < 1.3$, (c) $1.8 < |\etascale{det}| < 2.5$, and (d) $3.2 < |\etascale{det}| < 3.5$.
The solid line shows the MCJES, the long dashed line shows the GSC, and the short dashed line shows the GNNC\@.}
\label{fig:gnnc:comparisonResolution}
\end{figure}

\subsubsection{Flavour uncertainties}
 
The two flavour-dependence uncertainties in the JES are derived from simulation and account for relative flavour fractions and differing responses to quark- and gluon-initiated jets.
The flavour response uncertainty accounts for the fact that,
unlike the quark-initiated jet response $\mathcal{R}_{q}$, the gluon-initiated jet response $\mathcal{R}_g$ is found to differ significantly between generators.
This uncertainty is defined as
\begin{equation*}
\sigma_{\mathrm{response}} = f_{g} (\mathcal{R}_{g, \mathrm{\PYTHIA8}} - \mathcal{R}_{g,\mathrm{\HERWIG}}),
\end{equation*}
where $f_g$ is the fraction of gluon-initiated jets,
and $\mathcal{R}_{g, \mathrm{\PYTHIA8}}$ and $\mathcal{R}_{g,\mathrm{\HERWIG}}$ are the gluon-initiated jet response $\mathcal{R}_g$ in \PYTHIA8 and \HERWIG respectively.
The flavour composition uncertainty accounts for the fact that the jet response is different for quark- and gluon-initiated jets.
This is determined based on the fraction of gluon-initiated jets $f_g$, where $\mathcal{R}_q$ and $\mathcal{R}_g$ are the quark and gluon
jet responses measured in \PYTHIA8,
and $\sigma_g^f$ is the uncertainty in $f_g$ in the sample, with the uncertainty defined as
\begin{equation*}
\sigma_{\mathrm{composition}} = \sigma_{g}^{f} \frac{\mathcal{R}_{q} - \mathcal{R}_{g}}{f_{g} \mathcal{R}_{g} + (1-f_{g})\mathcal{R}_{q}}.
\end{equation*}

Figure~\ref{fig:gnnc:flavorUncertainties} shows a comparison of the flavour composition and flavour response uncertainties for the MCJES, GNNC and GSC\@.
After the MCJES calibration, $R_q - R_g$ becomes negative for jets above 100~\GeV, which appears as a dip in the flavour composition uncertainty.
Both the GSC and GNNC can reduce these uncertainties, with the GNNC providing a greater reduction.
For each $|\etascale{det}|$ bin, when compared with the GSC, the GNNC results in an average improvement of around 15--25\% in the $40 \le \pt < 300$~\GeV~ range for the flavour response uncertainty,
and up to 25\% improvements for the flavour composition uncertainty.

\begin{figure}[htbp]
\centering
\subfloat[]{\includegraphics[width=0.4\textwidth]{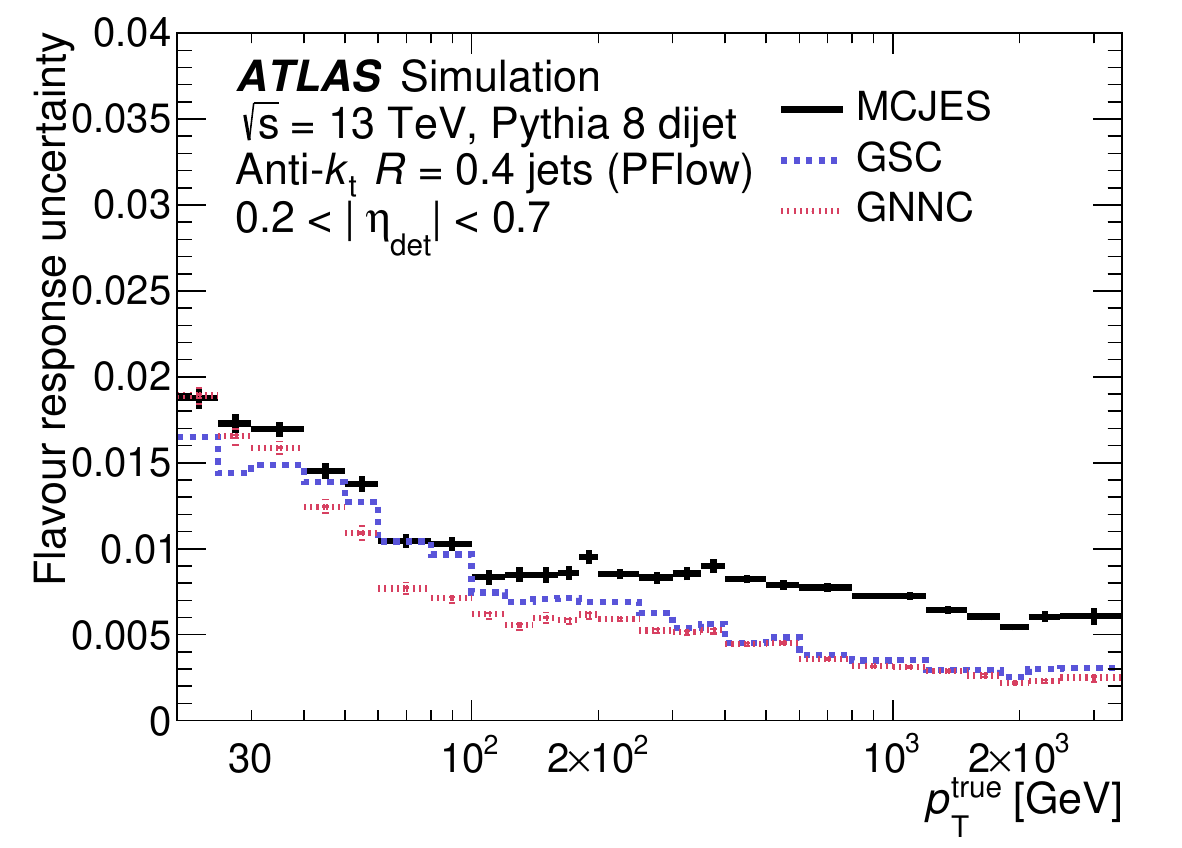}}
\subfloat[]{\includegraphics[width=0.4\textwidth]{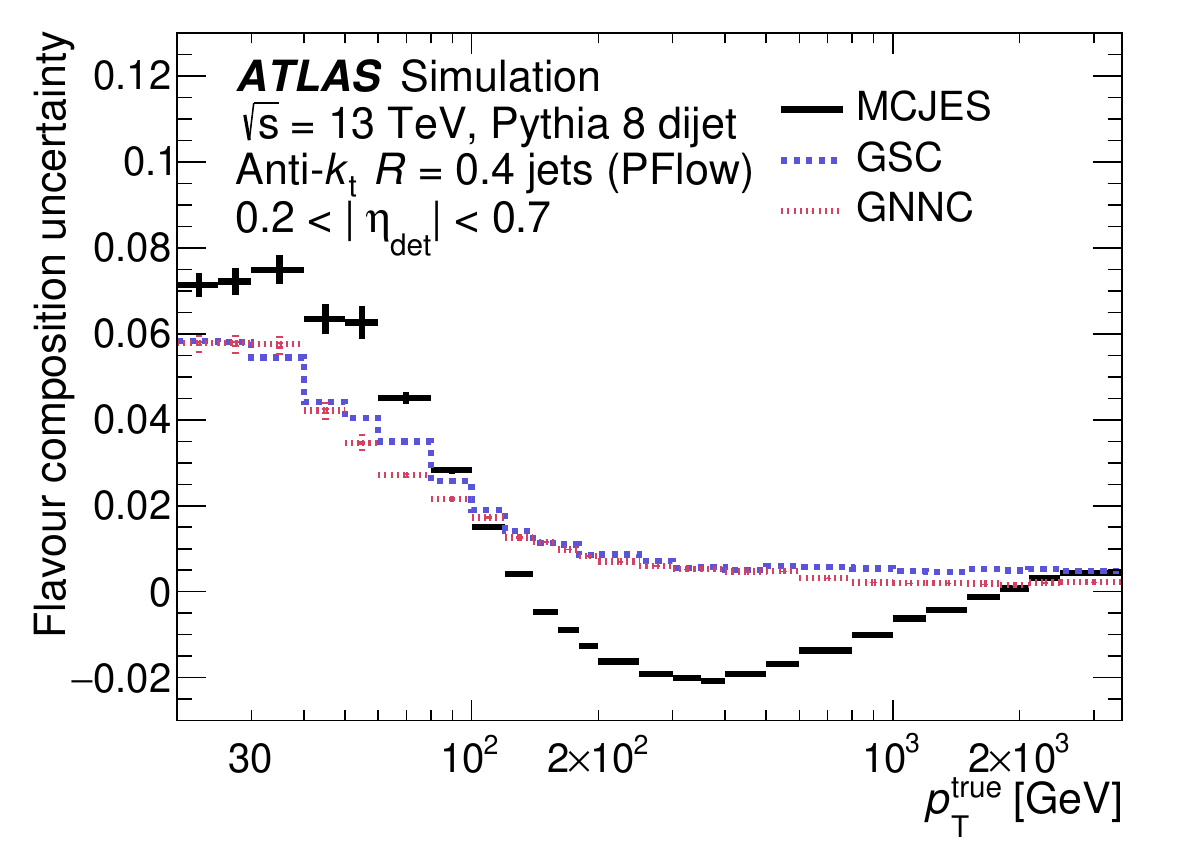}}
\caption{(a) The flavour response uncertainty, and (b) the flavour composition uncertainty for central jets. The solid line shows the MCJES,
the long dashed line shows the GSC, and the short dashed line shows the GNNC.}
\label{fig:gnnc:flavorUncertainties}
\end{figure}



\FloatBarrier
 
\section{\emph{In situ} analysis}
\label{sec:insitu}

The final calibration step accounts for differences in the jet response between simulation and data.
Such differences arise due to the imperfect simulation of detector response and detector material, and the modelling of physics processes involved: hard scatter,
underlying events, pile-up, jet formation and particle interactions with detector material.
For the remainder of these studies, a single jet calibration is studied,
using the sideband $\rho$ definition in Section~\ref{sec:area_pu} and the 3D residual calibration in Section~\ref{sec:residual_pu},
the absolute MC calibration implemented with p-splines in Section~\ref{sec:mcjes}, and the GNNC for the global calibration in Section~\ref{sec:gnnc}.
To fully understand the impact of these changes relative to the calibration procedure in Ref.~\cite{JETM-2018-05}, on the calibration and corresponding uncertainties, the \textit{in situ} calibration  is studied.
The \textit{in situ} calibration provides important validation of the new MC calibration of jets  by comparing the data-to-MC difference
between the \pt balance of a jet against a well-calibrated object or system.
In addition, novel studies are done to disentangle the physics effects and detector effects in the $\eta$-intercalibration to reduce the systematic uncertainties.
Furthermore, the $b-$jet JES is evaluated \textit{in situ} using PFlow jets, which is performed using $\gamma+$~jet events for the first time.
 
The \textit{in situ} calibration response $\cal{R}_{\insitu}$ is defined as the average ratio of the jet \pt to the transverse momentum of
the reference object \pTref, derived as a function of \pTref.
The $\cal{R}_{\insitu}$ response is susceptible to effects such as the radiation of additional partons or the loss of energy outside the reconstructed jet cone.
Dedicated event selections are applied to mitigate these effects.
A double ratio, insensitive to these secondary effects provided they are well-modelled in simulations, is defined
 
\begin{equation}\nonumber
\mathcal{C} = \frac{\cal{R}^{\text{data}}_{\insitu}}{\cal{R}^{\text{MC}}_{\insitu}}\ .
\end{equation}
The calibration factor to the jet four-momentum can be obtained by
a numerical inversion of this double ratio as a function of jet \pt, and as a function of $\etadet$ in $\eta$-intercalibration.
 
Two stages of \textit{in situ} analyses are done sequentially to assess the performance of MC calibrations.
First, a relative \textit{in situ} calibration referred to as the $\eta$-intercalibration is done, which corrects the energy scale of forward jets ($0.8<|\eta_\text{det}|<4.5$) to
match that of the central jets ($|\eta_\text{det}|<0.8$) using the \pt balance in a dijet system.
Second, an absolute calibration is done by measuring the \pt balance of a central jet against a well-calibrated $Z$ boson or a photon.
The missing-$E_\text{T}$ projection fraction (MPF) method~\cite{ATLAS-CONF-2015-057} is used in $Z/\gamma+$jet events to calculate the \pt balance between
the full hadronic recoil and a $Z$ boson or a photon. The method is less susceptible to effects of pile-up and the threshold of the jet reconstruction than
the direct balance method, allowing a reliable measurement of the low-\pt jet response below 100~\GeV.
The direct balance (DB) method measures the balance between a ($b-$)jet recoiled against a photon in $\gamma$+jet events. By using the DB instead of MPF,  the response of
the $b$-jet itself is studied without including the effects of the hadronic recoil.
 
For each \textit{in situ} analysis, main sources of systematic uncertainties arise from the MC model of physics processes, the measurement of the reference object
and the \pt balance due to the selected event topology. Uncertainties related to MC model of physics effects are addressed by taking the difference
between the predictions between two distinct MC event generators. The difference between jet response in simulations depends on hadronisation models that
cause different jet contents~\cite{ATL-PHYS-PUB-2022-021}. Uncertainties in the reference object are
estimated by propagating its own $\pm 1\sigma$ calibration uncertainties
through the analysis. Uncertainties due to the selected event topology are evaluated by varying the event selection
criteria and comparing the impact on the response ratios between
data and MC simulation. To reduce the statistical fluctuations when applying the systematic variations, a rebinning procedure similar to that used in previous publications~\cite{JETM-2018-05}
is employed to obtain statistically significant results using pseudo-experiments. This rebinning procedure is only performed in regions where no sharp variations in \pt response are observed to
ensure no real physics effects are removed.
 
Events must satisfy the common selection requirements in the \textit{in situ} analysis.
Each event is required to have at least one reconstructed primary vertex with at least two matched tracks of $\pt>500$~\MeV.
Jets arising from cosmic rays, non-collision background and calorimeter noise are vetoed by applying data-quality requirements~\cite{ATLAS-CONF-2015-029}.
In addition, jets with $20<\pt<60$~\GeV\ and $|\etadet|<2.4$ are required to satisfy the criteria of jet vertex tagging (JVT)~\cite{ATLAS-CONF-2014-018,PERF-2014-03}.
The JVT criteria rejects jets from pile-up interactions by matching jets with the primary vertex; it has a selection efficiency of 97\% for hard scatter jets at the nominal operating point.
 
\subsection{$\eta$-intercalibration}
The jet response in the forward region ($0.8<|\etadet|<4.5$) is typically less understood due to the more complicated detector structure.
The $\eta$-intercalibration provides a correction for forward jets ($0.8<|\etadet|<4.5$) to bring them to the same energy scale as central jets ($|\etadet|<0.8$).
This calibration uses events with a dijet topology, requiring two back-to-back jets in the transverse plane in different $\etadet$ regions.
In order to increase the statistical precision, there is no requirement on whether or not one of the two jets is in the central reference region: instead, all regions will be calibrated
relative to one another by solving a set of linear equations. This is referred to as the matrix method~\cite{JETM-2018-05}.
The momentum asymmetry is defined to measure the jet \pt balance between the two jets in two distinct detector regions (symbolically labelled \textit{left} and \textit{right} for simplicity)
\begin{equation}\nonumber
\asym = \frac{\pt^{\text{left}}-\pt^{\text{right}}}{\pTavg}\,,
\end{equation}

where \pTavg is the average of the transverse momentum of the \textit{left} and \textit{right} jets ($\pTavg=(\pt^{\text{left}}+\pt^{\text{right}})/2$).
For a narrow bin approximation in \pTavg, the relative response \Response of the \textit{left} and \textit{right} jets in terms of the calibration factor for each jet and the mean of \asym can be defined as
\begin{equation}\nonumber
\mathcal{R}  = \frac{c^{\mathrm{right}}}{c^{\mathrm{left}}} =  \frac{2 + \left\langle \mathcal{A} \right\rangle}{2 - \left\langle \mathcal{A} \right\rangle} \approx \frac{\left\langle p_{\mathrm{T}}^{\mathrm{left}} \right\rangle}{\left\langle p_{\mathrm{T}}^{\mathrm{right}} \right\rangle}
\end{equation}
 
where \Response is measured in terms of \etadet for \textit{left} and \textit{right} jets and \pTavg. The intercalibration factor $c$ is defined as $c=\frac{c^{\mathrm{right}}}{c^{\mathrm{left}}}$
and hence the relative response $\Response$ satisfies $\Response=1/c$.
 
Dijet events are selected using a combination of forward and central single-jet triggers, where each trigger is considered in the range of \pTavg\ that has an efficiency of at least 99\%.
Prescaled jet triggers are used to accommodate bandwidth limits, and each selected event is weighted accordingly. The trigger combination
method~\cite{TriggerCombination, JETM-2018-05} is used to maximise the statistical
precision. Each event must have at least two leading jets with $\pTavg>25$~\GeV~ and $|\etadet|<4.5$.
Events containing a third jet with $\pt^{\text{jet3}}/\pTavg>0.25$ are excluded.
The two leading jets must be back-to-back in the transverse plane satisfying a requirement on their azimuthal angle difference $\Delta \phi^{1,2}>2.5$.
 
The nominal calibration is estimated by taking the ratio of the simulated response in \POWPY{8} to the measured response in data.
The binning in \etadet and \pTavg is chosen to ensure enough of a sample size in scarce reference regions and to capture granular variations in detector response.
A two-dimensional Gaussian kernel is optimised to smooth statistical fluctuations while also capturing notable detector features.

The 2017 data sample is representative of the high pile-up conditions and thus discussed here.
The relative response, parameterised by \etadet in two \pTavg regions and by \pTavg in two \etadet regions between the 2017 data sample and MC simulations from \POWPY{8}
and \POWHER{7} with an angular ordered shower, is shown in Figure~\ref{fig:etaintercal_responses}.
The predicted response in the two MC simulations is found to capture the overall shape of the \etadet dependence.
However, the response predicted from simulations is consistently lower than that measured in data for the forward detector regions across all \pTavg bins.

\begin{figure}[htp!]
\begin{center}
 
\subfloat[]{\includegraphics[width=0.49\textwidth]{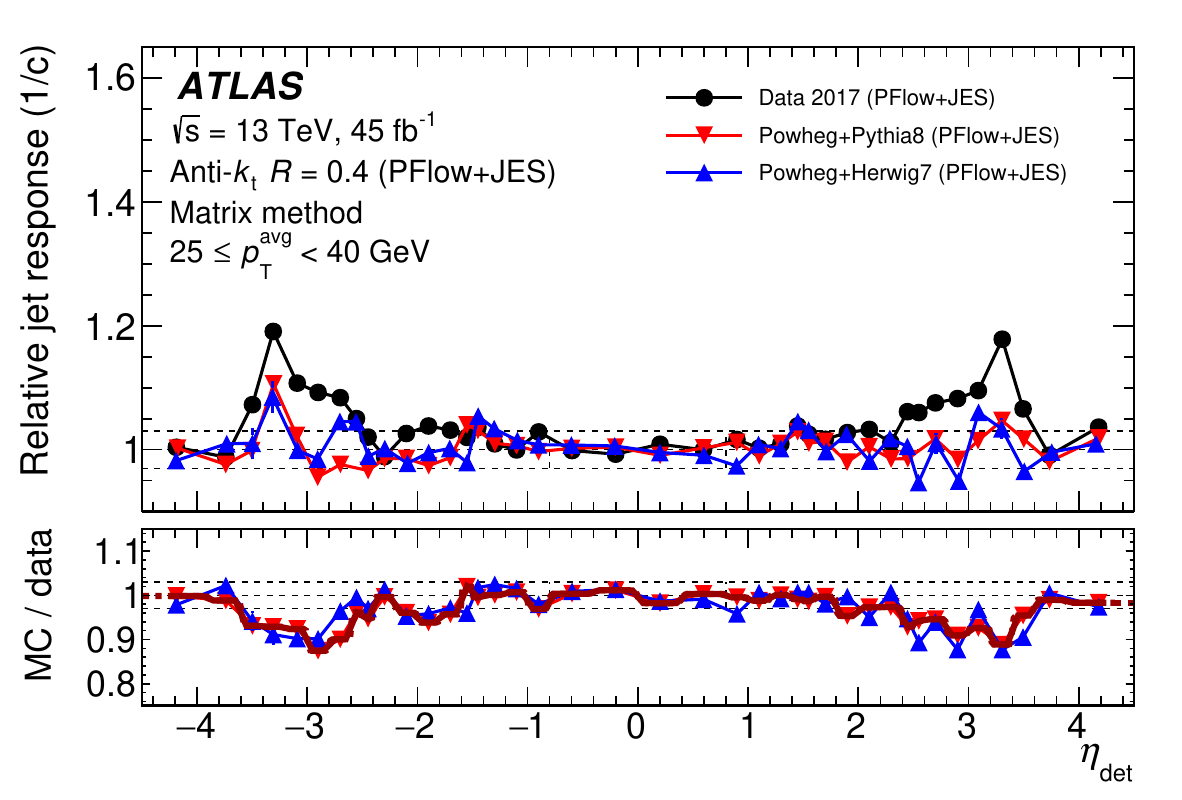}\label{subfig:etaintercal_vseta_lowpt}}
\subfloat[]{\includegraphics[width=0.49\textwidth]{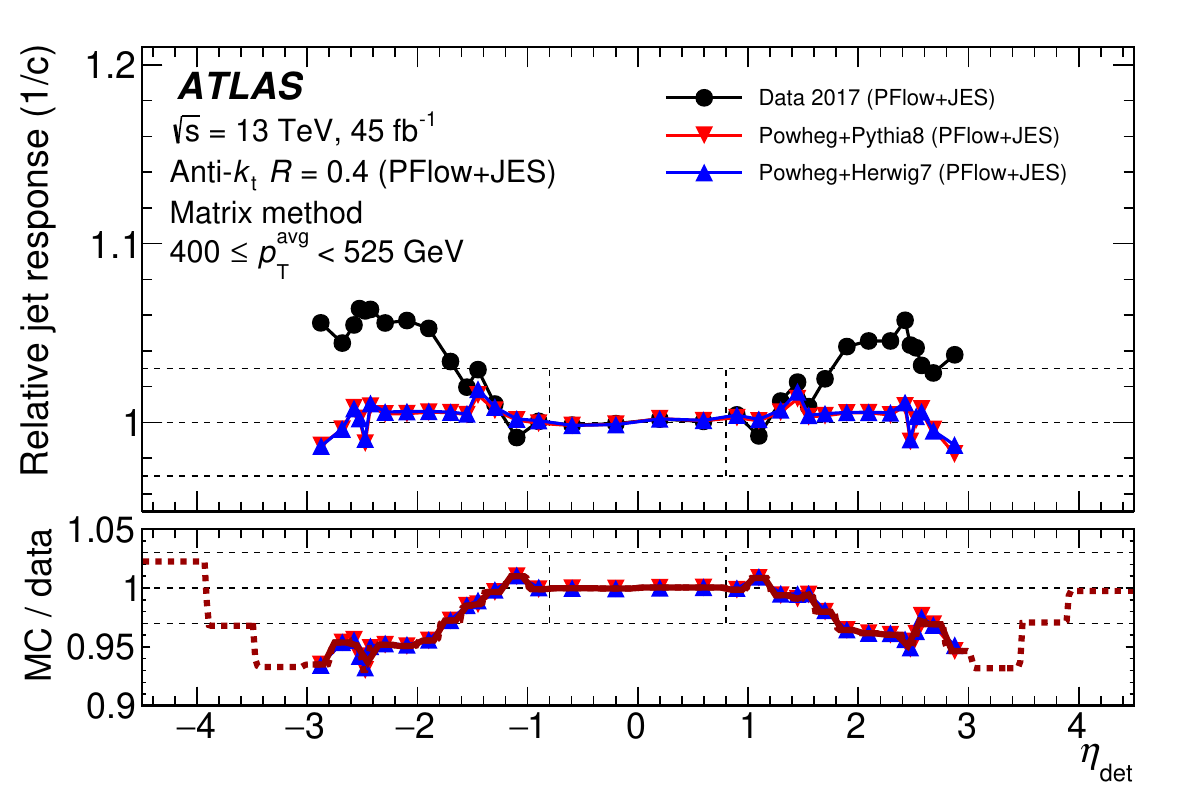}\label{subfig:etaintercal_vseta_highpt}}\\
\subfloat[]{\includegraphics[width=0.49\textwidth]{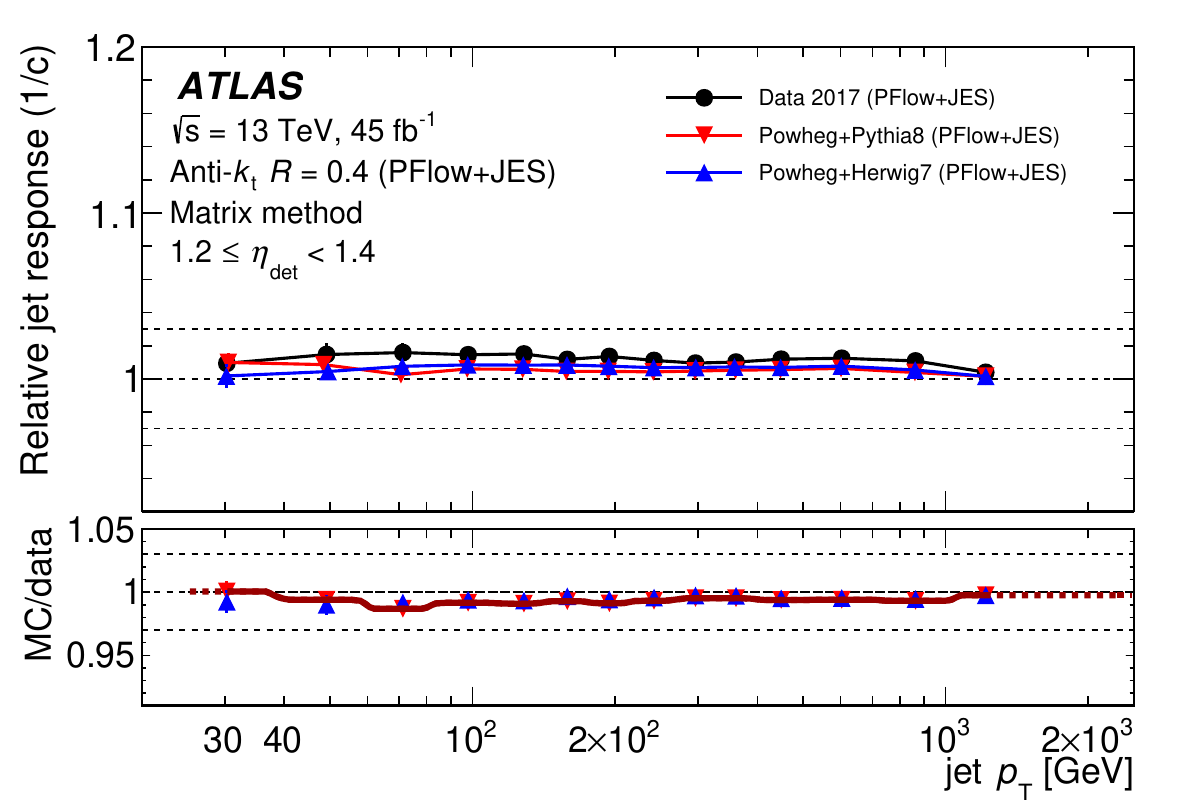}\label{subfig:etaintercal_vspt_forward}}
\subfloat[]{\includegraphics[width=0.49\textwidth]{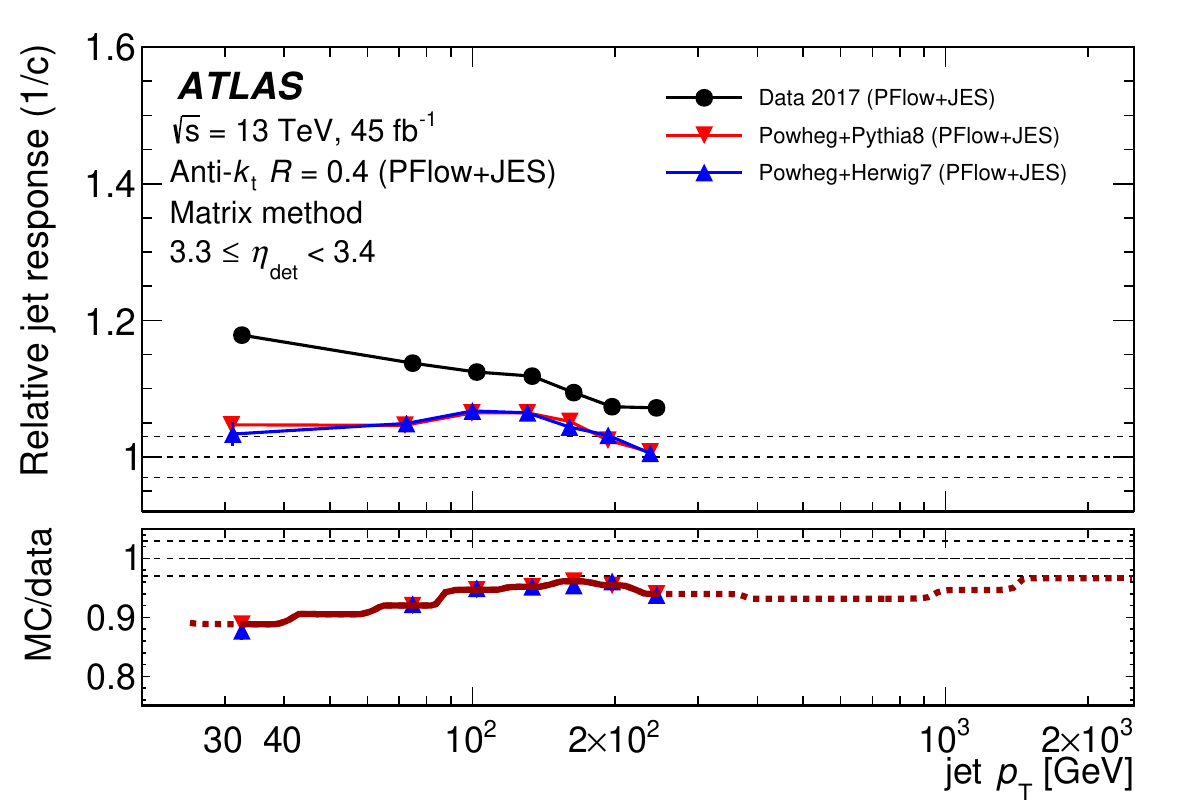}\label{subfig:etaintercal_vspt_veryforward}}\\
\caption{Relative jet response, $1/c$, calibrated with \EMPFlowJES  as a function of \etadet in the ranges
of \protect\subref{subfig:etaintercal_vseta_lowpt} $25~\GeV < \ptjet < 40~\GeV$
and \protect\subref{subfig:etaintercal_vseta_highpt} $400~\GeV < \ptjet < 525~\GeV$, and as a function of \pt in ranges
of \protect\subref{subfig:etaintercal_vspt_forward} $1.2 < \etadet < 1.4$ and \protect\subref{subfig:etaintercal_vspt_veryforward} $3.3 < \etadet < 3.4$.
The top panel presents the measured relative response for data (dots), \POWPY{8}  (triangles) and
\POWHER{7} (triangles). The bottom panel presents the MC-to-data response ratios represented by triangles and the smoothed \insitu corrections are represented by overlayed
curves, in which the solid line shows the derived calibration and the dashed line shows the extrapolated calibration to sparse detector regions using the two dimensional Gaussian Kernels. Two
perpendicular lines are drawn at $\etadet=\pm 0.8$ to indicate the central ($|\etadet|<0.8$) and the forward ($0.8<|\etadet|<4.5$) detector region. Three horizontal dashed lines are drawn at 0.97, 1, and 1.03 to
provide reference points for the viewer.}
\label{fig:etaintercal_responses}
\end{center}
\end{figure}

Uncertainties can arise due to the inaccurate description of physics, detector response and the dijet topology on the momentum balance. They are evaluated in terms of \pTavg and \etadet.
Uncertainties arising from the MC mismodelling are estimated by taking the difference between the smoothed residual correction between~\POWPY{8} and~\POWHER{7} with angular ordered shower. Other uncertainties due to the mismodelling in physics and event topology are estimated by modifying the requirement on the  third jet veto, the $\Delta\phi^{1,2}$ separation, and JVT.
 
Further studies are performed at particle and reconstruction level separately to disentangle physics and detector effects. The particle level can be used to study physics effects affecting the dijet balance due to additional parton radiations or out-of-cone corrections. It is performed using the same procedure as the reconstruction level except that no JVT requirements are applied. The matrix method is used as the nominal method while the central reference method~\cite{JETM-2018-05} is used as a cross-check. These physics effects induce a smooth and non trivial structure of the relative response $1/c$ with a slight asymmetry between positive and negative \etadet\ in the forward region as shown in Figure~\ref{fig:truthLevelTests_intercalib}. A similar structure with sharper variations due to convolution with detector effects is also present at the reconstruction level shown in Figure~\ref{subfig:etaintercal_vseta_lowpt}.
The systematic uncertainty $\Delta c$ on the intercalibration factor at particle level is
$\Delta c = \left| \frac{c^{\mathrm{syst}}}{c^{\mathrm{nominal}}} - 1 \right|$,
where $c^{\mathrm{syst}}$ is the intercalibration coefficient obtained with a different selection of the events, either a different selection on  $\pt^\mathrm{jet~3}/\pt^{\mathrm{avg}}$ or $\Delta\phi^{1,2}$.  By comparing Figures~\ref{fig:truthLevelTests_intercalib} and \ref{fig:truthLevelTests_intercalibSystematics}, the magnitude of these physics effects at particle level is similar to the magnitude of the systematic uncertainties designed to cover them, which are evaluated by varying the selection criteria for $\pt^\mathrm{jet~3}/\pt^{\mathrm{avg}}$ and $\Delta\phi^{1,2}$ at reconstruction level in data and MC simultaneously. Therefore, these uncertainties are not underestimated.
 
Variations in parton showering and hadronisation models can affect dijet balance that convolves both the physics and detector effects. The MC modelling uncertainty derived at particle level as a function of \etadet\ only considers the physics effects on the dijet balance and excludes impacts on the detector response which were evaluated in the jet flavour response uncertainty using various MC simulations discussed in Section~\ref{sec:gnnc}. Such a procedure will significantly reduce the MC modelling uncertainty shown in Figure~\ref{fig:truthLevelTests_MCmodeling} and avoid possible double counting of uncertainties.
 
\begin{figure}
\begin{center}
\subfloat[]{\includegraphics[scale=0.29]{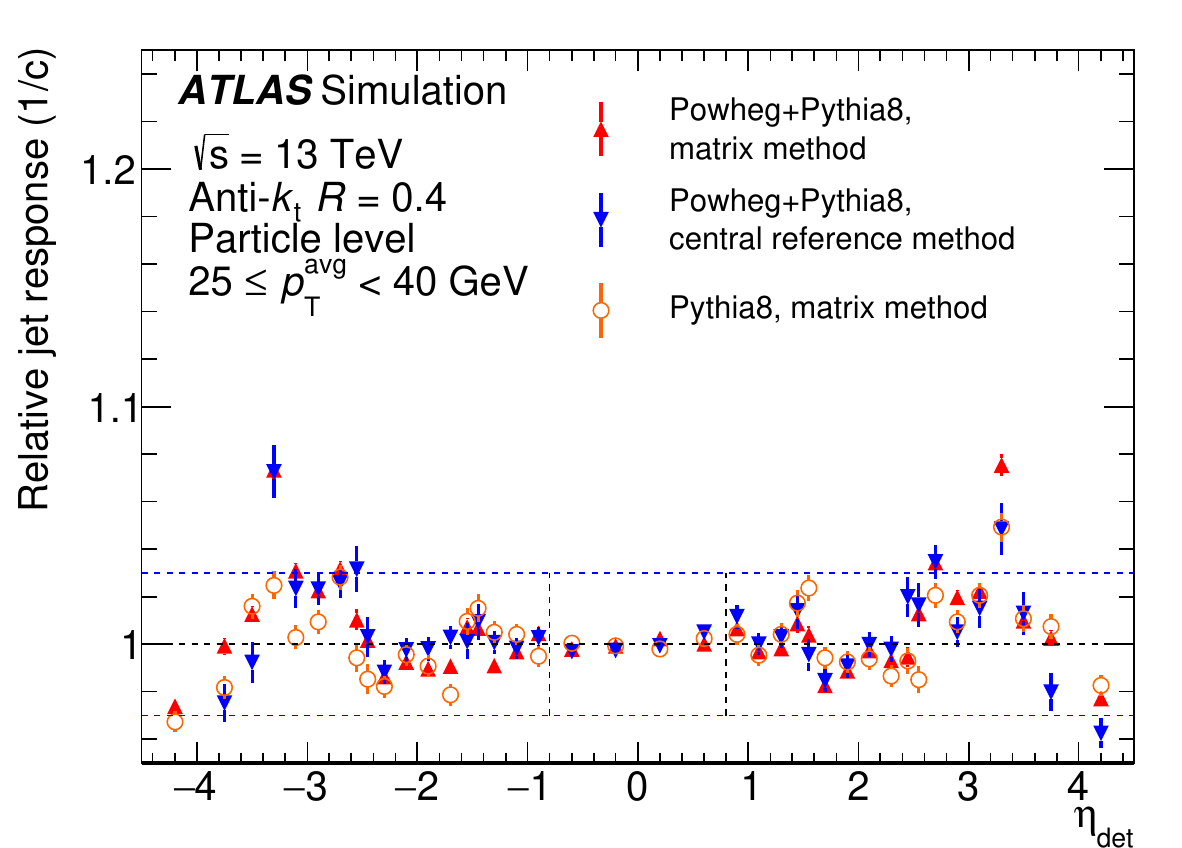}\label{fig:truthLevelTests_intercalib}}
\subfloat[]{\includegraphics[scale=0.29]{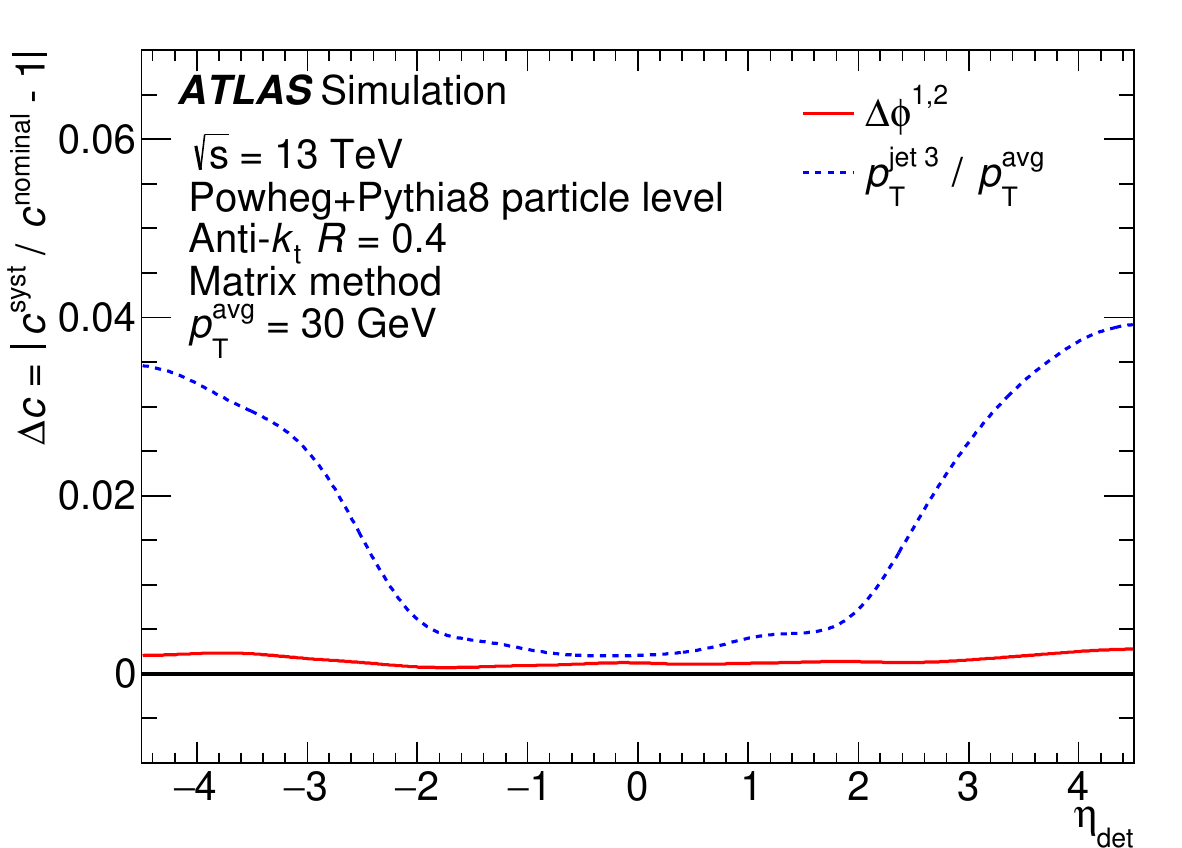}\label{fig:truthLevelTests_intercalibSystematics}}
\subfloat[]{\includegraphics[scale=0.29]{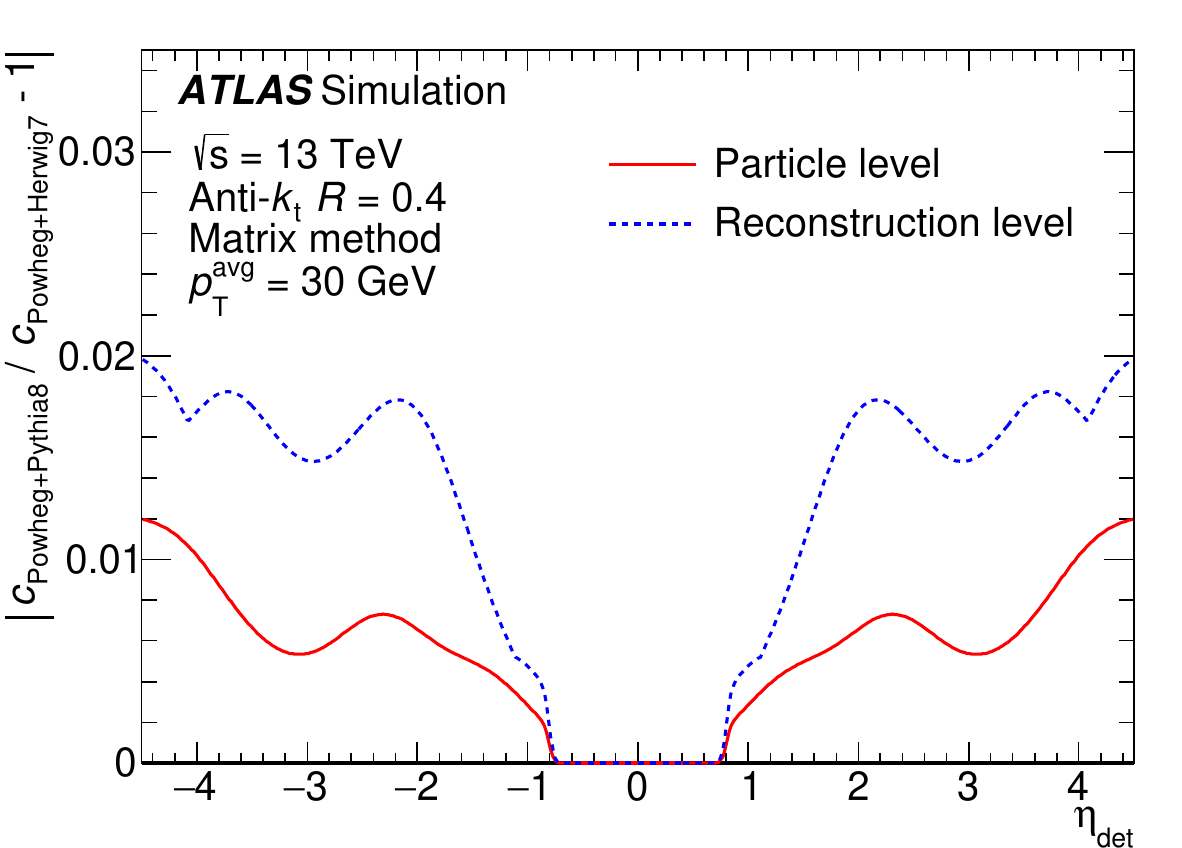}\label{fig:truthLevelTests_MCmodeling}}
\end{center}
\caption{Intercalibration coefficients and uncertainties derived at both the particle and reconstruction level, using MC simulated events reconstructed with the conditions of 2017 data-taking period. \protect\subref{fig:truthLevelTests_intercalib} Intercalibration coefficients obtained for different generators and different methods. \POWPY{8} is used as the nominal MC
generator. Three horizontal dashed lines are drawn at 0.97, 1, and 1.03 to provide reference points for the viewer. \protect\subref{fig:truthLevelTests_intercalibSystematics} Systematic uncertainties obtained with the matrix method and with \POWPY{8} at particle level.
The up and down variations are symmetrised by taking the maximum of either of them. A 2D smoothing is applied ( in $\pt^{\mathrm{avg}}$ and in $\eta$) with a
Gaussian kernel. \protect\subref{fig:truthLevelTests_MCmodeling} MC modelling uncertainty, evaluated either at reconstruction level in the dashed line or at particle level in the solid line.
A smoothing is applied after the computation of $\left| \frac{c_{\mathrm{\POWPY{8} }}}{c_{\mathrm{\POWHER{7} }}} - 1 \right|$.}
\end{figure}

Figure~\ref{fig:etaintercal_systs} shows the fractional uncertainties derived as a function of \etadet for two representative \pt values. The systematic uncertainty for $|\etadet|<0.8$ is set to zero as they are determined from the absolute \textit{in situ} JES measurements such as $Z/\gamma$+jet analysis. The fractional uncertainties increase with \etadet\
for $|\etadet|>0.8$ and illustrate a significant decrease with increasing \pt. Dominant uncertainties arise from the choice of event generators and variations in the selection criteria
on $\pt^\mathrm{jet~3}/\pt^{\mathrm{avg}}$. The total systematic uncertainty is significantly reduced by using MC modelling uncertainty estimated at particle level instead of reconstruction level.
It is worth noting that systematic variations in the selection criterion such as $\pt^\mathrm{jet~3}/\pt^{\mathrm{avg}}$
are performed simultaneously in data and simulation at reconstruction level while Figure~\ref{fig:truthLevelTests_intercalibSystematics} shows only the relative impact at particle level.
If there is a difference between up and down variations, then the systematic uncertainty is taken to be the larger absolute value.
Systematic uncertainties are symmetrised  around $\etadet=0$ between the positive and negative \etadet\ values using the most conservative approach, as whether the asymmetry of the systematic uncertainty in \etadet\ arises from statistical fluctuations or detector effects is unknown.

\begin{figure}[htp!]
\begin{center}
 
\subfloat[]{\includegraphics[width=0.49\textwidth]{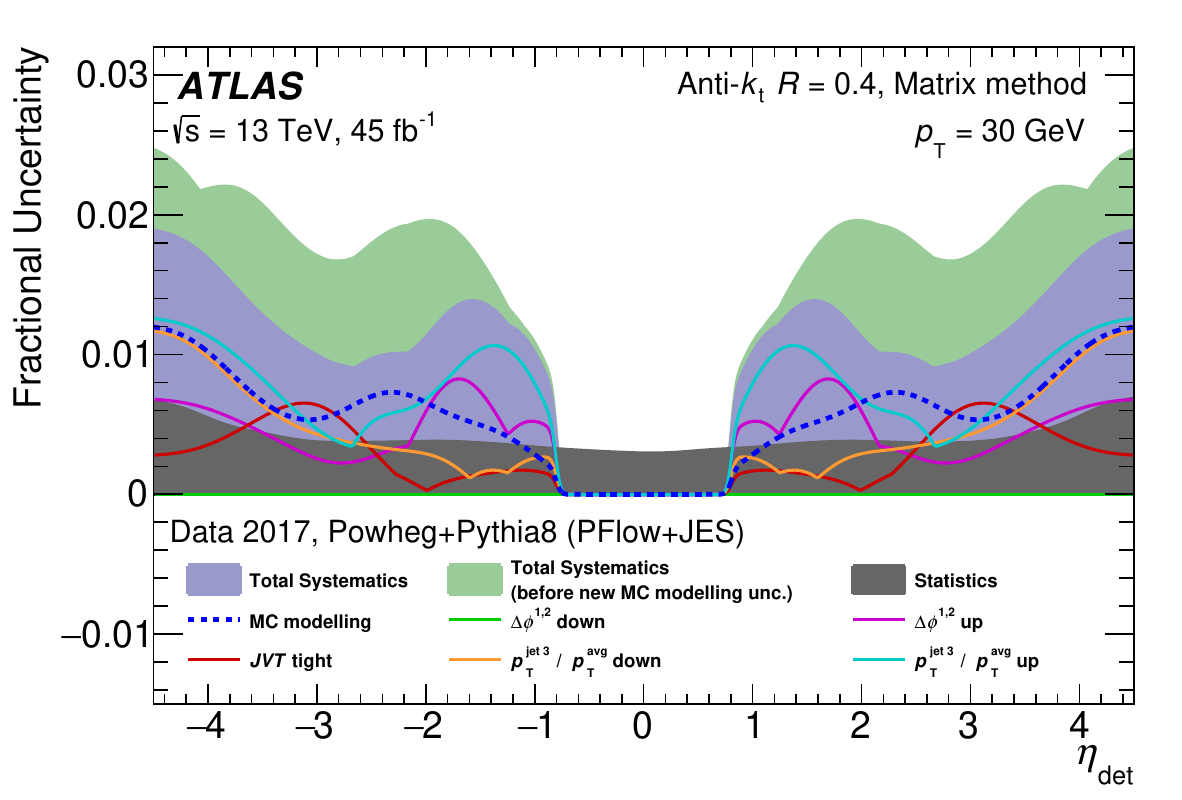}\label{subfig:etaintercal_syst_50}}
\subfloat[]{\includegraphics[width=0.49\textwidth]{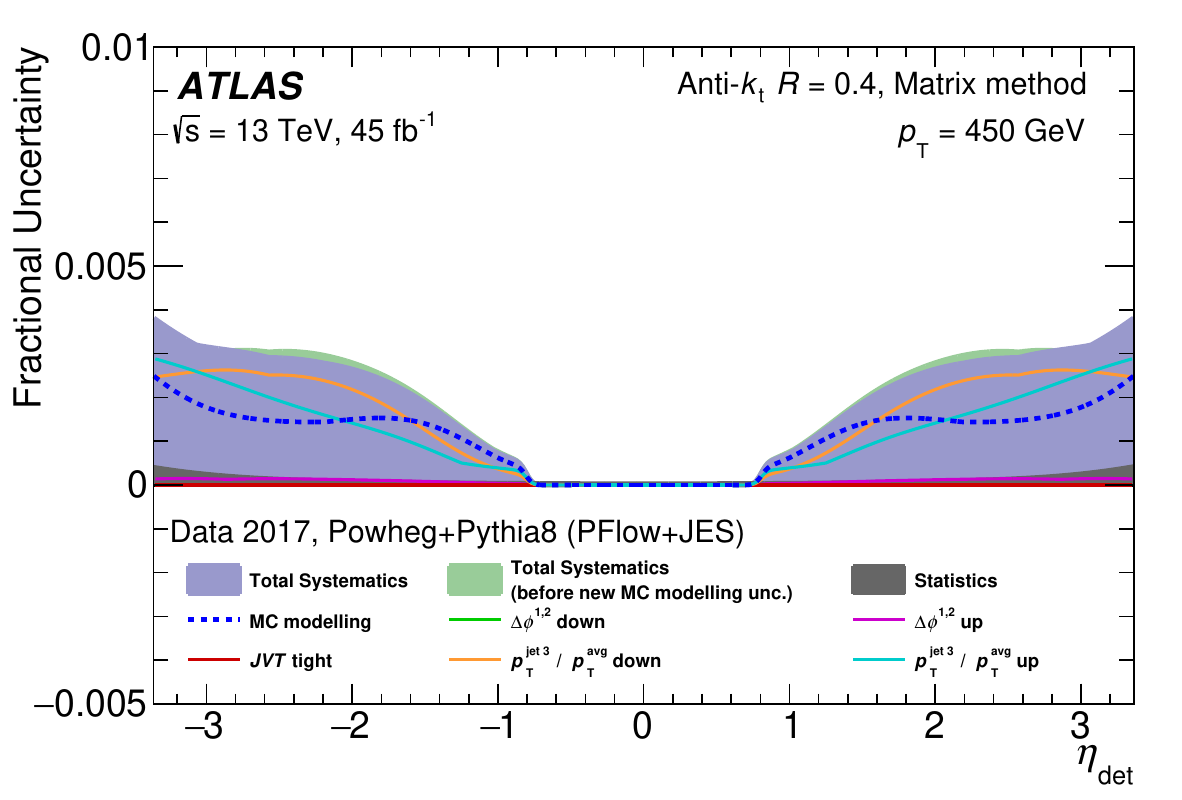}\label{subfig:etaintercal_syst_100}}\\
\caption{Systematic uncertainties in the $\eta$ intercalibration as a function of $\etadet$ for \EMPFlowJES jets of \protect\subref{subfig:etaintercal_syst_50} $\pt=35~\GeV$, \protect\subref{subfig:etaintercal_syst_100} $\pt=450~\GeV$. The total systematic uncertainty is represented by the middle shaded band, which is a quadrature sum of different components of systematic uncertainties marked by coloured lines. The statistical uncertainty is indicated by the lowest shaded band. The highest shaded band shows the quadrature sum of different components of systematic uncertainties using MC modelling uncertainty estimated at reconstruction level. A smoothing procedure is applied to the systematic uncertainty to suppress statistical fluctuations. The precision is limited by the MC modelling uncertainty estimated at particle level and variations in the selection criteria for $\pt^\mathrm{jet~3}/\pt^{\mathrm{avg}}$.}
\label{fig:etaintercal_systs}
\end{center}
\end{figure}

\FloatBarrier
\subsection{$Z/\gamma+$jet balance}
 
The next step in the jet calibration brings the absolute jet energy scale in data to the same scale in simulation by exploiting the \pt balance between
the hadronic recoil and a well-calibrated object such as a $Z$ boson or a photon.
The jet used in the \textit{in situ} analysis is required to be from the central detector region ($|\eta|<0.8$),
in which the derived correction can be applied to jets in the forward region via the $\eta$ intercalibration.
The $Z/\gamma+$jet balance measurement is built upon the precise determination of the energy of the photon or $e/\mu$ pair from a $Z$ boson decay.
These measurements benefit from the accurate knowledge of the energy scale and resolution of the leptons.
The calibration of electrons and photons is accurately known through measurements using $Z\rightarrow ee$ data and other final states~\cite{EGAM-2018-01}
while the muon calibration is determined to high precision through studies of $J/\Psi \rightarrow \mu \mu$, $Z\rightarrow \mu \mu$ and $\Upsilon\rightarrow \mu\mu$~\cite{MUON-2018-03}.
 
Three independent measurements consisting of $Z$+jet for the $Z\rightarrow ee$ and $Z\rightarrow \mu\mu$ decay channels,
and $\gamma$+jet are used to do the absolute \textit{in situ} calibration.
The $Z+$jet measurement provides enough of a sample size at low and medium jet \pt covering $17<\pt<800$~\GeV~ with limited precision above 800~\GeV.
The $\gamma+$jet analysis provides a complimentary measurement at medium and high \pt covering $30<\pt<2000$~\GeV,
with limited precision below 100~\GeV~due to prescaled low-\pt triggers, jets misidentified as photons, and MC event generator choices.
 
The MPF method measures the \pt balance between the reference objects and the full hadronic recoil in $Z/\gamma+$jet events.
This technique allows the calorimeter response to the hadronic showers to be computed directly.
It has low susceptibility to pile-up and underlying event which is uniform across the detector and thus cancelled out in the MPF method. According to conservation of transverse momentum,
the transverse momentum of all of the hadronic activity in a $Z/\gamma+$jet event, \pTrecoil, should be equal and opposite to the transverse momentum of the reference boson, \pTref, at particle level,
such that
 
\begin{equation}
\vec{p}_\text{T,truth}^\text{\,ref} + \vec{p}_\text{T,truth}^{\text{\,recoil}} = 0.
\label{eq:conservofmom}
\end{equation}
 
At the detector level, the well-calibrated objects have a response of one while
the calorimeter response to the hadronic recoil $r_{\text{MPF}}$ is lower than unity, resulting in possible
missing transverse momentum \vmet\ in the event. Therefore, equation~\ref{eq:conservofmom} can be written as:
\begin{equation}\nonumber
\ptrefvec + r_{\text{MPF}}\,\, \ptrecvec = - \vmet.
\end{equation}
 
Projecting the vector terms along the direction of the reference boson using a unit vector $\hat{n}_\text{ref}$ in the transverse plane, then $r_{\text{MPF}}$ is
only dependent on the missing transverse momentum and the transverse momentum of the reference boson.
The average of $r_{\text{MPF}}$, $\Response_\text{MPF}$, is measured as a function of \ptref , the \pt of the reference $Z/\gamma$ boson:

\begin{equation}\nonumber
\Response_{\text{MPF}} = \left< 1 + \frac{\hat{n}_\text{ref} \cdot \vmet}{\ptref } \right>\,
\end{equation}
 
where $\vmet$ is computed using particle-flow objects calibrated at the EM scale.
 
$Z+$jet events are selected using either the lowest-\pt unprescaled dielectron or dimuon trigger.
The lowest-\pt\ threshold in the dielectron trigger corresponds to 15~\GeV~ for each electron while the lowest-\pt\ threshold
in the dimuon trigger corresponds to 14~\GeV~\cite{TRIG-2018-01,TRIG-2018-05} for each muon.
Both the leptons are required to have $\pt>20$~\GeV\ to have fully efficient triggers.
Electrons or muons must satisfy loose identification and isolation criteria~\cite{EGAM-2018-01,MUON-2018-03}.
Electrons are required to fall within $|\eta|=2.47$ and are rejected if they fall in the calorimeter crack region $1.37<|\eta|<1.52$. Muons must fall within $|\eta|=2.4$.
The oppositely charged electron and muon pair is required to have an invariant mass around the $Z$ boson mass, $66<m_{ee/\mu\mu}<116$~\GeV.
$\gamma+$jet events are selected using a combination of prescaled and unprescaled single photon triggers in which the lowest prescaled trigger \et threshold is 10~\GeV.
Photon candidates entering the analysis are required to have $\ET^{\gamma}>25$~\GeV\ and $|\eta^{\gamma}| < 1.37$ and to satisfy the
tight identification and isolation selection criteria~\cite{EGAM-2018-01}. The jet is removed if it falls within $\Delta R=0.4$ (0.35) of a photon (lepton).

Further selection criteria are imposed in the $Z/\gamma+$jet measurements to reduce the impact from pile-up and additional parton radiations.
To suppress contamination from pile-up, jets are required to satisfy the cleaning criteria and to satisfy the JVT requirement.
Events must contain a jet with \pt greater than 10~\GeV\ that falls within $|\eta|=0.8$.
To suppress effects from additional parton radiations, further requirements are imposed on the azimuthal angle between the reference boson and the leading jet
$\Delta \phi^{\text{ref, jet}} > 2.9$ and \pt of the subleading jet $\pt < \text{max}(0.3 \times \pt^{\text{ref}},12)~\GeV$, where the subleading jet falls within $|\eta|=4.5$.
 
The MPF response as a function of reference boson \pt is shown in Figure~\ref{fig:MPF_response_zjet} and Figure~\ref{fig:MPF_response_gammajet}
using $Z+$jet and $\gamma$+jet events for data and two distinct MC samples.
The MC sample used to derive the nominal calibration for $Z+$jet ($\gamma$+jet) corresponds to \MADPY ~ (\Pythia{8}). The alternative MC sample corresponds to
\Sherpa to determine the uncertainty from the MC event modelling.
The dip in the MPF response
at low $\ptref$ arises due to two opposing effects: the jet reconstruction threshold which tends to increase the response at the lowest jet \pt values between 17~\GeV and 20~\GeV
and the apparent rise in MPF response as a function of \pt. The MC-to-data response ratio are rather consistent between $Z+$jet and $\gamma+$jet.
 
\begin{figure}[htp!]
\begin{center}
 
\subfloat[]{\includegraphics[width=0.49\textwidth]{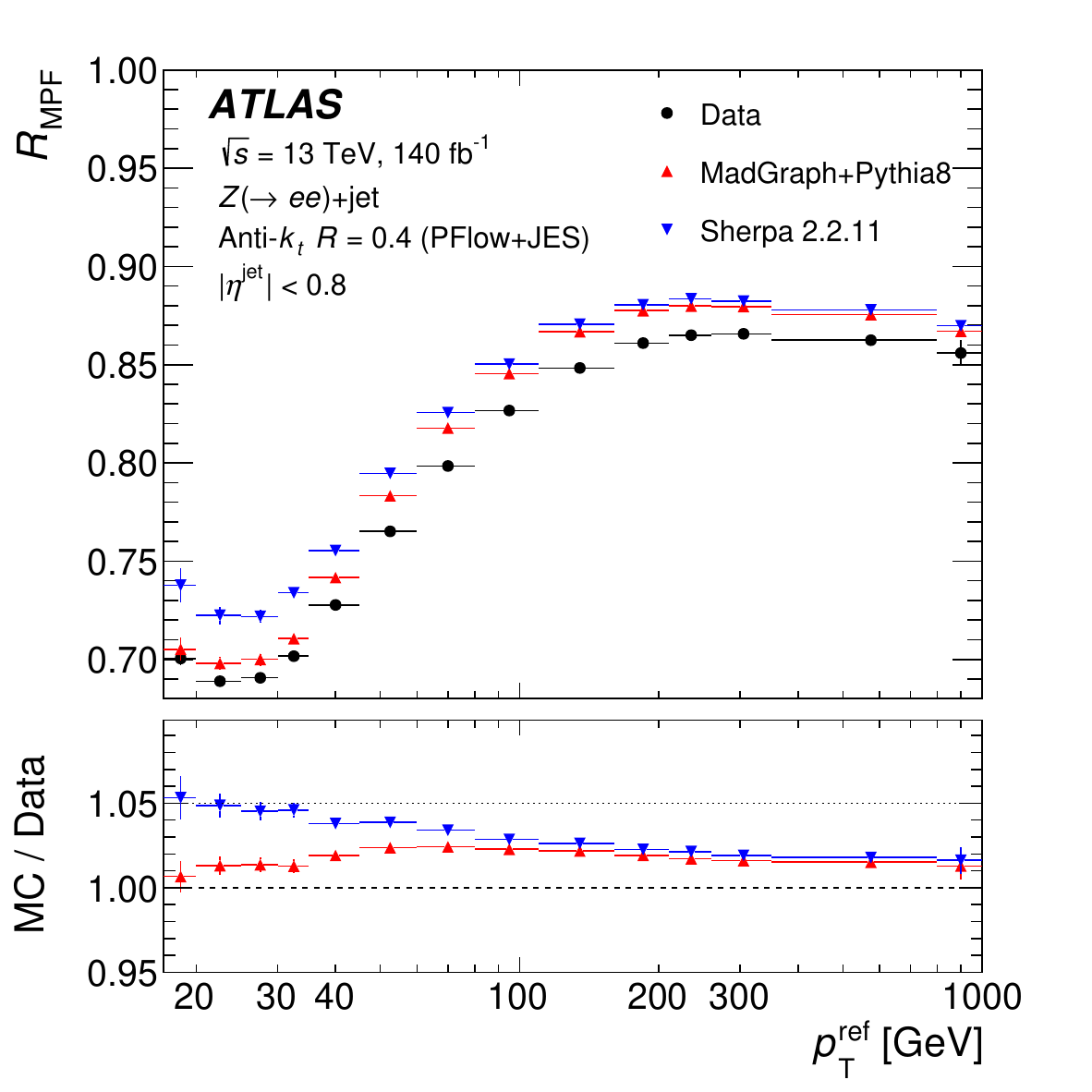}\label{subfig:mpf_response_zjet_e}}
\subfloat[]{\includegraphics[width=0.49\textwidth]{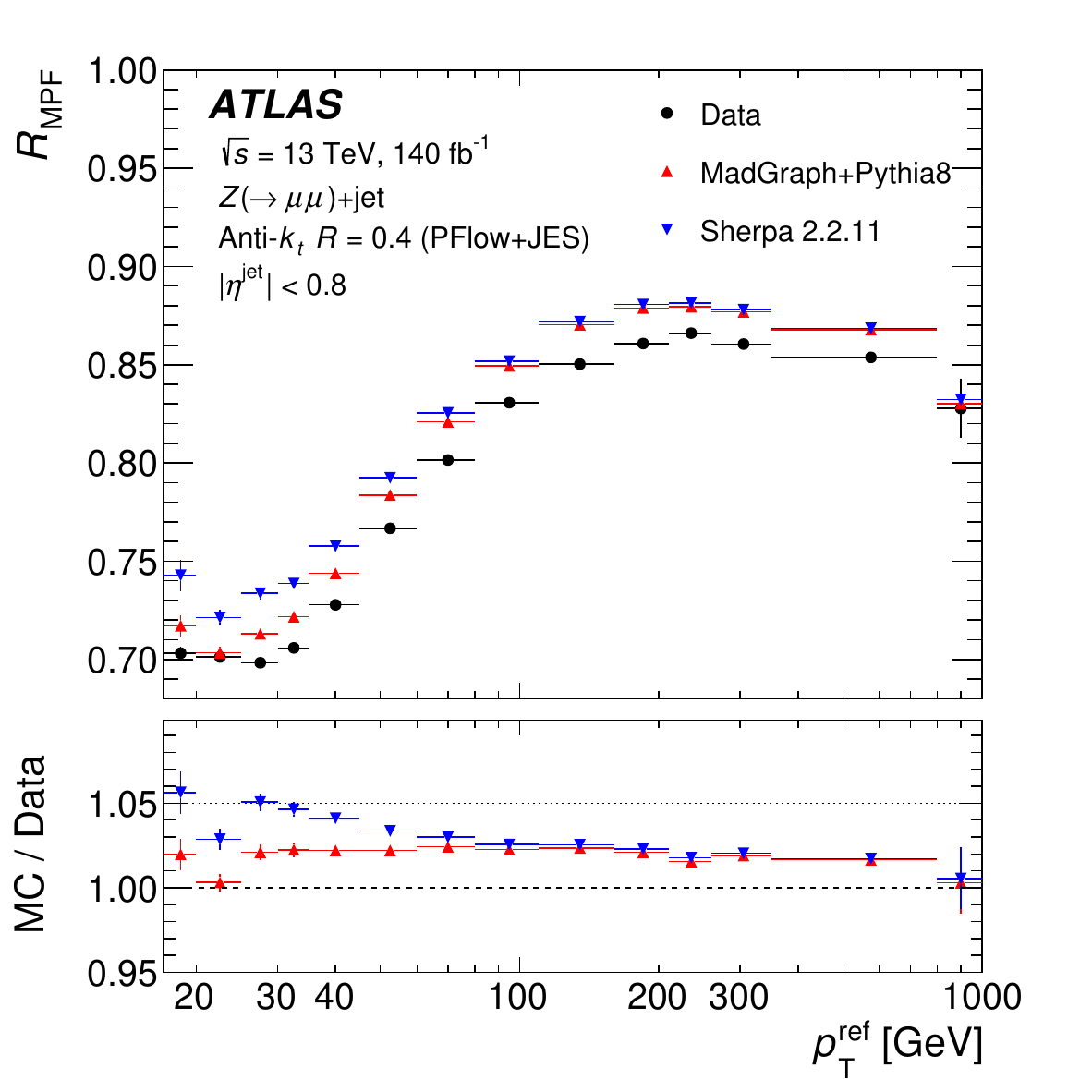}\label{subfig:mpf_response_zjet_mu}}\\
\caption{
The MPF response as a function of \ptref measured in data and simulations in $Z$+jet events for \protect\subref{subfig:mpf_response_zjet_e} $Z\rightarrow ee$ and \protect\subref{subfig:mpf_response_zjet_mu} $Z\rightarrow \mu\mu$. The data are represented by the black dots. The \MADPY predictions are represented by the triangles while the \Sherpa predictions are represented by the inverted triangles. The MC-to-data response ratios are shown in the bottom panel. The error bars correspond to the statistical uncertainties.}
\label{fig:MPF_response_zjet}
\end{center}
\end{figure}
 
\begin{figure}[htp!]
\begin{center}
\includegraphics[width=0.4\textwidth]{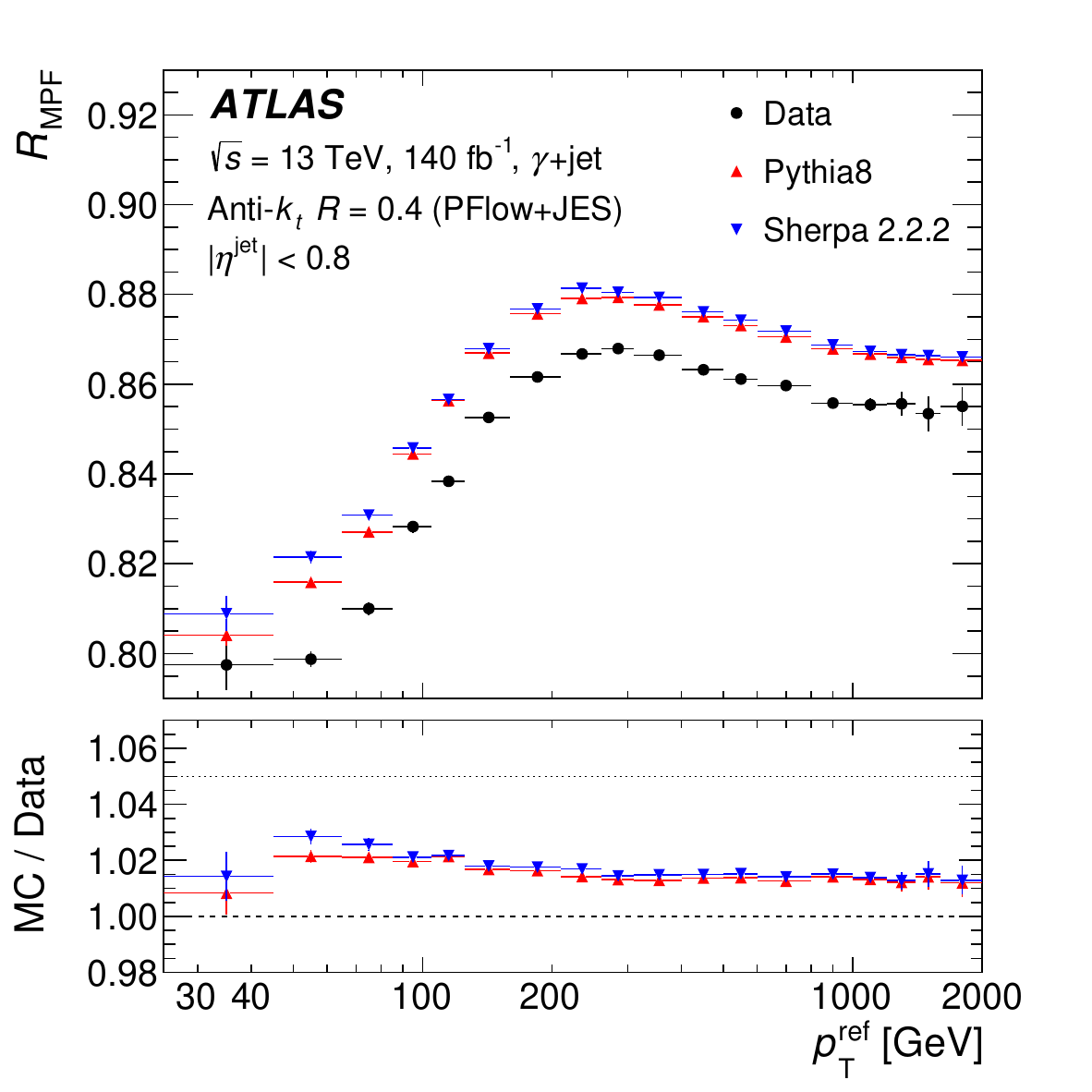}
\caption{The MPF response as a function of \ptref measured in data and simulations for $\gamma+$jet. The data are represented by the black dots. The \Pythia{8} predictions are represented by the triangles while the \Sherpa predictions are represented by the inverted triangles. The MC-to-data response ratios are shown in the bottom panel. The error bars correspond to the statistical uncertainties.}
\label{fig:MPF_response_gammajet}
\end{center}
\end{figure}
 
Several sources of systematic uncertainties are considered.
Uncertainties due to the energy scale and resolution of the reference objects $e/\mu/\gamma$ are derived from existing calibrations for each object and propagated through the corresponding analysis.
The impact of additional parton radiation on the response measurement is evaluated by varying the selection criteria for the subleading jet veto and $\Delta \phi^{\text{ref, jet}}$.
Uncertainties arising from pile-up suppression are estimated by comparing the response measurement between tighter and looser JVT working points.
Uncertainties arising from photon purity in $\gamma+$jet events are assessed using the same methodology documented in~\cite{STDM-2017-12},
in which one of the jets is misreconstructed as a photon.
The pseudo-experiments are implemented in the estimate of uncertainties to reduce statistical fluctuations.
 
The uncertainties for the calibration are presented for the $Z\rightarrow ee$ and $Z\rightarrow \mu\mu$ measurements in Figure~\ref{fig:mpf_uncertainties_zjet} and for the $\gamma+$jet measurement in Figure~\ref{fig:mpf_uncert_gammajet}. Uncertainties are dominated by the modelling in MC simulations in the low- and medium-\pT regions, and by the energy scale of the photon/electron for $\pt>100$~\GeV.

\begin{figure}[htp!]
\begin{center}
 
\subfloat[]{\includegraphics[width=0.45\textwidth]{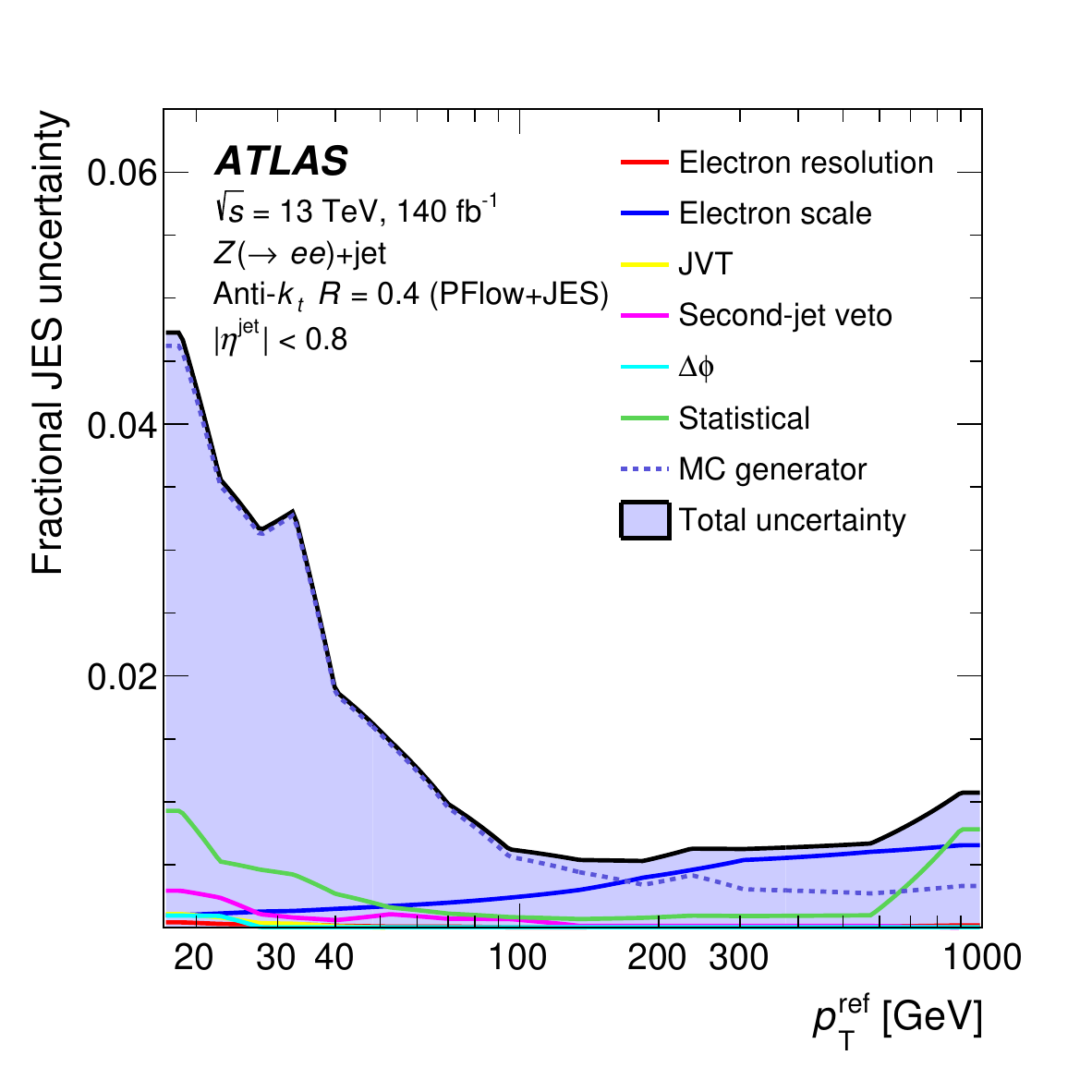}\label{subfig:mpf_uncert_zjet_ee}}
\subfloat[]{\includegraphics[width=0.45\textwidth]{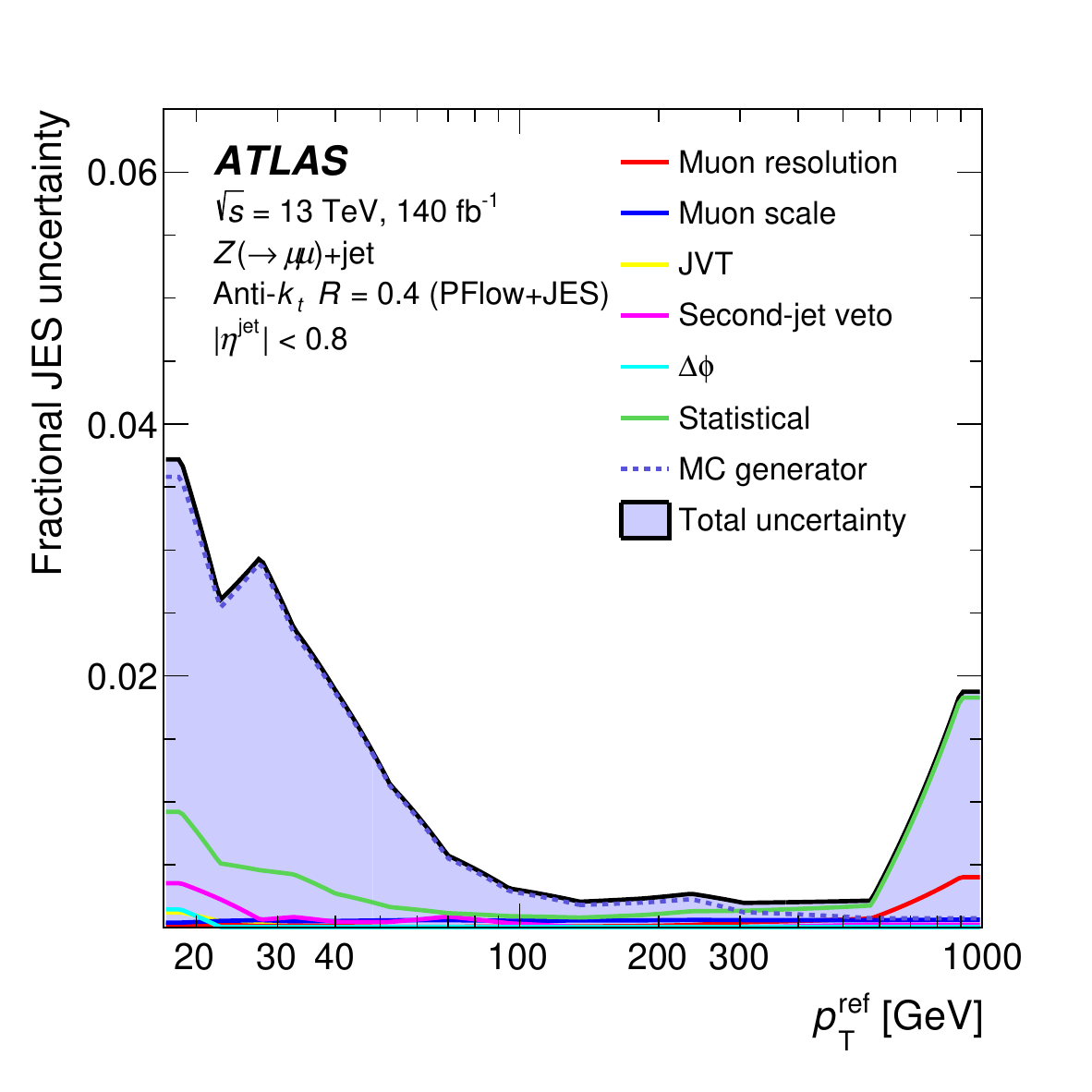}\label{subfig:mpf_uncert_zjet_mumu}}\\
\caption{Systematic uncertainty in the MPF response ratios as a function of \pt for jets calibrated up to, and including, the $\eta$ intercalibration for \protect\subref{subfig:mpf_uncert_zjet_ee} $Z(\rightarrow ee)+$jet events and \protect\subref{subfig:mpf_uncert_zjet_mumu} $Z(\rightarrow \mu\mu)+$jet events. Uncertainties arise from JVT, the subleading jet veto and $\Delta \phi^{\text{ref, jet}}$ requirement in the analysis selection. Uncertainties due to electron and muon energy scale and resolution are propagated through the analysis. The statistical uncertainty of the MC-to-data response ratios and the uncertainties due to choice of event generators are shown. Each uncertainty is smoothed to suppress statistical fluctuations.}
\label{fig:mpf_uncertainties_zjet}
\end{center}
\end{figure}
 
\begin{figure}[htp!]
\begin{center}
\includegraphics[width=0.45\textwidth]{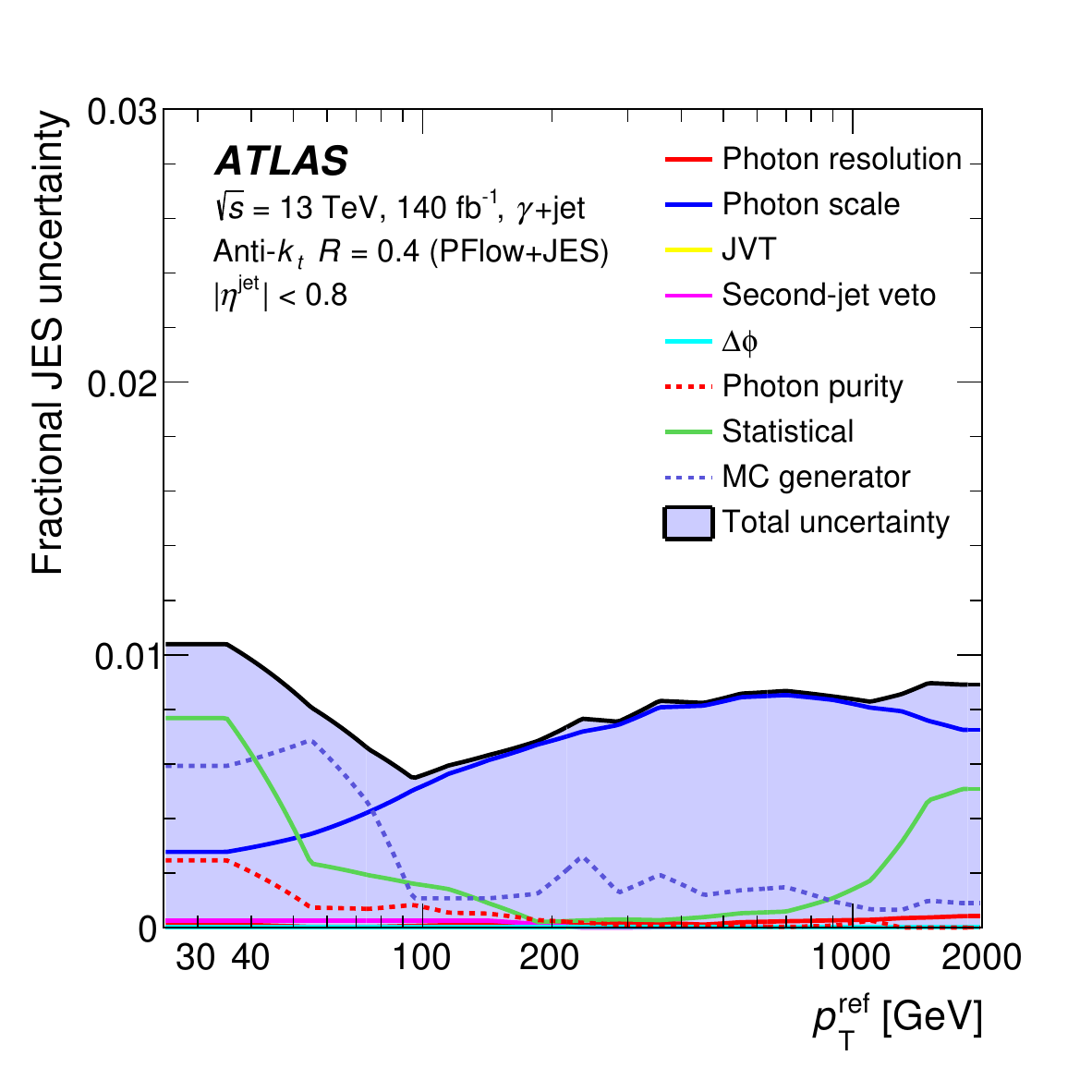}
\caption{Systematic uncertainty in the MPF response ratios as a function of \pt for jets calibrated up to, and including, the $\eta$ intercalibration in $\gamma$+jet events. Uncertainties arise from JVT, the subleading jet veto and $\Delta \phi^{\text{ref, jet}}$ requirement in the analysis selection. Uncertainties due to photon energy scale and resolution are propagated through the analysis. The statistical uncertainty of the MC-to-data response ratios and the uncertainties due to choice of event generators and photon purities are shown. Each uncertainty is smoothed to suppress statistical fluctuation.}
\label{fig:mpf_uncert_gammajet}
\end{center}
\end{figure}
 
The derived calibrations are stable over a range of pile-up conditions in Run 2. Figure~\ref{fig:MPF_response_gamjet_pileup}
shows the MC-to-data response ratios as a function of $\mu$ and $N_\text{PV}$ in $\gamma+$jet event for $45<\pTref<65$~\GeV. The \textit{in situ} calibration is consistent as a function of $\mu$ or $N_\text{PV}$, demonstrating the expected stability.

\begin{figure}[htp!]
\begin{center}
 
\subfloat[]{\includegraphics[width=0.49\textwidth]{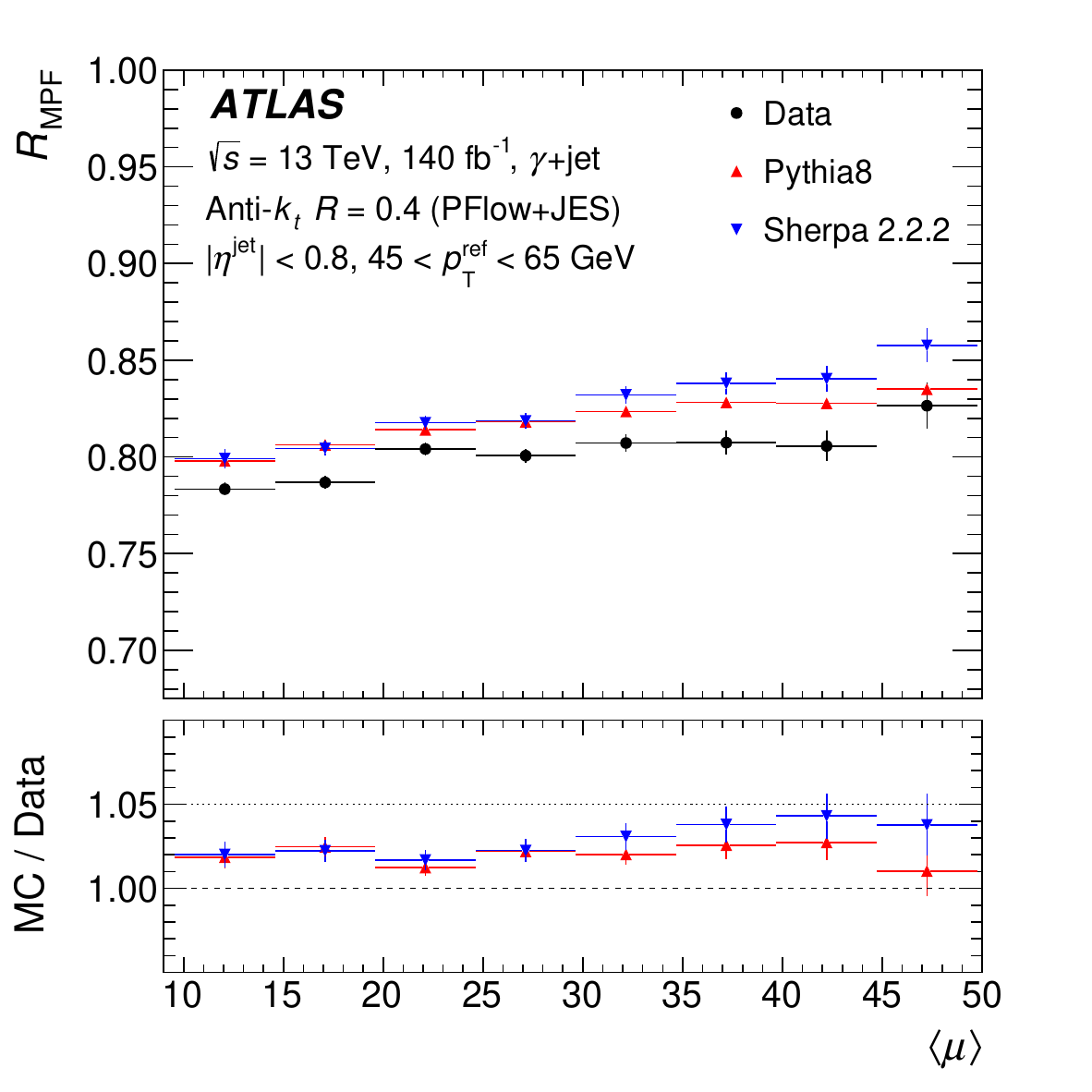}\label{subfig:mpf_response_gamjet_mu_45-65}}
\subfloat[]{\includegraphics[width=0.49\textwidth]{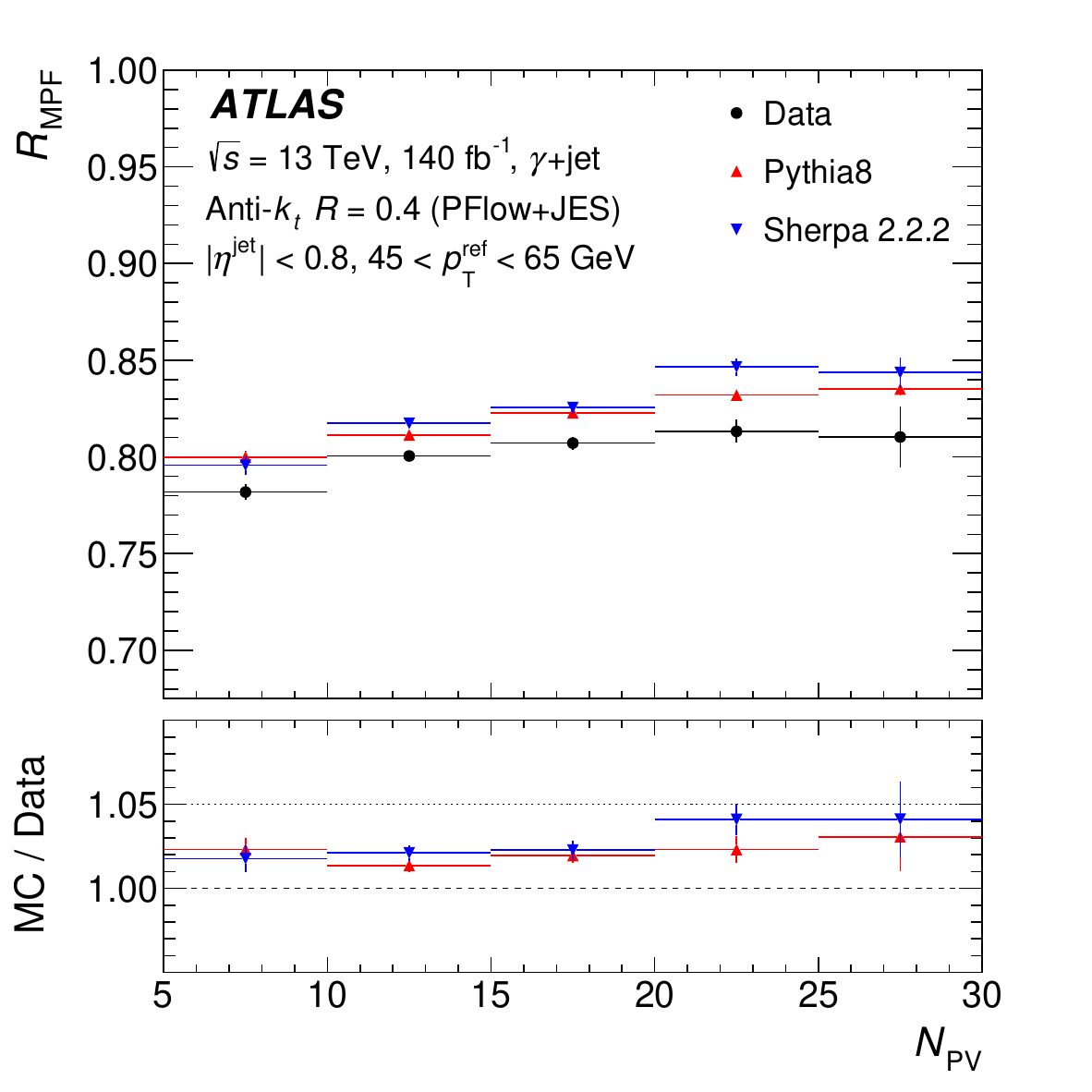}\label{subfig:mpf_response_gamjet_npv_45-65}}\\
\caption{
The MPF response for $45<\ptref<65$~\GeV~ measured in data and simulations as a function of \protect\subref{subfig:mpf_response_gamjet_mu_45-65} $\mu$ and
\protect\subref{subfig:mpf_response_gamjet_npv_45-65} $N_\text{PV}$. The data are represented by the black dots. The \Pythia{8} predictions are represented by the triangles while the \Sherpa predictions are represented by the inverted triangles. The MC-to-data response ratios are shown in the bottom panel. The error bars correspond to the statistical uncertainties.}
\label{fig:MPF_response_gamjet_pileup}
\end{center}
\end{figure}
\FloatBarrier
\subsection{ $b$-quark jet energy scale in $\gamma$+jet balance}
 
The measurement of the top-quark mass is limited by the $b$-quark jet energy scale ($b$JES) and a measurement of the $b$JES can potentially improve the precision.
The direct balance (DB) technique is used in $\gamma$+jet events to measure the balance of a ($b$-tagged) jet against a well-calibrated photon.
It represents the first measurement determining the $b-$tagged jet energy scale using the PFlow jets in this event topology.
\ptref is defined in terms of the reference object \pt,  $\ptref=\pt\times\cos\Delta\phi$, where \pt is the transverse momentum of
the photon, and $\Delta\phi$ is the azimuthal angle difference between the photon and the leading jet.
 
The selections are similar to the $\gamma$+jet selection in the MPF method unless stated otherwise.
Events must have a jet with $\pt>20$~\GeV\ instead of $\pt>10$~\GeV\ in the central detector region ($|\etadet|<0.8$).
The higher jet \pt threshold arises due to the tighter requirement on the jet transverse momentum in the $b-$tagging algorithm used.
To suppress additional radiation, the DB technique requires the subleading jet $\pt^\text{j2} < \text{max}(0.1 \times \ptref,15)~\GeV$ and $\Delta \phi^{\gamma,\text{jet}} > 2.8$.
Jets in the inner tracker coverage ($|\eta|<2.5$) containing $b-$hadrons are identified ($b-$tagged) by a multivariate algorithm (DL1r)
using information of impact parameters of tracks and displaced vertices~\cite{ATLAS:2022qxm}.
The $b-$tagging working points with an average efficiency of 60\%, 70\%, 77\% and 85\% are used.
The events are classified into inclusive and $b$-tagged categories.
The $b-$tagged category is predominantly composed of the $b-$ and $c-$ quark jets while the light quark and
gluon jets dominates the inclusive categories. A jet is labelled as $b-$ ($c-$) quark jets if any $b$($c$) parton or hadron
at particle level is found to be within a cone of $\Delta R<0.3$ around a reconstructed jet, otherwise it is labelled as light quark or gluons.
A summary of the jet flavour composition for inclusive and $b-$tagged jets
is documented in Table~\ref{tab:btaggingWP_composition}.
The 85\% $b-$tagging working points are dominated by the presence of $c-$quark jets and the measurement
can be used to constrain the $c$JES in $H\rightarrow c\bar{c}$ analysis~\cite{HIGG-2021-12} for instance.

Figure~\ref{fig:DB_response_gamjet} shows the DB response as a function of the reference photon \pt for the inclusive and $b$-tagged jets using $b-$tagging working
points with an average efficiency of 77\%. The MC simulations are in reasonable agreement with data.
The MC-to-data response ratios are found to be slightly below one for $b-$tagged jets and above one for inclusive jets in almost all bins.
The difference between DB response between~\Pythia{8} and \Sherpa arises due to different $b-$quark fragmentation and decay models.
Checks on the apparent rise of the DB response around 150~\GeV~ for $b-$tagged jets are done such as the quality of the DB response fit,
$b-$tagging scale factors applied in simulations, the jet flavour composition between neighbouring \pt bins and a looser second jet veto with $\pt^\text{j2} <0.2 \times \ptref$.
None of the checks mentioned above is responsible for the DB response rise around 150~\GeV. Hence these checks suggest that the feature is due to statistical fluctuations.
Figure~\ref{fig:DB_systs_gamjet} shows the uncertainties for the $b$-tagged case with a precision between 1\% and 5\% and inclusive jets with a precision up to 1\% for the chosen \pt\ range.
For $b-$tagged jets, uncertainties are dominated by the event generator modelling everywhere,
while for inclusive jets the precision is limited by the event generator modelling, photon purity and the subleading jet veto at lower \pt and photon energy scale for $\pt>70$~\GeV.

\begin{table}[htb]
\begin{center}
\scalebox{0.8}{
\begin{tabular}{|l|ll|ll|ll|ll|ll|}
\hline
\multirow{2}{*}{Hadron Label}&\multicolumn{2}{c|}{Inclusive jet} &\multicolumn{2}{c|}{WP 60\%}&\multicolumn{2}{c|}{WP 70\%}&\multicolumn{2}{c|}{WP 77\%}&\multicolumn{2}{c|}{WP 85\%}\\
&\Pythia{8}&\Sherpa&\Pythia{8}&\Sherpa&\Pythia{8}&\Sherpa&\Pythia{8}&\Sherpa&\Pythia{8}&\Sherpa \\
\hline
$b$               & 1.9\%&2.0\%  &92.1\%&85.8\%&81.3\%&79.2\%&59.2\%&61.7\%&36.3\%&39.2\%\\
\hline
$c$               &14.2\%&10.6\% &2.3\% &1.6\%&13.4\%&9.2\%&35.4\%&27.0\%&57.4\%&49.7\%\\
\hline
light $q$ or gluon&83.9\%&87.4\% &5.6\% &12.6\%&5.3\%&11.6\%&5.4\%&11.3\%&6.3\%&11.1\%\\
\hline
\end{tabular}
}
\caption{The average fractions of jet flavours for various $b-$tagging working points and inclusive jet.}
\label{tab:btaggingWP_composition}
\end{center}
\end{table}

\begin{figure}[htp!]
\begin{center}
\subfloat[]{\includegraphics[width=0.49\textwidth]{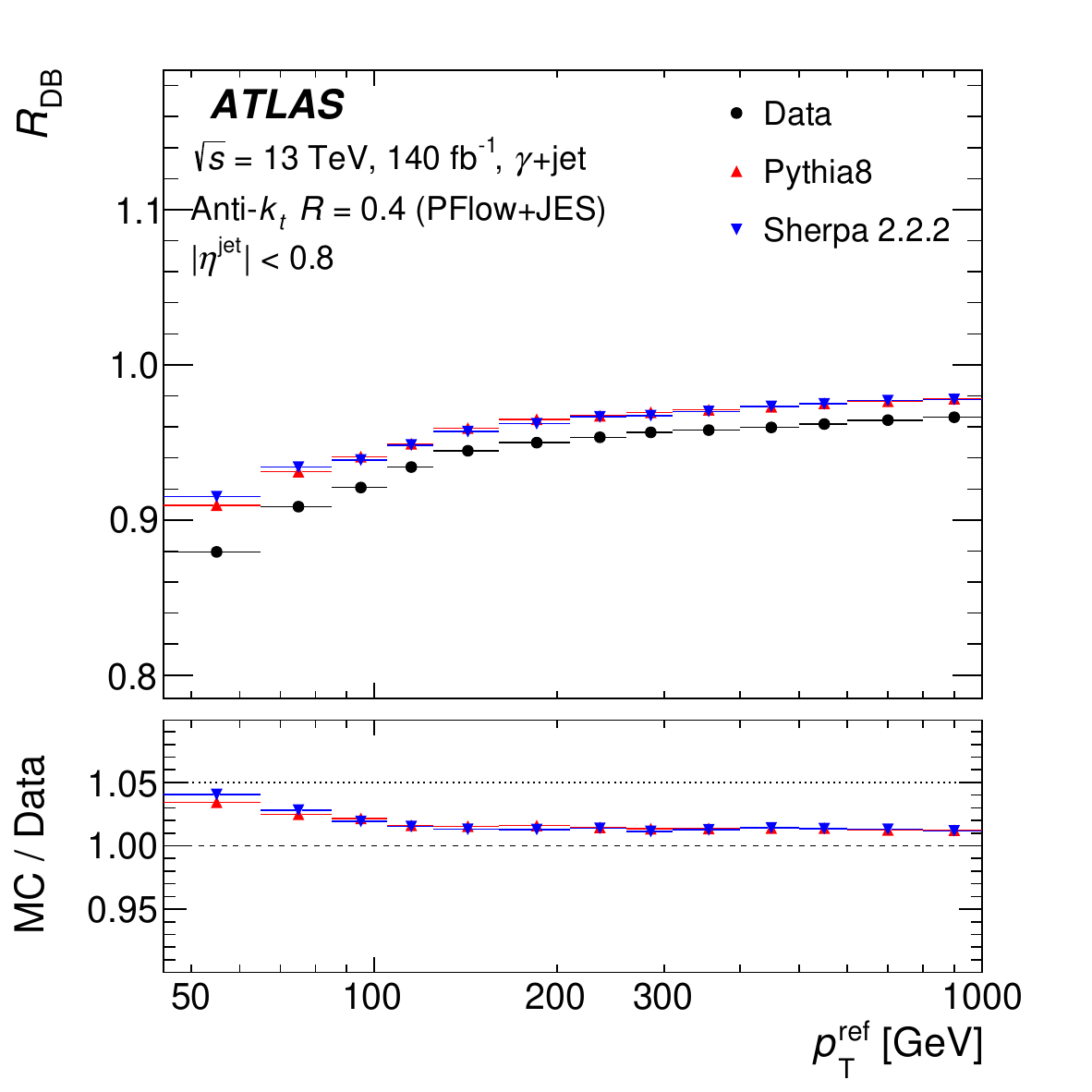}\label{subfig:db_response_gamjet_nonb}}
\subfloat[]{\includegraphics[width=0.49\textwidth]{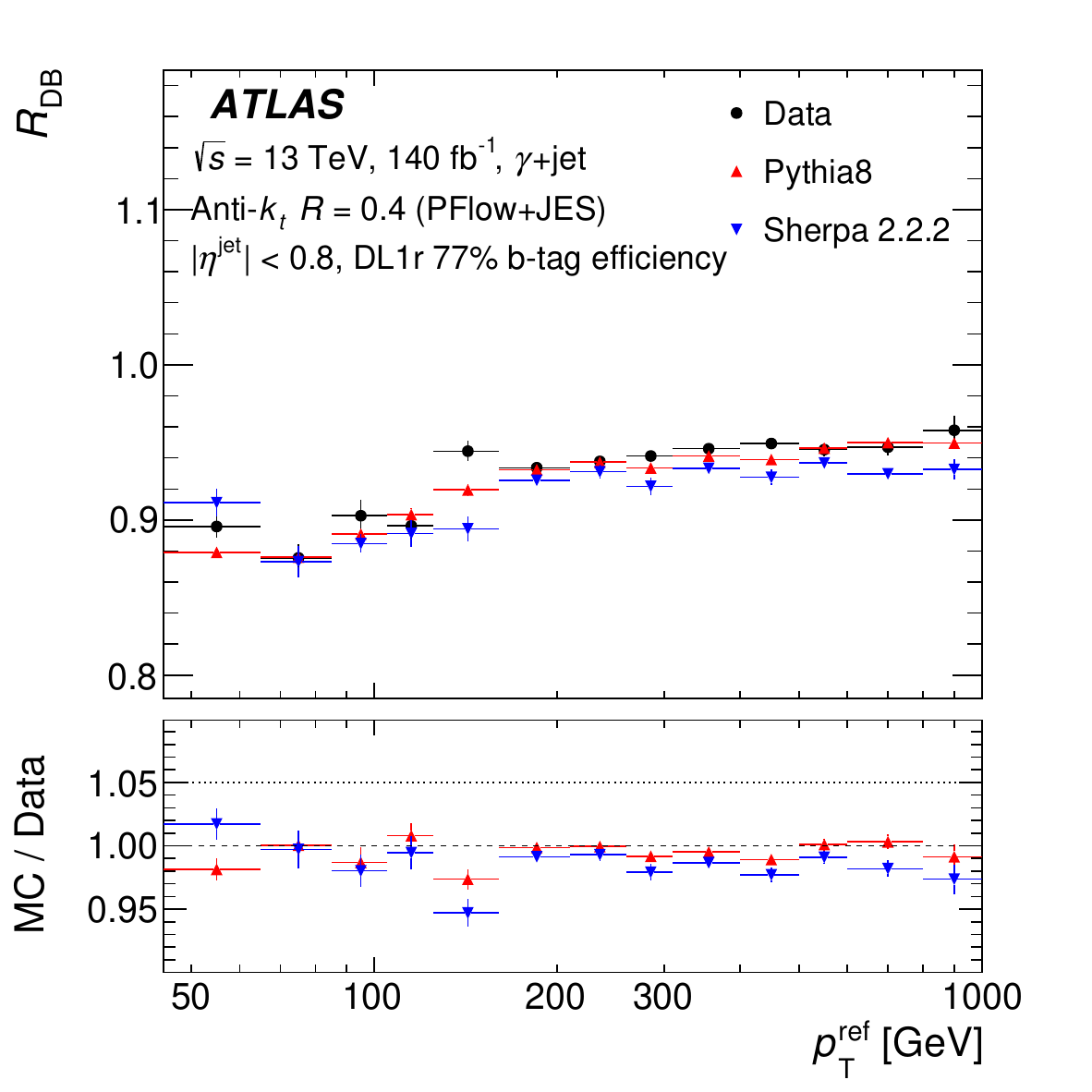}\label{subfig:db_response_gamjet_b}}\\
 
\caption{The DB response as a function of \ptref measured in data and simulations for \protect\subref{subfig:db_response_gamjet_nonb} $\gamma+$jet and \protect\subref{subfig:db_response_gamjet_b} $\gamma+b$-tagged jet. The data are represented by the black dots and the \Pythia{8} are represented by the triangles, and the (\Sherpa) predictions are represented by the inverted triangles. The MC-to-data response ratios are shown in the bottom panel. The error bars correspond to the statistical uncertainties.}
\label{fig:DB_response_gamjet}
\end{center}
\end{figure}
 
\begin{figure}[htp!]
\begin{center}
\subfloat[$\gamma+$jet]{\includegraphics[width=0.48\textwidth]{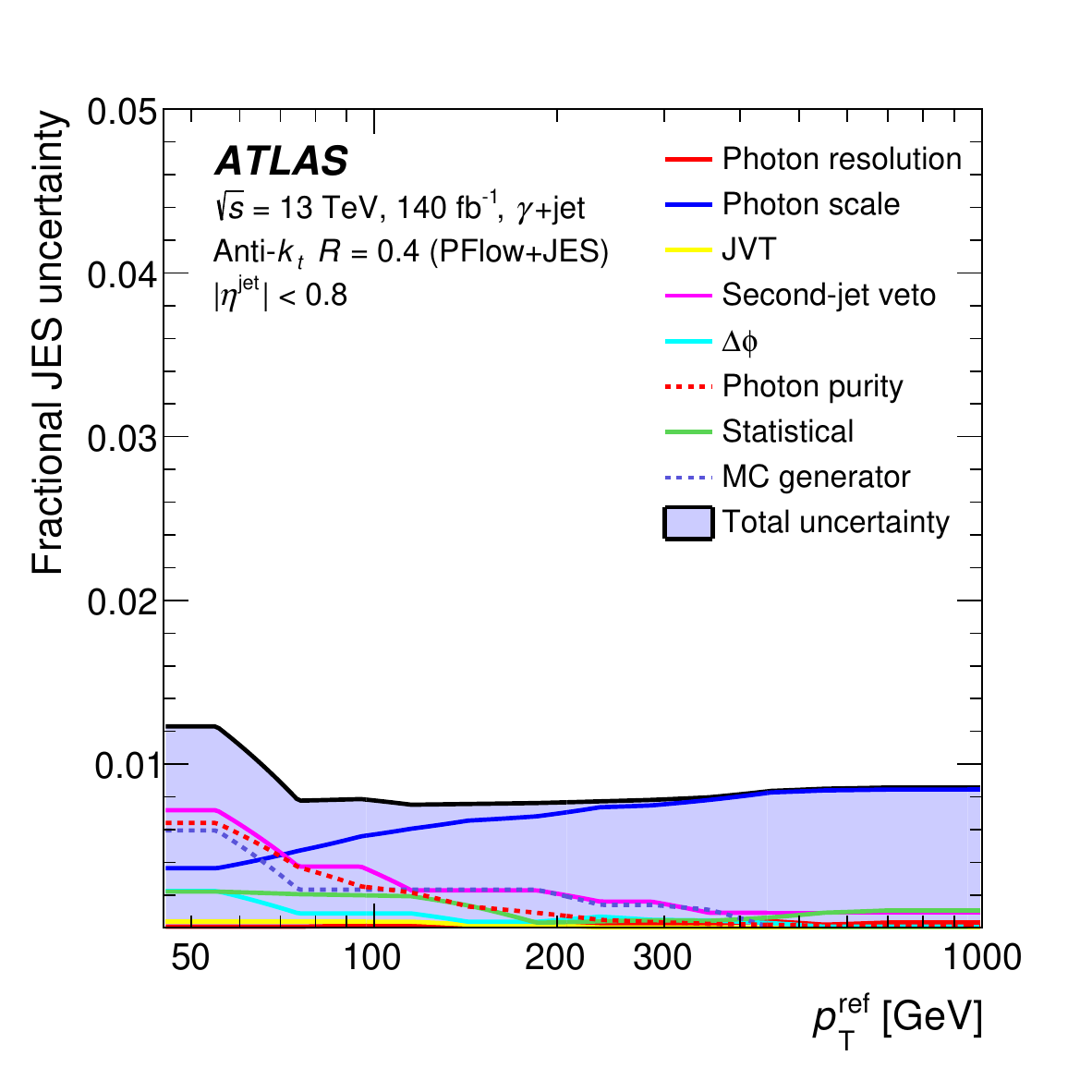}\label{subfig:db_response_gamjet_nonb_uncer}}
\subfloat[$\gamma+b$-tagged jet]{\includegraphics[width=0.48\textwidth]{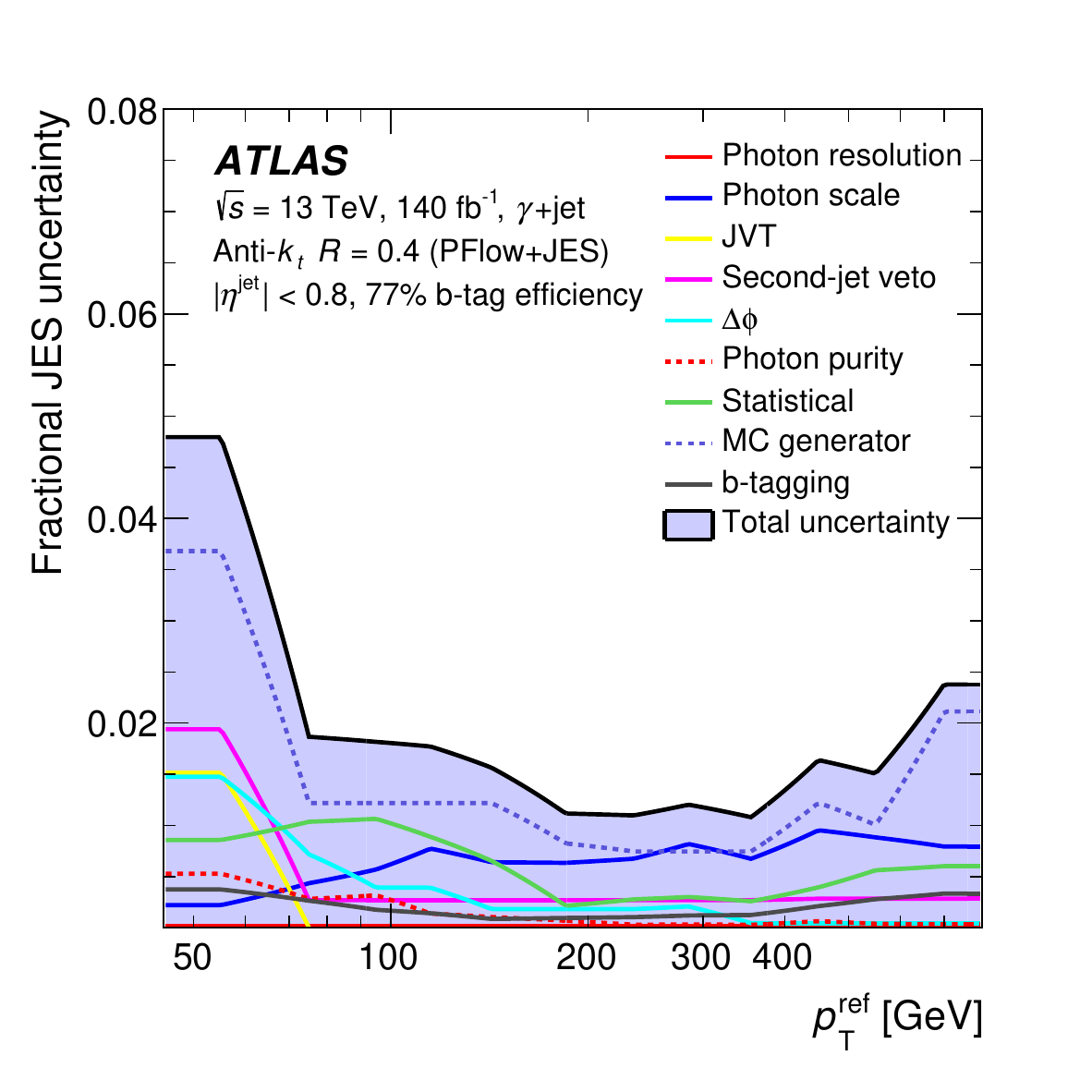}\label{subfig:db_response_gamjet_b_uncer}}
\caption{Summaries of uncertainties in the MC-to-data response ratio as a function of \ptref for \protect\subref{subfig:db_response_gamjet_nonb_uncer} $\gamma+$jet and \protect\subref{subfig:db_response_gamjet_b_uncer} $\gamma+b$-tagged jet. Uncertainties can arise from photon purity, $b-$tagging, JVT, $\Delta\phi$ and subleading jet veto requirement. Uncertainties due the photon energy scale and resolution are propagated through the analysis.
}
 
\label{fig:DB_systs_gamjet}
\end{center}
\end{figure}

A new observable $\tilde{R}_{b\text{JES}}$ is defined as a double ratio of $b-$tagged response to the inclusive jet response to further measure the energy scale differences
between the $b$-tagged and inclusive jets,
\begin{equation}\nonumber
\tilde{R}_{b\text{JES}} = \frac{\cal{R}^{\text{MC}}_{\text{b-tagged}}/\cal{R}^{\text{data}}_{\text{b-tagged}}}{\cal{R}^{\text{MC}}_{\text{inclusive}}/\cal{R}^{\text{data}}_{\text{inclusive}}}\, .
\end{equation}
As the nominal jet calibration is determined relative to the inclusive jet, such a double ratio can be applied on top of the nominal jet calibration to correct $b$JES\@.
The value of $\tilde{R}_{b\text{JES}}$ is determined to be below one using both the MC samples with
a slightly higher response in \Pythia{8} than \Sherpa shown in Figure~\ref{fig:DB_response_gamjet_bJES}.
The difference between the two event generators arises from different
fragmentation and decay models. The ratio, $\tilde{R}_{b\text{JES}}$, is also determined inclusively for
photon \ptref between 85 and 1000~\GeV for various $b$-tagging working points
in Table~\ref{tab:btaggingWP_bJES} to increase statistical precision for \Pythia{8} and \Sherpa, respectively.
It is foreseen to provide MC specific calibrations for the $b$JES to reduce
the effects arising from MC modelling. The ratio, $\tilde{R}_{b\text{JES}}$, was measured with unprecedented precision up to 1\%.
This in turn will improve precision in measurements of top mass.

\begin{figure}[htp!]
\begin{center}
\subfloat[]{\includegraphics[width=0.49\textwidth]{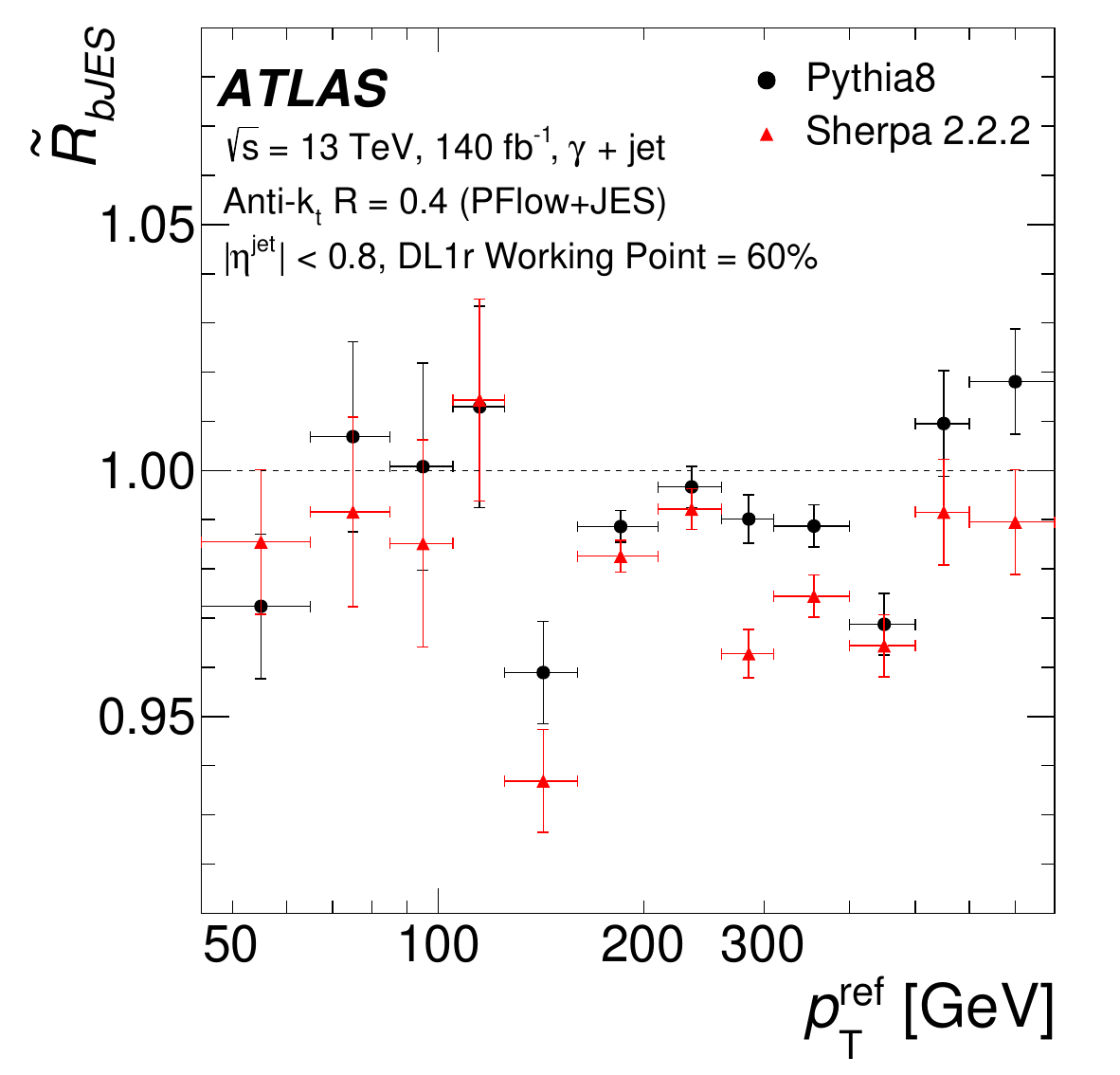}\label{subfig:db_bjes_60}}
\subfloat[]{\includegraphics[width=0.49\textwidth]{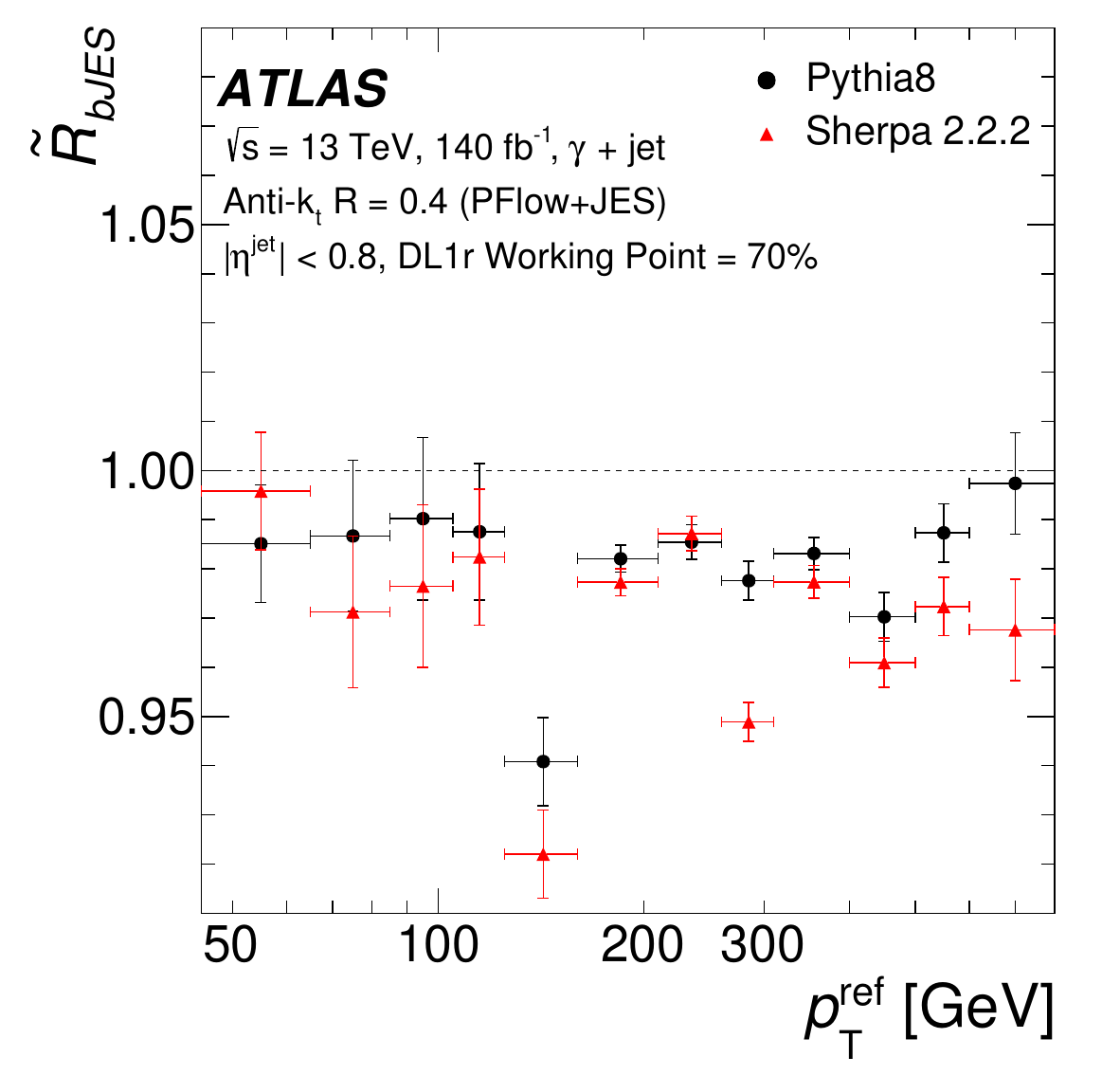}\label{subfig:db_bjes_70}}\\
\subfloat[]{\includegraphics[width=0.49\textwidth]{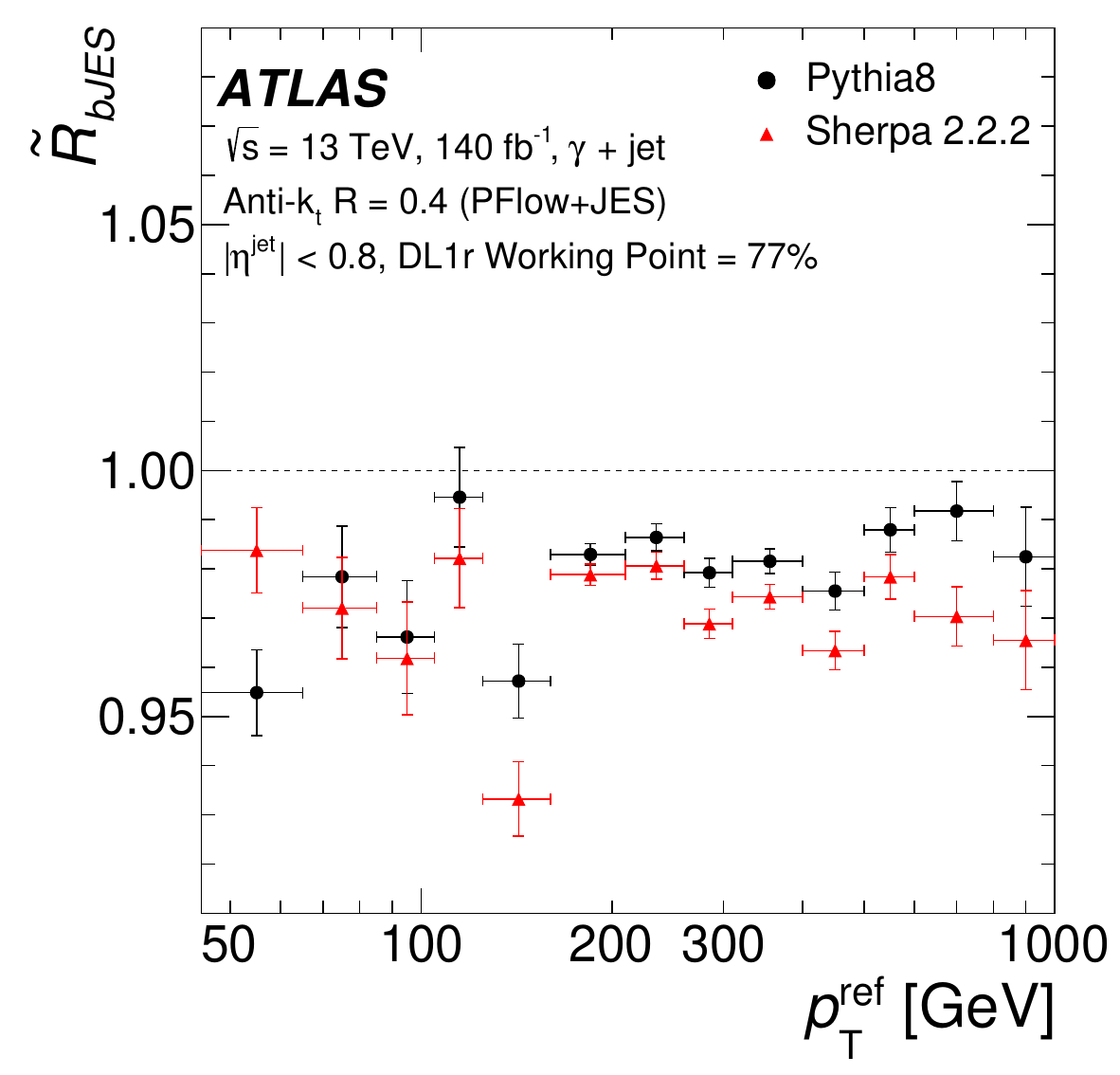}\label{subfig:db_bjes_77}}
\subfloat[]{\includegraphics[width=0.49\textwidth]{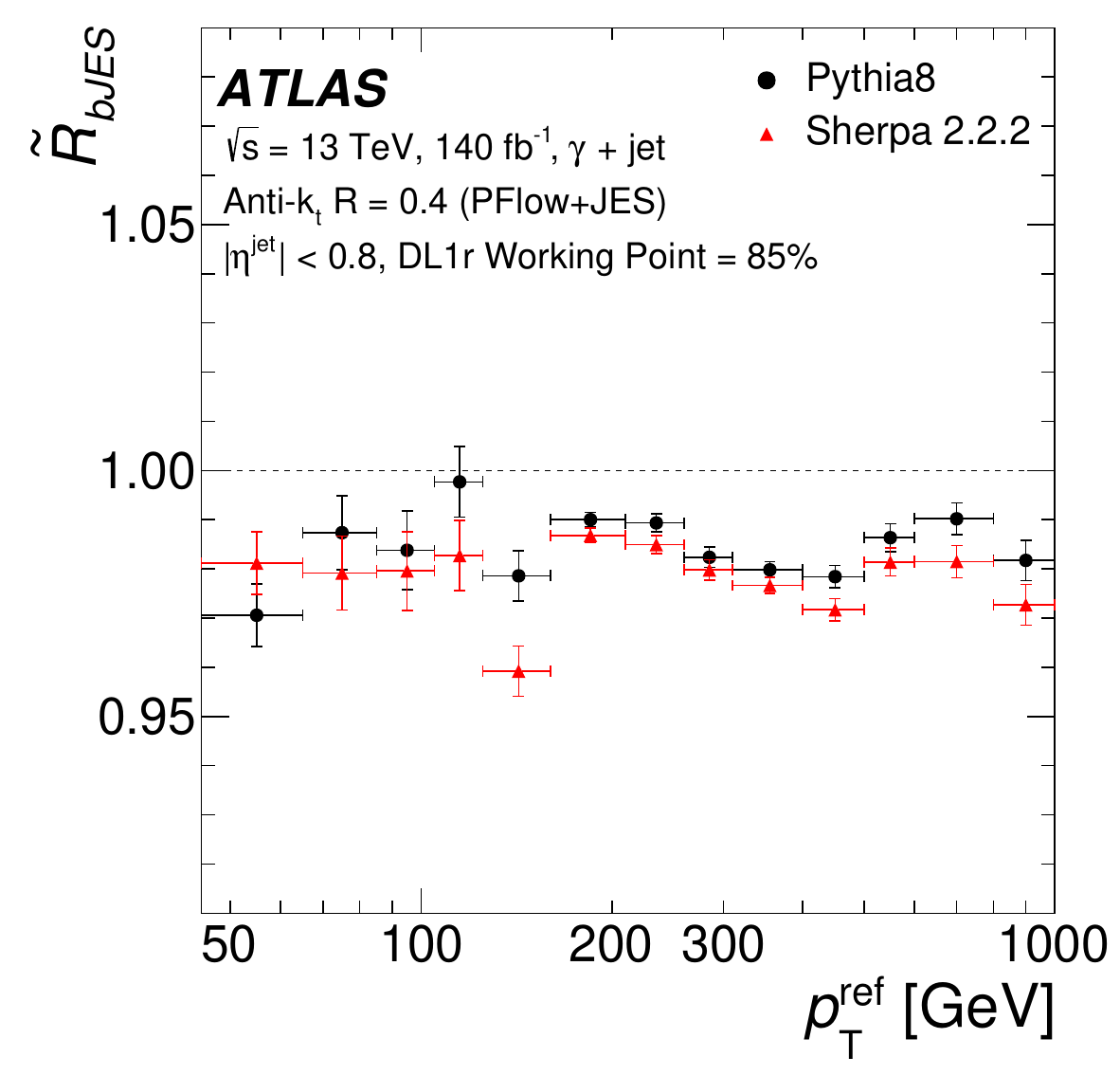}\label{subfig:db_bjes_85}}\\
\caption{$\tilde{R}_{b\text{JES}}$ as a function of reference photon \pt determined using either \Pythia{8} or \Sherpa for b-tagging working point
with an efficiency of \protect\subref{subfig:db_bjes_60} 60\%, \protect\subref{subfig:db_bjes_70} 70\%, \protect\subref{subfig:db_bjes_77} 77\% and
\protect\subref{subfig:db_bjes_85} 85\%. The error bars correspond to the statistical uncertainties.}
\label{fig:DB_response_gamjet_bJES}
\end{center}
\end{figure}

\begin{table}[htb]
\begin{center}
\begin{tabular}{|l|c|c|}
\hline
WP & \Pythia8 & Sherpa \\
\hline
60\% & $0.990\pm 0.010~\text{(stat.)} \pm 0.013~\text{(syst.)}$ & $0.984\pm 0.010~\text{(stat.)} \pm 0.013~\text{(syst.)}$ \\
70\% & $0.984\pm 0.010~\text{(stat.)} \pm 0.011~\text{(syst.)}$ & $0.974\pm 0.010~\text{(stat.)} \pm 0.012~\text{(syst.)}$ \\
77\% & $0.978\pm 0.006~\text{(stat.)} \pm 0.011~\text{(syst.)}$ & $0.966\pm 0.006~\text{(stat.)} \pm 0.011~\text{(syst.)}$ \\
85\% & $0.989\pm 0.004~\text{(stat.)} \pm 0.007~\text{(syst.)}$ & $0.979\pm 0.004~\text{(stat.)} \pm 0.007~\text{(syst.)}$ \\
\hline
\end{tabular}
\caption{$\tilde{R}_{b\text{JES}}$ obtained for various $b-$tagging working points using \Pythia{8} and \Sherpa separately for $85<\pt<1000$~\GeV.}
\label{tab:btaggingWP_bJES}
\end{center}
\end{table}


\FloatBarrier

\section{Conclusion}
\label{sec:conclusion}

The determination of the jet energy scale (JES) is presented using data recorded by the ATLAS experiment in $pp$ collisions at $\sqrt{s} = 13~\TeV$.
The calibration scheme used for \antikt jets reconstructed using radius parameter $R = 0.4$ consists of two steps:
a Monte-Carlo-based calibration that corrects jets to the truth jet scale, and an \insitu calibration correcting the scale of jets in data.
 
The simulation-based calibration implements several new strategies to improve the pile-up stability at higher \pt, closure, energy resolution, and modelling uncertainties of the jets.
Biases related to the determination of the pile-up \pt density were a dominant source of uncertainty for jets with \pt below 30~\GeV.
The new procedure presented, combined with improvements to the multi-parton interactions model in Monte Carlo simulation, reduces this uncertainty by a factor of seven.
Following this, a new residual calibration is applied, which reduces the effects of pile-up by simultaneously correcting for \avint, \npv, and \pt.
For the absolute MCJES, a new fit method based on splines is used, leading to better closure for jets with \pt below 30~\GeV.
Finally, for the global calibration, which improves the resolution of jets and reduces the difference between the energy scale for quark- and gluon-initiated jets,
a new method using a DNN is used, which allows information from correlated observables to be used for this calibration step.
This DNN results in an average improvement of the JER of around 15\% improvement compared with previous methods, with maximum improvement of over 40\%.

Following these simulation-based calibration steps, the full Run 2 data sample is used to do a residual \insitu calibration to correct the data-MC differences and constrain the uncertainties.
Dijet events are used to calibrate jets in the forward region relative to the central region as a function of jet transverse momentum and pseudorapidity.
The precision is improved by up to a factor of two in the forward detector region at low \pt by evaluating the MC modelling uncertainty at particle level instead of reconstruction level.
Central jets are calibrated by exploiting the balance between jets recoiling against either a photon or a $Z$ boson. Unprecedented precision up to 1\% is achieved
in the \textit{in situ} analysis. For the first time, the energy scale of $b$-tagged jets relative to  inclusive jets is determined with precision
up to 1\% in $\gamma$+jet events. This result is important for improving precision in analysis sensitive to $b-$JES
such as the top quark mass and $H\rightarrow b\bar{b}$ measurements.


\section*{Acknowledgements}


We thank CERN for the very successful operation of the LHC, as well as the
support staff from our institutions without whom ATLAS could not be
operated efficiently.
 
We acknowledge the support of
ANPCyT, Argentina;
YerPhI, Armenia;
ARC, Australia;
BMWFW and FWF, Austria;
ANAS, Azerbaijan;
CNPq and FAPESP, Brazil;
NSERC, NRC and CFI, Canada;
CERN;
ANID, Chile;
CAS, MOST and NSFC, China;
Minciencias, Colombia;
MEYS CR, Czech Republic;
DNRF and DNSRC, Denmark;
IN2P3-CNRS and CEA-DRF/IRFU, France;
SRNSFG, Georgia;
BMBF, HGF and MPG, Germany;
GSRI, Greece;
RGC and Hong Kong SAR, China;
ISF and Benoziyo Center, Israel;
INFN, Italy;
MEXT and JSPS, Japan;
CNRST, Morocco;
NWO, Netherlands;
RCN, Norway;
MEiN, Poland;
FCT, Portugal;
MNE/IFA, Romania;
MESTD, Serbia;
MSSR, Slovakia;
ARRS and MIZ\v{S}, Slovenia;
DSI/NRF, South Africa;
MICINN, Spain;
SRC and Wallenberg Foundation, Sweden;
SERI, SNSF and Cantons of Bern and Geneva, Switzerland;
MOST, Taiwan;
TENMAK, T\"urkiye;
STFC, United Kingdom;
DOE and NSF, United States of America.
In addition, individual groups and members have received support from
BCKDF, CANARIE, Compute Canada and CRC, Canada;
PRIMUS 21/SCI/017 and UNCE SCI/013, Czech Republic;
COST, ERC, ERDF, Horizon 2020 and Marie Sk{\l}odowska-Curie Actions, European Union;
Investissements d'Avenir Labex, Investissements d'Avenir Idex and ANR, France;
DFG and AvH Foundation, Germany;
Herakleitos, Thales and Aristeia programmes co-financed by EU-ESF and the Greek NSRF, Greece;
BSF-NSF and MINERVA, Israel;
Norwegian Financial Mechanism 2014-2021, Norway;
NCN and NAWA, Poland;
La Caixa Banking Foundation, CERCA Programme Generalitat de Catalunya and PROMETEO and GenT Programmes Generalitat Valenciana, Spain;
G\"{o}ran Gustafssons Stiftelse, Sweden;
The Royal Society and Leverhulme Trust, United Kingdom.
 
The crucial computing support from all WLCG partners is acknowledged gratefully, in particular from CERN, the ATLAS Tier-1 facilities at TRIUMF (Canada), NDGF (Denmark, Norway, Sweden), CC-IN2P3 (France), KIT/GridKA (Germany), INFN-CNAF (Italy), NL-T1 (Netherlands), PIC (Spain), ASGC (Taiwan), RAL (UK) and BNL (USA), the Tier-2 facilities worldwide and large non-WLCG resource providers. Major contributors of computing resources are listed in Ref.~\cite{ATL-SOFT-PUB-2021-003}.


\clearpage

\printbibliography
 
\clearpage
\input{atlas_authlist}

\end{document}

%% file: atlas_authlist.tex
 
\begin{flushleft}
\hypersetup{urlcolor=black}
{\Large The ATLAS Collaboration}

\bigskip

\AtlasOrcid[0000-0002-6665-4934]{G.~Aad}$^\textrm{\scriptsize 102}$,
\AtlasOrcid[0000-0002-5888-2734]{B.~Abbott}$^\textrm{\scriptsize 120}$,
\AtlasOrcid[0000-0002-1002-1652]{K.~Abeling}$^\textrm{\scriptsize 55}$,
\AtlasOrcid[0000-0001-5763-2760]{N.J.~Abicht}$^\textrm{\scriptsize 49}$,
\AtlasOrcid[0000-0002-8496-9294]{S.H.~Abidi}$^\textrm{\scriptsize 29}$,
\AtlasOrcid[0000-0002-9987-2292]{A.~Aboulhorma}$^\textrm{\scriptsize 35e}$,
\AtlasOrcid[0000-0001-5329-6640]{H.~Abramowicz}$^\textrm{\scriptsize 151}$,
\AtlasOrcid[0000-0002-1599-2896]{H.~Abreu}$^\textrm{\scriptsize 150}$,
\AtlasOrcid[0000-0003-0403-3697]{Y.~Abulaiti}$^\textrm{\scriptsize 117}$,
\AtlasOrcid[0000-0003-0762-7204]{A.C.~Abusleme~Hoffman}$^\textrm{\scriptsize 137a}$
% ...
\bigskip
\\

% ...
$^{29}$Physics Department, Brookhaven National Laboratory, Upton NY; United States of America.\\
% ...
$^{35}$$^{(a)}$Facult\'e des Sciences Ain Chock, R\'eseau Universitaire de Physique des Hautes Energies - Universit\'e Hassan II, Casablanca;$^{(b)}$Facult\'{e} des Sciences, Universit\'{e} Ibn-Tofail, K\'{e}nitra;$^{(c)}$Facult\'e des Sciences Semlalia, Universit\'e Cadi Ayyad, LPHEA-Marrakech;$^{(d)}$LPMR, Facult\'e des Sciences, Universit\'e Mohamed Premier, Oujda;$^{(e)}$Facult\'e des sciences, Universit\'e Mohammed V, Rabat;$^{(f)}$Institute of Applied Physics, Mohammed VI Polytechnic University, Ben Guerir; Morocco.\\
% ...
$^{49}$Fakult\"{a}t Physik , Technische Universit{\"a}t Dortmund, Dortmund; Germany.\\
% ...
$^{55}$II. Physikalisches Institut, Georg-August-Universit\"{a}t G\"ottingen, G\"ottingen; Germany.\\
% ...
$^{102}$CPPM, Aix-Marseille Universit\'e, CNRS/IN2P3, Marseille; France.\\
% ...
$^{117}$Department of Physics, New York University, New York NY; United States of America.\\
% ...
$^{120}$Homer L. Dodge Department of Physics and Astronomy, University of Oklahoma, Norman OK; United States of America.\\
% ...
$^{137}$$^{(a)}$Departamento de F\'isica, Pontificia Universidad Cat\'olica de Chile, Santiago;$^{(b)}$Millennium Institute for Subatomic physics at high energy frontier (SAPHIR), Santiago;$^{(c)}$Instituto de Investigaci\'on Multidisciplinario en Ciencia y Tecnolog\'ia, y Departamento de F\'isica, Universidad de La Serena;$^{(d)}$Universidad Andres Bello, Department of Physics, Santiago;$^{(e)}$Instituto de Alta Investigaci\'on, Universidad de Tarapac\'a, Arica;$^{(f)}$Departamento de F\'isica, Universidad T\'ecnica Federico Santa Mar\'ia, Valpara\'iso; Chile.\\
% ...
$^{150}$Department of Physics, Technion, Israel Institute of Technology, Haifa; Israel.\\
$^{151}$Raymond and Beverly Sackler School of Physics and Astronomy, Tel Aviv University, Tel Aviv; Israel.\\
% ...
$^{a}$ Also Affiliated with an institute covered by a cooperation agreement with CERN.\\
$^{b}$ Also at An-Najah National University, Nablus; Palestine.\\
$^{c}$ Also at APC, Universit\'e Paris Cit\'e, CNRS/IN2P3, Paris; France.\\
$^{d}$ Also at Borough of Manhattan Community College, City University of New York, New York NY; United States of America.\\
$^{e}$ Also at Center for High Energy Physics, Peking University; China.\\
$^{f}$ Also at Center for Interdisciplinary Research and Innovation (CIRI-AUTH), Thessaloniki ; Greece.\\
% ...

\end{flushleft}
